\documentclass{article}

\usepackage[english]{babel}
\usepackage{bm}

\usepackage[letterpaper,top=2cm,bottom=2cm,left=3cm,right=3cm,marginparwidth=1.75cm]{geometry}

\usepackage[utf8]{inputenc}
\usepackage{amsmath}
\usepackage{amssymb}
\usepackage{graphicx}
\usepackage[colorlinks=true, allcolors=blue]{hyperref}
\usepackage{upgreek}

\usepackage{natbib}
\usepackage{subcaption}
\DeclareCaptionSubType * [alph]{table}
\usepackage{setspace}
\usepackage[acronym]{glossaries}
\glsdisablehyper
\newacronym{mfpca}{MFPCA}{multivariate functional principal component analysis}
\newacronym[longplural={functional principal component analyses}]{fpca}{FPCA}{functional principal component analysis}
\newacronym{mfpc}{MFPC}{multivariate functional principal component}
\newacronym{fpc}{FPC}{functional principal component}
\newacronym{iid}{i.i.d.}{independent and identically distributed}
\newacronym{gamm}{GAMM}{generalized additive mixed model}
\newacronym{bamlss}{BAMLSS}{Bayesian additive models for location, scale, and shape}
\newacronym{mcmc}{MCMC}{Markov Chain Monte Carlo}
\newacronym{kl}{KL}{Karhunen-Lo\`eve}
\newacronym{famm}{FAMM}{functional additive mixed model}
\newacronym{rrmse}{rrMSE}{root relative mean squared error}
\newacronym{mse}{MSE}{mean squared error}

\title{Generalized Multivariate Functional Additive Mixed Models for Location, Scale, and Shape}
\author{Alexander Volkmann, Nikolaus Umlauf, Sonja Greven}

\begin{document}
\maketitle

\begin{abstract}
We propose a flexible regression framework to model the conditional distribution of multilevel generalized multivariate functional data of potentially mixed type, e.g.\ binary and continuous data. We make pointwise parametric distributional assumptions for each dimension of the multivariate functional data and model each distributional parameter as an additive function of covariates. The dependency between the different outcomes and, for multilevel functional data, also between different functions within a level is modelled by shared latent multivariate Gaussian processes. For a parsimonious representation of the latent processes, (generalized) multivariate functional principal components are estimated from the data and used as an empirical basis for these latent processes in the regression framework. Our modular two-step approach is very general and can easily incorporate new developments in the estimation of functional principal components for all types of (generalized) functional data. Flexible additive covariate effects for scalar or even functional covariates are available and are estimated in a Bayesian framework. We provide an easy-to-use implementation in the accompanying \texttt{R} package \textbf{gmfamm} on CRAN and conduct a simulation study to confirm the validity of our regression framework and estimation strategy. The proposed multivariate functional model is applied to four dimensional traffic data in Berlin, which consists of the hourly numbers and mean speed of cars and trucks at different locations.
\end{abstract}

\section{Introduction}

Repeated measurements of variables of interest are becoming more easily available to researchers, with one observation often combining several variables of different characteristics such as binary, count, or continuous data. A range of different approaches are available to model such multivariate longitudinal data of mixed types with \cite{verbeke2014analysis} giving an excellent overview and current literature working on improving or extending these approaches, e.g.\ \cite{li2017joint, rustand2023fast, kock2023truly}. Given that the data can often be measured on a nearly continuous, fine grid, such as for data obtained from wearable devices \citep{leroux2023fast}, or the repeated measurements can represent non-temporal dependencies, such as tract distance for diffusion tensor imaging \citep{cao2019modeling}, understanding these data as generated from an underlying smooth multivariate process can often be insightful. Viewing the repeated scalar measurements as realizations of a multivariate functional observation over some domain $\mathcal{I}$ then allows to compare functions and identify differences in their functional forms,  or analyze variation in the functional data sample, see e.g.\ \cite{ramsay2005} for an introduction to functional data analysis. We develop a flexible additive regression approach for such (generalized) multivariate functional data, which can be of mixed type, exhibit a multilevel structure, and are potentially only sparsely observed.

When the underlying continuous process is not Gaussian for each $t \in \mathcal{I}$, however, the analysis of functional data becomes more intricate. Previous approaches in univariate non-Gaussian functional data analysis for \gls{fpca} \citep{hall2008modelling, van2009bayesian, serban2013multilevel, gertheiss2017note, leroux2023fast, ye2023functional, weishampel2023classification, zhou2023analysis}, registration \citep{wrobel2019registration, bauer2021registration}, or functional regression \citep{chen2013marginal, goldsmith2015generalized, scheipl2016generalized} typically assume a latent continuous (often Gaussian) process related to the smooth mean of the non-Gaussian functional data via a known link function. Other authors have proposed to use copulas to model the temporal dependence \citep{staicu2012modeling, dey2023covariance} or approaches more robust against distributional assumptions building on Kendall's $\tau$ \citep{zhong2022functional, zhong2022robust}. For generalized multivariate functional data of mixed type, however, few approaches exist. \cite{huang2014joint} and \cite{li2014hierarchical} model bivariate functional data with exponential family distributions using known link functions and correlated random effects. Such an approach can become computationally quite expensive when a higher number of functional outcomes are jointly analyzed. \cite{li2018three} extend this approach by including an ordinal component to model different distributional assumptions for the bivariate functional outcome depending on the individual's activity pattern. \cite{cao2019modeling} apply a Gaussian copula model to skewed continuous functional data but do not provide an extension to discrete functional responses. \cite{jiang2022analysis} propose to model multivariate non-Gaussian continuous functional data through a shared univariate Gaussian process in a normal transformation linear model. While their semiparametric approach does not assume given link functions or covariance structures, the dependency between the multivariate functional data is restricted to one shared latent factor and the approach does not extend to discrete, e.g.\ binary, functional data.

We propose to use a multivariate latent Gaussian process to model the dependency within and between the functional dimensions of a generalized multivariate functional observation of possibly mixed type. Our approach is similar to \cite{tidemann2016modeling}, who suggest to model (bivariate) binary and continuous multivariate functional data by either using a prespecified basis such as B-splines for the latent process or to estimate multivariate eigenfunctions from the data in a preliminary step. Their approach, however, in addition to only considering binary and continuous data, requires the estimation of all auto- and cross-covariance functions of the multivariate latent Gaussian process. This can become computationally intensive quite quickly when the multivariate functional data contain several dimensions, and cannot be applied to functional data that feature additional dependencies such as a multilevel structure. Instead, we show that the analysis can benefit from recent developments in \gls{mfpca}. While most work has focused on advancing \gls{mfpca} for Gaussian or continuous functional data, e.g.\ \cite{happ2018multivariate, li2020fast,jiang2022bayesian, song2022sparse, nolan2023efficient, tengteng2023estimation, zhang2023interpretable}, we are not aware of an \gls{mfpca} approach dedicated to generalized multivariate functional data of mixed type. Notably, however, \cite{happ2018multivariate} provide a flexible framework allowing the estimation of the \glspl{mfpc} from separate univariate \glspl{fpca}.  This avoids the computationally intensive and potentially difficult explicit estimation of cross-covariance functions. \cite{lim2020generalization} have shown that the \gls{mfpca} approach proposed by \cite{happ2018multivariate} can be successfully applied to multivariate count functional data by assuming a latent Gaussian process. We extend this idea to multivariate functional data of mixed type, allowing us to build on and combine the wide range of \gls{fpca} methods for univariate non-Gaussian functional data.

Combining the \gls{mfpca} approach of \cite{happ2018multivariate} with the latent multivariate Gaussian process assumption allows us to establish a general and flexible approach to analyzing generalized multivariate functional data of mixed type with a potential multilevel structure. For the pointwise distributional assumptions of the generalized functional outcomes, we build on \gls{bamlss} \citep{umlauf2018bamlss}, which opens up the analysis to a wide range of parametric distributions \citep{rigby2020distributions}, suitable for continuous, binary, count, and discrete functional data. This also allows to flexibly model all distributional parameters such as mean, variance, or skewness parameters as functions of covariates. Building on the \glspl{famm} established by \cite{scheipl2015functional}, our generalized multivariate functional regression framework  comprises possibly nonlinear effects or smooth interaction terms of scalar or functional covariates -- possibly differing by dimension -- as implemented in the well developed additive model framework \citep{wood2017generalized}. As a result, the model proposed here extends the multivariate functional additive mixed model for multivariate Gaussian functional data introduced by \cite{volkmann2023multivariate} in two ways: the Gaussian assumption can be replaced by a more suitable distribution if needed and all distributional parameters instead of only the mean can be modeled by additive predictors. We employ a Bayesian estimation strategy to efficiently handle the (potentially) large number of parameters to estimate. Compared to existing approaches, our proposed model is thus more general  with respect to 
\begin{enumerate}
    \item[a)]the possible distributions and their combination for the different dimensions of generalized multivariate functional outcomes,
    \item[b)] the possible dependence structure between the dimensions as well as between potentially multilevel observations,
\end{enumerate}
and more flexible with respect to 
\begin{enumerate}
    \item[c)] the modeling options of covariate information, including functional random effects, nonlinear additive, or smooth interaction effects, as well as
    \item[d)] flexible effects on other distributional parameters than the mean.
\end{enumerate}

We showcase our approach in an analysis of data on traffic flows in the German capital. Scattered across the municipal area of Berlin, there is a large number of traffic detectors measuring the hourly number and mean speed of vehicles passing the detectors, giving a rich data base on traffic movements and changes since 2015. In our analysis, we focus on commuter traffic into the city center of Berlin and investigate seasonality and changes over the recorded years. The outcome of interest is a four dimensional generalized functional observation corresponding to the number and mean speed of cars and trucks measured over the last hour. Modeling these count and (potentially skewed) positive continuous outcomes together gives a more holistic understanding of the processes involved than would be obtained with univariate analyses. For example, the mean speed of cars might be closely connected to the number of cars passing the detectors as well as the number of (often slower) trucks on the road \citep{kock2023truly}.

In the following, Section \ref{sec:Model} introduces the proposed generalization of multivariate functional additive mixed models, while Section \ref{sec:Estimation} covers the estimation approach. Section \ref{sec:Simulation} then illustrates the suitability of the approach in a simulation study and Section \ref{sec:Application} presents the analysis of the Berlin traffic data. Section \ref{sec:Discussion} closes with a discussion and outlook.

\section{Model}
\label{sec:Model}

Let $\bm{Y}_i(t) = \left(Y_i^{(1)}(t), ..., Y_i^{(K)}(t)\right)^{\top}, i = 1,..., n$ be a (generalized) multivariate functional observation of $K$ dimensions gathering $K$ univariate functions $Y_i^{(k)}(t), k = 1,.., K, t \in \mathcal{I}$ on the common bounded interval $\mathcal{I}$. At each point $t$, we assume that given some covariate set $\mathcal{X}_i$ and some set of latent processes $\Uplambda_i$, the observed $Y_{it}^{(k)} = Y_i^{(k)}(t)$ are conditionally independent and follow a parametric distribution $\mathcal{D}^{(k)}$ with $R^{(k)}$ distributional parameters $\theta_r^{(k)}(t), r = 1,..., R^{(k)}$ depending on $\mathcal{X}_i$ and $\Uplambda_i$, so that 
\begin{equation}
    \begin{gathered}
        Y_{it}^{(k)} \mid \mathcal{X}_i, \Uplambda_i \sim \mathcal{D}^{(k)}\left(\theta_{i1}^{(k)}(t), ..., \theta_{iR^{(k)}}^{(k)}(t)\right),\\
    \theta_{ir}^{(k)}(t) = h_{r}^{(k)}\left(\eta_{ir}^{(k)}(t)\right).
    \end{gathered}
    \label{eq:DistributionalAss}
\end{equation}
The distributional parameters are linked to additive predictors $\eta_{ir}^{(k)}(t)$ containing the information on $\mathcal{X}_i$ and $\Uplambda_i$ using known monotonic and twice differentiable functions $h_r^{(k)}$. For notational convenience, we drop the conditional dependency on $\mathcal{X}_i$ and $\Uplambda_i$. We then assume that the $\theta_{ir}^{(k)}(t) \in L^2(\mathcal{I})$ are smooth functions over $t$ and that the first distributional parameter $\theta_{i1}^{(k)}(t)$ is related to the location of the distribution. Note that most common distributions such as the normal, Bernoulli, or Poisson distributions can be parameterized so that  $\theta_{i1}^{(k)}(t)$ corresponds to the conditional mean $\mathbb{E}(Y_i^{(k)}(t) \mid \mathcal{X}_i, \Uplambda_i)$. The introduced notation thus allows us to gather $K$ univariate generalized functions of (potentially) mixed type, that is with different pointwise distributional assumptions, in $\bm{Y}_i(t)$. 

We assume that the dependency (conditional on $\mathcal{X}_i$) between the univariate functions is fully captured by the set of latent processes $\Uplambda_i$ and is limited to the location parameters $\theta_{i1}^{(1)},..., \theta_{i1}^{(K)}$. In other words, distributional parameters $\theta_{i2}^{(k)}, ..., \theta_{iR^{(k)}}^{(k)}, k = 1,...,K$ such as scale parameters related to the spread of the pointwise distributions are not connected to distributional parameters of other dimensions, except via shared covariates in $\mathcal{X}_i$. For the location parameters, however, we allow dependence via independent replicates of a latent centered multivariate Gaussian process $\bm{\Lambda}_{0i}(t) = (\Lambda_{0i}^{(1)}(t),..., \Lambda_{0i}^{(K)}(t))^{\top} \sim \mathcal{GP}(\bm{0}(t), \bm{\mathcal{K}}_0(t, \cdot))$ for each $i = 1,..., n$ with zero-mean function $\bm{0}(t)$ and covariance kernel $\bm{\mathcal{K}}_0(t, \cdot)$. The univariate functions $\Lambda_{0i}^{(k)}(t)\in L^2(\mathcal{I})$ are square-integrable functions  with inner product $\langle f^{(k)}, g^{(k)}\rangle = \int_{\mathcal{I}}f^{(k)}(t) g^{(k)}(t)dt$ for $f^{(k)}, g^{(k)}\in L^2(\mathcal{I})$, and $\bm{\Lambda}_{0i}(t) \in L_K^2(\mathcal{I}) = L^2(\mathcal{I}) \times ... \times L^2(\mathcal{I})$, which is a Hilbert space with scalar product $\langle \langle \bm{f}, \bm{g}\rangle \rangle = \sum_{k = 1}^{K} w_k \langle f^{(k)}, g^{(k)}\rangle$ for $\bm{f},\bm{g} \in L_K^2(\mathcal{I})$ with positive weights $w_k$ typically being 1. The associated covariance operator $\bm{\Gamma}_0$ allows the decomposition of the covariance functions via Mercer's theorem as $Cov(\Lambda_{0i}^{(k)}(s), \Lambda_{0i}^{(k')}(t)) = \sum_{m=1}^{\infty}\nu_{0m}\psi_{0m}^{(k)}(s)\psi_{0m}^{(k')}(t), s,t \in \mathcal{I}, k,k' = 1,..., K$ with $\nu_{0m}$ the eigenvalues and $\bm{\psi}_{0m}(t) = (\psi_{0m}^{(1)}(t),..., \psi_{0m}^{(K)}(t))^\top$ the eigenfunctions of $\bm{\Gamma}_0$. We denote the covariance functions as cross-covariances for $k \neq k'$ and auto-covariances for $k = k'$. 

For data with more complex sampling regimes such as nested or crossed study designs, the set of latent processes $\Uplambda_i$ and thus the model for $\theta_{i1}^{(k)}$ can be extended by additionally including mutually independent multivariate Gaussian processes $\bm{\Lambda}_{u}(t) \sim  \mathcal{GP}(\bm{0}(t), \bm{\mathcal{K}}_{u}(t, \cdot)), \bm{\Lambda}_{u}(t)\in L_K^2(\mathcal{I}), u = 1,..., U$ as in \cite{volkmann2023multivariate}. Each $\bm{\Lambda}_{u}(t)$ for a given $u$ then represents a grouping layer such as a hospital in a multicenter study with patients in different hospitals, capturing dependency between multivariate functional observations of patients in the same hospital and allowing a multilevel structure. For each level, e.g. hospital $j = 1,..., J_u$ in the grouping layer $u$, we assume an \gls{iid} copy of the multivariate Gaussian process $\bm{\Lambda}_{u}(t)$ denoted as $\bm{\Lambda}_{uj}(t)$ and introduce an indicator $z_{uij}$ which is $1$ if observation $i$ belongs to level $j$ of grouping layer $u$ and is $0$ otherwise. This allows us to account for similarities within the grouping levels and capture variation across groups.

We set up the following additive model structure
\begin{equation}
\begin{gathered}
    \eta_{i1}^{(k)}(t) = \sum_{l = 1}^{L_1^{(k)}} f_{1l}^{(k)}(\mathcal{X}_i, t) + \Lambda_{0i}^{(k)}(t) + \sum_{u = 1}^{U}\sum_{j=1}^{J_u}\Lambda_{uj}^{(k)}(t)z_{uij}, \\
    \eta_{ir}^{(k)}(t) = \sum_{l = 1}^{L_r^{(k)}} f_{rl}^{(k)}(\mathcal{X}_i, t), \quad r =2,..., R^{(k)}
\end{gathered}
\label{eq:additive_eta}
\end{equation}
with additive functions $f_{rl}^{(k)}(\mathcal{X}_i, t), l = 1,..., L_r^{(k)}, r = 1,...,R^{(k)}$ flexibly modeling the additive predictor $\eta_{ir}^{(k)}(t)$ depending on the covariate set $\mathcal{X}_i$ and $\Uplambda_i$. The number of additive functions $L_r^{(k)}$ can differ for each distributional parameter $r$ of each univariate functional outcome $k$ and is chosen by the analyst. We use the general notation $f_{rl}^{(k)}(\mathcal{X}_i, t)$ to indicate that each additive term can depend on single covariates in $\mathcal{X}_i$ as well as consist of interactions between subsets of $\mathcal{X}_i$ and can represent e.g.\ (functional) intercepts, linear effect terms, smooth (interaction) effects, or even more complex model terms such as spatio-temporal effects, see \cite{umlauf2018bamlss}. This flexibility is achieved by using basis function representations
\begin{align}
\label{eq:Basisexpansion}
    f_{rl}^{(k)}(\mathcal{X}_{i}, t) = \bm{b}_{rl}^{(k)}(\mathcal{X}_{i}, t)^\top\bm{\beta}_{rl}^{(k)}
\end{align}
to model potentially nonlinear covariate effect terms. Here, $\bm{b}_{rl}^{(k)}(\mathcal{X}_{i}, t)$ 
denotes a vector of $d_{rl}^{(k)}$ known basis functions evaluated at the included subset of 
covariates of $\mathcal{X}_{i}$ and $t$, and $\bm{\beta}_{rl}^{(k)}$ is the corresponding coefficient 
vector (see Section \ref{subsubsec:TensorProds}). The latent multivariate Gaussian process
$\bm{\Lambda}_{0i}(t)$ is represented using a truncated multivariate \gls{kl} expansion with
$k$-th dimension
\begin{align}
\label{eq:KLexpansion}
    \Lambda_{0i}^{(k)}(t) \approx \sum_{m = 1}^{M_0} \rho_{0im}\psi_{0m}^{(k)}(t),
\end{align}
where $M_0$ denotes the number of eigenfunctions $\bm{\psi}_{0m}(t) = (\psi_{0m}^{(1)}(t),..., \psi_{0m}^{(K)}(t))^\top$ in the approximation and $\rho_{0im}$ is the $m$-th random score for subject $i$ shared across the $K$ dimensions. Note that for $M_0 \to \infty$, this approximation would become exact. As $\bm{\Lambda}_{0i}(t)$ is a Gaussian process, the random scores are \gls{iid} $\bm{\rho}_{0i} = (\rho_{0i1},...\rho_{0iM})^{\top} \sim N(\bm{0}, diag(\nu_{01}, ..., \nu_{0M}))$ with $\bm{0}$ a vector of zeros and $diag(\cdot)$ a diagonal matrix. Like the number of basis functions $d_{rl}^{(k)}$, the truncation order $M_0$ is chosen by the analyst, depending on the data problem at hand. We will provide a guideline for setting $M_0$ in Section \ref{sec:Estimation}. Similarly, the additional $U$ random grouping effects 
\begin{align*}
    \Lambda_{uj}^{(k)}(t) \approx \sum_{m = 1}^{M_u} \rho_{ujm}\psi_{um}^{(k)}(t), \quad k=1,...,K,
\end{align*}
are also represented using finite \gls{kl} expansions with individual truncation orders $M_u$.

While the latent processes in $\Uplambda_i$ could also be represented using rich basis function representations giving functional random effects similar in structure to \eqref{eq:Basisexpansion}, the finite \gls{kl} expansion has several advantages. The leading eigenfunctions $\bm{\psi}_{0m}(t)$ capture the directions of most variation in the random process, which not only generates insight into the variation of the functional data and provides interpretable basis functions, but also allows to intuitively quantify the goodness of fit of the \gls{kl} approximation via the ratio of explained variation. In fact, the finite \gls{kl} expansion can be seen as an optimal rank $M_0$ approximation with respect to the integrated squared error \citep[cf.][]{ramsay2005}. The finite \gls{kl} expansion also constitutes a parsimonious basis representation as the eigenfunctions intrinsically model the dependency between the $K$ different outcomes. This drastically reduces the number of parameters to estimate given that the scalar scores $\rho_{0im}$ serve as weights for multivariate basis functions and are therefore not dimension-specific. Furthermore, the \gls{kl} theorem \citep[cf.][]{happ2018multivariate} allows the assumption of independent, or for non-Gaussian random processes at least uncorrelated, random scores. Using an alternative basis representation \eqref{eq:Basisexpansion} would require the estimation of $\sum_{k=1}^{K}d_{1l}^{(k)}$ correlated random basis coefficients and their (co)variances, which is clearly more computationally expensive than estimating the (typically much) lower number $M$ of independent random scores, i.e.\ with diagonal covariance matrix. Note that in practice, however, the eigenfunctions are generally unknown and have to be estimated as well.

\section{Estimation Approach}
\label{sec:Estimation}

We simplify the estimation of covariate and latent effects in model \eqref{eq:DistributionalAss} and \eqref{eq:additive_eta} by applying a two-step approach, where the first step estimates the eigenfunctions using \gls{mfpca} and with these as a basis, the second step estimates the covariate effects and random scores. Few approaches exist which use a \gls{kl} expansion for the dependency structure and simultaneously estimate both eigenfunctions and scores in a reduced rank model, e.g.\ \cite{goldsmith2015generalized, jiang2022bayesian}. However, they do not allow multivariate non-Gaussian functional data with potentially mixed types and are computationally intensive. On the other hand, previous work such as \cite{tidemann2016modeling, volkmann2023multivariate}, and \cite{volkmann2023flexible} has shown that estimating the multivariate eigenfunctions in a preliminary step is a practicable way to obtain a parsimonious basis representation replacing other common basis choices such as splines, yielding good results in fitting even non-standard multivariate functional data.

We emphasize the modular nature of our approach: Estimating the \gls{mfpc} basis for the latent processes is detached from the second step of estimating the generalized functional additive model. This allows the user to select and combine suitable \gls{fpca} methods from a wide toolbox, which can be subsequently updated as new methods are proposed in the literature. In particular, while direct estimation of \glspl{mfpc} has been proposed \citep[e.g.][]{tidemann2016modeling} and could be used in the first step, we exploit the flexibility of the \gls{mfpca} approach proposed by \cite{happ2018multivariate}. They estimate the \glspl{mfpc} based on univariate scores and eigenfunctions obtained via univariate \glspl{fpca}. The specific estimation methods used for the univariate \glspl{fpca}, however, are a matter of choice and can differ between the different dimensions with different distributions, thus further modularizing the estimation process. This has the advantage that advances made for univariate (generalized) \gls{fpca} can be directly transferred to the estimation of \glspl{mfpc} such as improvements of computational efficiency or accounting for multilevel structure, e.g.\ \cite{cui2023fast, zhou2023analysis}. Consequently, we present a schematic outline of the estimation of the eigenfunctions in the following, while Sections \ref{sec:Simulation} and \ref{sec:Application} will offer more detailed accounts of the methods applied in the simulations and the application.

\subsection{Estimation of Eigenfunctions}

\cite{happ2018multivariate} establish a relationship between the finite multivariate \gls{kl} representation of $\bm{\Lambda}_{0i}(t)$ given by \eqref{eq:KLexpansion} and the finite univariate \gls{kl} representations $\Lambda_{0i}^{(k)}(t) = \sum_{m' = 1}^{M_0^k}\xi_{0im'}^{(k)}\phi_{0m'}^{(k)}(t)$ for all $k = 1,...,K$, where $\xi_{0im'}^{(k)}$ are the univariate scores and $\phi_{0m'}^{(k)}(t)$ are the univariate eigenfunctions of the auto-covariance operator $\Gamma_0^{(k)}$ with auto-covariance kernel $\mathcal{K}_0^{(k,k)}(t, \cdot)$, allowing the representation $Cov(\Lambda_{0i}^{(k)}(s), \Lambda_{0i}^{(k)}(t)) = \sum_{m'=1}^{\infty}\upsilon_{0m'}^{(k)}\phi_{0m'}^{(k)}(s)\phi_{0m'}^{(k)}(t), s,t \in \mathcal{I}$ (Mercer's theorem) with $\upsilon_{0m'}^{(k)}$ the corresponding univariate eigenvalues. Note that the univariate eigenfunctions of dimension $k$ are orthonormal with respect to the univariate scalar product $\langle\cdot,\cdot\rangle$ and the multivariate eigenfunctions are orthonormal with respect to $\langle\langle\cdot,\cdot\rangle\rangle$. This relationship between univariate and multivariate \gls{kl} representation then allows to estimate the multivariate eigenfunctions $\bm{\psi}_{0m}(t), m = 1,...,M$ based on the $K$ sets of univariate eigenfunctions $\phi_{0m'}^{(k)}(t), m' = 1,...,M_0^k$, using an explicit formula in \cite{happ2018multivariate} and the covariance between the univariate scores $\xi_{0im'}^{(k)}$. \cite{volkmann2023multivariate} have shown that this approach can also be applied to each mutually independent latent process in the set $\Uplambda_i$.

\subsubsection{Estimation of Univariate Eigenfunctions}
Separate univariate (generalized) \glspl{fpca} on each dimension offer estimates of the univariate eigenfunctions, which we denote as univariate \glspl{fpc} $\hat{\phi}_{m'}^{(k)}(t)$. A wide range of different approaches for univariate (generalized) \gls{fpca} exist, which are applicable for different distributional assumptions and study designs. We focus here on a branch of research that provides some of the most general, computationally efficient, and well implemented \gls{fpca} approaches currently available. In particular,
\cite{xiao2016fast} provide a computationally efficient eigendecomposition of the covariance matrix of Gaussian functional data observed on a regular grid, with \cite{cui2023fast} extending the approach to multilevel data for a nested study design (functional observations as repeated visits of independent subjects). For this special case of sampling regimes contained in \eqref{eq:additive_eta}, \cite{cui2023fast} derive estimates for the between-subject covariance $\mathcal{K}_1^{(k,k)}(t, \cdot)$ and the within-subject covariance $\mathcal{K}_0^{(k,k)}(t, \cdot)$, respectively, and apply the fast covariance estimation approach of \cite{xiao2016fast} for computationally efficient estimates of the corresponding univariate eigenfunctions. \cite{leroux2023fast} and \cite{zhou2023analysis} generalize these fast \gls{fpca} approaches to non-Gaussian functional data by first binning the data and estimating scalar generalized mixed models for each bin. The estimated random effects are then treated as noisy and discretized realizations of the latent processes and serve as starting points for the fast \gls{fpca} approaches by \cite{xiao2016fast} and \cite{cui2023fast}. Binning the data has the additional advantage that a common grid of scalar observation points is created, which allows the handling of sparse and irregularly sampled functional data. On the other hand, \cite{leroux2023fast} and \cite{zhou2023analysis} point out that the resulting score estimates are biased due to the binning process. As a result, they propose to re-estimate the scores in a global generalized functional mixed model over the entire functional domain using the estimated \glspl{fpc} as basis functions. Adjusting for covariate information in this estimation approach of the univariate eigenfunctions is straightforward, as additive effects analogous to the multivariate functional regression models can be included in both the scalar generalized mixed models for each bin, dropping potential dependency on $t$ in these pointwise models, as well as the global generalized functional mixed models.

\subsubsection{Estimation of Multivariate Eigenfunctions}

With estimates of the univariate \glspl{fpc} and scores given, the multivariate eigenfunctions and eigenvalues can be estimated using the (weighted) covariance between the univariate scores. Let $\bm{\Xi}_0$ be an $n \times M_0^{+}$ matrix with rows $(\xi_{0i1}^{(1)}, \dots, \xi_{0iM_0^{(1)}}^{(1)}, \dots, \xi_{0i1}^{(K)}, \dots, \xi_{0iM_0^{(K)}}^{(K)})$ containing the univariate scores of observation $i$, where $M_0^{+} = \sum_{k=1}^{K}M_0^{(k)}$ is the total number of estimated univariate \glspl{fpc}. Note that the number of rows $n$ of $\bm{\Xi}_0$ is replaced by the number of grouping levels for the \gls{mfpca} of other latent processes in $\Uplambda_i$. We perform a matrix eigenanalysis for the matrix $(n-1)^{-1}\bm{D}\bm{\Xi}_0^{\top}\bm{\Xi}_0\bm{D}$ with multivariate scalar product weights contained in the diagonal matrix $\bm{D} = diag(\bm{w}_{1}^{1/2},...,\bm{w}_{K}^{1/2})$ and $\bm{w}_{k} = (w_{k}, ... ,w_{k}) \in \mathbb{R}^{M_0^{(k)}}$ to obtain eigenvectors $\hat{\bm{c}}_{0m}$ with corresponding eigenvalues $\hat{\nu}_{0m}$, $m = 1,..., M_0^{+}$. The multivariate eigenfunctions are then estimated as linear combination $\hat{\psi}_{0m}^{(k)}(t) = \sum_{m' = 1}^{M_0^{(k)}}[\hat{\bm{c}}_{0m}]_{m'}\hat{\phi}_{0m'}^{(k)}(t)$, where $[\hat{\bm{c}}_{0m}]_{m'}$ denotes the $m'$-th element of eigenvector $\hat{\bm{c}}_{0m}$ \citep{happ2018multivariate}.

Given that the number $M_0^{+}$ of \glspl{mfpc} that can be estimated using the approach by \cite{happ2018multivariate} directly depends on the number of univariate \glspl{fpc}, we suggest to preserve as much information as possible in the univariate \glspl{fpca} and choose high univariate truncation orders (if necessary) accordingly \cite[compare e.g.][]{golovkine2023estimation}. Furthermore, we recommend to consider the estimated univariate eigenvalues before perfoming the \gls{mfpca}. The sum $\sum_{m = 1}^{M^{(k)}}\upsilon_{0m}^{(k)}$ provides an estimate of the variation of the respective process on each dimension and thus reflects the contribution of the dimension to the overall variation in the data. When these contributions differ greatly between the dimensions, the leading \glspl{mfpc} can be dominated by single dimensions, capturing little of the covariance between the dimensions. If the full set of estimated \glspl{mfpc} is used as a basis, this is inconsequential as we are not primarily concerned with the interpretation of the \glspl{mfpc}. However, if the basis is truncated for a more parsimonious model, such imbalances might lead to considerably poorer model fits for some dimensions. An alternative to standardizing the data before the analysis is to use unequal weights in the definition of the multivariate scalar product \cite[compare][]{happ2018multivariate, volkmann2023multivariate}. We propose to use the inverse sum of univariate eigenvalues as weights to increase the contributions of dimensions with lower overall variation in the estimation of the \glspl{mfpc}. This weighting scheme can also be applied to each random process separately, giving $U + 1$ differently defined weighted scalar products and process-specific diagonal weight matrices $\bm{D}_u, u = 0,..., U$. We argue that, although this complicates the comparison of eigenfunctions across processes, the main focus is to construct an empirical, parsimonious basis which represents the variation in different dimensions equally well.

As the resulting \glspl{mfpc} are ordered with respect to the (weighted) covariance they explain, we can additionally introduce a truncation order $M < M^{+}$ to decrease the number of scores to be estimated in the full model. This ensures that the main modes of variation are accounted for in the model but noisier, higher order modes of variation are excluded, offering an avenue for regularization of the model fit as well as increased computational efficiency. Different approaches are possible such as defining an elbow-criterion for the eigenvalues or using the amount of explained total variation in the data to choose the truncation order. For example, \cite{volkmann2023multivariate} propose to use a truncation order that provides a prespecified amount of univariate variation explained on each dimension or, alternatively, truncating based on the amount of total multivariate variation explained, accumulating the leading \glspl{mfpc} across all $U + 1$ processes. In extreme cases, this can lead to considerable model simplifications if no \gls{mfpc} is included for a process, dropping the entire process from model. To avoid such a behavior and assure that all random processes are adequately represented in the model, we instead propose to keep a prespecified amount of total multivariate variation explained based on the estimated $\hat{\nu}_{um}$ for each process $u$ separately, leading to process specific truncation orders $M_u, u = 0,...,U$.

\subsection{Estimation of the Generalized Functional Additive Model}
 
For the multivariate functional regression model, we apply a Bayesian estimation framework, where the estimated \glspl{mfpc} are treated as empirical basis functions to parsimoniously represent the functional random processes. This can be understood as an empirical Bayes approach with prior information about the covariance between different dimensions and the functional form of the random processes estimated from the data. We follow the \gls{famm} framework \citep{scheipl2015functional} in using basis representations to model the observed functional data points in a scalar regression model. This allows to formalize a wide array of assumptions about functional form and smoothness of covariate effects specified via penalty matrices in normal priors. Our approach extends \cite{scheipl2016generalized}, who have previously shown that the (generalized) \gls{famm} framework can handle univariate generalized functional data to multivariate generalized functional data. In the following, we provide a short introduction to the \gls{famm} framework for the basis representation of functional effects, as well as an introduction to the \gls{bamlss} framework for the Bayesian model fitting approach used for estimation.

\subsubsection{Basis Representation of Effects}
\label{subsubsec:TensorProds}

\cite{scheipl2015functional} have shown that it is possible to represent linear and smooth (functional) covariate effects of scalar and functional covariates using tensor products of appropriate bases. Their framework allows functional data to be observed at function- and here also dimension-specific time points $t_{i}^{(k)} = 1,..., N_i^{(k)}$ so that irregular scalar observations of the functional data can be combined in matrix notation. Let $\bm{y} = \left(\bm{y}^{(1)\top}, ..., \bm{y}^{(K)\top}\right)^{\top}$ with $\bm{y}^{(k)} = \left(\bm{y}_{1}^{(k)}, ..., \bm{y}_{n}^{(k)}\right)^{\top}$ and $\bm{y}_{i}^{(k)} = \left(y_{i1}^{(k)}, ..., y_{i{N_i^{(k)}}}^{(k)}\right)^{\top}$ contain the scalar observations of $\bm{y}(t)$ at these time points, and denote $\bm{t}_i^{(k)} = \left(t_{i1}^{(k)}, ..., t_{i{N_i^{(k)}}}^{(k)}\right)^{\top}$, $\bm{t}^{(k)} = \left(\bm{t}_1^{(k)\top}, ..., \bm{t}_n^{(k)\top}\right)^{\top}$ of length $N^{(k)} = \sum_{i= 1}^{n}N_i^{(k)}$, and $\bm{t} = \left(\bm{t}^{(1)\top}, ..., \bm{t}^{(K)\top}\right)^{\top}$ of length $N = \sum_{k= 1}^{K}N^{(k)}$. Further define the row tensor product $\odot$ of $s \times v$ matrix $\bm{V}$ and $s \times w$ matrix $\bm{W}$ as the $s\times vw$ matrix $\bm{V} \odot \bm{W} = (\bm{V}\otimes \bm{1}_w^{\top}) \cdot (\bm{1}_v^{\top} \otimes \bm{W})$ with $\bm{1}_c$ a $c$-vector of ones, the Kronecker product $\otimes$, and $\cdot$ element-wise multiplication. Then, all of the proposed flexible covariate effects can be represented in the matrix notation of \eqref{eq:Basisexpansion}, now for all observations in $\bm{y}$, as
\begin{align}
\label{eq:tensorProdRep}
f_{rl}^{(k)}(\mathcal{X}, \bm{t}^{(k)}) = \bm{B} \bm{\beta}_{rl}^{(k)} = (\bm{B}_{rlx} \odot \bm{B}_{rlt})\bm{\beta}_{rl}^{(k)},
\end{align} 
where $f_{rl}^{(k)}(\mathcal{X}, \bm{t}^{(k)})$ is the $N^{(k)}$ vector of covariate effects of the respective covariate subset of $\mathcal{X} = \bigcup_{i = 1}^{n}\mathcal{X}_i$ at time points $\bm{t}^{(k)}$ for distributional parameter $r = 1,..., R^{(k)}$ and covariate effect term $l = 1,..., L_r^{(k)}$. The $N^{(k)}\times d_{rlx}^{(k)}$ matrix $\bm{B}_{rlx}$ contains evaluations of a suitable marginal basis for the respective covariate subset, while the $N^{(k)}\times d_{rlt}^{(k)}$ matrix $\bm{B}_{rlt}$ contains evaluations of a suitable marginal basis over the scalar observation times in $\bm{t}^{(k)}$. For a time-constant covariate effect, for example, $\bm{B}_{rlt} = \bm{1}_{N^{(k)}}$ or, for a smooth effect over time, it might contain an evaluated spline basis of $d_{rlt}^{(k)}$ basis functions. Similarly, a linear covariate effect of a single covariate $x \in \mathcal{X}$ can be represented using the corresponding covariate vector $\bm{B}_{rlx} = \bm{x}$, while a spline basis allows to model the covariate effect smoothly over different values of $x$. This basis approach supports a variety of effects such as smooth, spatial, time-varying or random effects terms. Note that tensor products can also be extended to higher dimensions with more than two margins as well as functional covariates, but we restrict this introduction to the more common two-dimensional and scalar cases. The tensor product representation \eqref{eq:tensorProdRep} is thus very flexible with \cite{scheipl2015functional} providing a more detailed overview of available modeling options.

Note that the flexibility of the estimated effects depends on the number of specified marginal basis functions $d_{rlx}^{(k)}$ and $d_{rlt}^{(k)}$. We, therefore, apply a Bayesian P-spline approach by using a large number of marginal B-spline basis functions and imposing random walk priors on $\bm{\beta}_{rl}^{(k)}$ \citep{lang2004bayesian}. This corresponds to a penalized estimation of smooth effects using difference penalties for adjacent B-spline coefficients in a likelihood framework. For the tensor product effects, the penalty term can be constructed from the marginal penalty matrices $\tilde{\bm{P}}_{rlx}$ and $\tilde{\bm{P}}_{rlt}$ associated with each marginal basis as
\begin{align}
\label{eq:quadraticPen}
    \bm{\beta}_{rl}^{(k)\top} \left(\frac{1}{\tau_{rlx}^{(k)2}}\tilde{\bm{P}}_{rlx}\otimes \bm{I}_{d_{rlt}^{(k)}} + \frac{1}{\tau_{rlt}^{(k)2}}\bm{I}_{d_{rlx}^{(k)}}\otimes \tilde{\bm{P}}_{rlt}\right)\bm{\beta}_{rl}^{(k)} = \bm{\beta}_{rl}^{(k)\top} \left(\frac{1}{\tau_{rlx}^{(k)2}}\bm{P}_{rlx} + \frac{1}{\tau_{rlt}^{(k)2}}\bm{P}_{rlt}\right)\bm{\beta}_{rl}^{(k)}
\end{align}
with $\bm{I}_c$ the $c\times c$ identity matrix \citep{scheipl2015functional}.  The terms $(\tau_{rlx}^{(k)2})^{-1}$ and $(\tau_{rlt}^{(k)2})^{-1}$ can be interpreted as smoothing parameters and allow different degrees of smoothness in the marginal directions, that is, the covariate direction of $\mathcal{X}$ and the functional direction of $t$. We point out that this construction can also be applied to linear covariate effects or time-constant effects by setting the respective penalty matrices $\bm{P}_{rlx}$ or $\bm{P}_{rlt}$ to a matrix of zeros and dropping the corresponding smoothing parameter from the model. As an analogy to the quadratic penalty \eqref{eq:quadraticPen}, we assume equivalent partially improper Gaussian priors for $\bm{\beta}_{rl}^{(k)}$ with variance parameters $\tau_{lx}^{(k)2}$ and $\tau_{rlt}^{(k)2}$ as described in more detail in Section \ref{subsubsec:Bayes}.

In order to adequately represent the latent processes using tensor products, we slightly adapt \eqref{eq:tensorProdRep} to account for the inherently multidimensional nature of the \glspl{mfpc}, as the scores affect all $K$ dimensions simultaneously. In particular, let $\bm{\rho}_0 = (\bm{\rho}_{0(1)}^{\top}, ..., \bm{\rho}_{0(n)}^{\top})^\top$ contain all observation-specific scores $\bm{\rho}_{0(i)} = (\rho_{0i1}, ..., \rho_{0iM_0})^{\top}$ of the finite \gls{kl} expansion \eqref{eq:KLexpansion} of $\bm{\Lambda}_{0i}(t)$, which accounts for the dependency within and between all functional observations $i=1,...,n$. The corresponding $N$-vector $\bm{\Lambda}_0(\bm{t})$ subsumes the evaluations of all $n$ independent copies of $\bm{\Lambda}_{0i}(t)$ over all $N = \sum_{k= 1}^{K}N^{(k)}$ scalar observations in $\bm{t}$  as 
\begin{equation*}
    \bm{\Lambda}_0(\bm{t}) = \bm{\Psi}_0\bm{\rho}_0 = (\bm{\delta}_0 \odot \tilde{\bm{\Psi}}_0)\bm{\rho}_0,
\end{equation*}
where the $i$-th column of the $N\times n$ indicator matrix $\bm{\delta}_0$ contains a $1$ if the corresponding element of $\bm{t}$ is from the $i$-th functional observation and $0$ otherwise. The $N\times M_0$ matrix $\tilde{\bm{\Psi}}_0$ holds the evaluations of the $M_0$ \glspl{mfpc} $\psi_{0m}^{(k)}(s)$ for time points $t_{is}^{(k)}$ in $\bm{t}$ for the truncated \gls{kl} expansion using $M_0$ basis functions. Other random processes in $\Uplambda_i$ accounting for further grouping structure in the data can be represented analogously using the respective $M_u$ \glspl{mfpc} in $\tilde{\bm{\Psi}}_u$ and scores in $\bm{\rho}_{u} = \left(\bm{\rho}_{1u}^{\top}, ..., \bm{\rho}_{J_uu}^{\top}\right)^{\top}$, $\bm{\rho}_{ju} = (\rho_{j1u}, ..., \rho_{jM_uu})^{\top}$, together with adapting the indicator matrix to reflect the $j=1,..., J_u$ grouping levels of $\bm{\Lambda}_u(t), u = 1,..., U$. As the scores are independent across all grouping levels $j$ (and for $\bm{\Lambda}_0(t)$ for all $i$) as well as for all \glspl{mfpc}, simple random effects prior assumptions can be used (see Section \ref{subsubsec:Bayes}).

\subsubsection{Bayesian Model Fitting}
\label{subsubsec:Bayes}

We use Newton-Raphson and derivative-based \gls{mcmc} algorithms to estimate the model parameters in a Bayesian framework. As previously stated, we condition on the estimated eigenfunction basis, i.e.\ the set $\Uppsi$ of all $M_0 + \sum_{u = 1}^{U} M_u$ \glspl{mfpc}. Further, the model specification in \eqref{eq:DistributionalAss} and \eqref{eq:additive_eta} assumes conditional independence of all $s = 1,..., n_i^{(k)}$ scalar observations $y_{is}^{(k)}$ of the $i = 1,..., n$ functional observations and the $k = 1,..., K$ different dimensions given all latent processes. We can thus specify the likelihood as
\begin{equation}
    \begin{aligned}
    \mathcal{L}(\bm{\beta}, \bm{\rho}_0, \bm{\rho_1},..., \bm{\rho_U}\mid \bm{y}, \mathcal{X}, \Uppsi) &= \prod_{i=1}^{n} \prod_{k=1}^{K} \prod_{s=1}^{n_i^{(k)}} \mathcal{L}_{iks}\left(\bm{\beta}, \bm{\rho}_{0(i)}, \bm{\rho}_{1},..., \bm{\rho}_U\mid y_{is}^{(k)}, \mathcal{X}_i, \Uppsi\right)\\
    & = \prod_{i=1}^{n} \prod_{k=1}^{K} \prod_{s=1}^{n_i^{(k)}} d_k\left(y_{is}^{(k)} \mid \theta_{i1s}^{(k)},..., \theta_{iR^{(k)}s}^{(k)}\right)
\end{aligned}
\label{eq:likelihood}
\end{equation}
with $d_k$ the probability density function or probability mass function of the parametric distribution $\mathcal{D}^{(k)}$ with distributional parameters $\theta_{irs}^{(k)}, r = 1,..., R^{(k)}$.

We assign partially improper Gaussian priors for all covariate effects $\bm{\beta}_{rl}^{(k)}, l = 1,..., L_r^{(k)}$ with multiple variance parameters $\bm{\tau}_{rl}^{(k)2} = \left(\tau_{rlx}^{(k)2}, \tau_{rlt}^{(k)2}\right)$, allowing anisotropic smoothing in the directions of the marginal basis functions for the covariates in $\mathcal{X}$ and the functional domain $t$, respectively, given as
\begin{align*}
    p\left(\bm{\beta}_{rl}^{(k)}\mid \bm{\tau}_{rl}^{(k)2}\right) \propto \left| \frac{1}{\tau_{rlx}^{(k)2}} \bm{P}_{rlx} + \frac{1}{\tau_{rlt}^{(k)2}} \bm{P}_{rlt}\right |^{\frac{1}{2}}\exp\left(-\frac{1}{2}\bm{\beta}_{rl}^{(k)\top}\left[\frac{1}{\tau_{rlx}^{(k)2}} \bm{P}_{rlx} + \frac{1}{\tau_{rlt}^{(k)2}} \bm{P}_{rlt}\right]\bm{\beta}_{rl}^{(k)}\right).
\end{align*}
For covariate terms which have a penalty term only in one marginal direction, such as linear functional effects or (scalar) random effects terms, this prior simplifies to a multivariate normal with one variance parameter. Vague normal priors $\bm{\beta}_{rl}^{(k)} \sim N(\bm{0}, 1000^2\bm{I})$ are used for linear or parametric terms, which are not subject to smoothing and therefore are not modeled with an associated variance parameter. To improve readability for the prior specification of the random scores, let $\bm{\rho}_0^{(m)} = (\rho_{01m},..., \rho_{0nm})^\top$ with corresponding variance parameter denoted as $\nu_{0m}, m = 1,..., M_0$, and $\bm{\nu}_0 = (\nu_{01},...,\nu_{0M_0})^\top$. Then, the independence assumption between the levels of a grouping layer such as the functional observations $i= 1,..., n$ for $\bm{\Lambda}_0(t)$ and between the respective \glspl{mfpc} leads to the product of multivariate normal priors
\begin{align*}
    p(\bm{\rho}_0\mid \bm{\nu}_0) = \prod_{m = 1}^{M_0}p\left(\bm{\rho}_0^{(m)}\mid \nu_{0m}\right) \propto \prod_{m = 1}^{M_0}\left( \frac{1}{\nu_{0m}}\right )^{\frac{n}{2}}\exp\left(-\frac{1}{2\nu_{0m}}\bm{\rho}_0^{(m)\top}\bm{\rho}_0^{(m)}\right)
\end{align*}
and analogously for $\bm{\rho}_u, u = 1, ..., U$ of the other $U$ independent latent processes with variance parameters $\bm{\nu}_u = (\nu_{u1}, ..., \nu_{uM_u})^\top$, where $n$ is replaced by the number of levels $J_u$ in the grouping layer $u$. Independent inverse Gamma hyperpriors $IG(0.001, 0.001)$ are used for all variance parameters so that we obtain inverse Gamma full conditionals for the variance parameters.

For a more concise notation, let $\bm{\vartheta}$ denote the vector of all model parameters, containing all coefficient vectors, score vectors and corresponding variance parameters. Then, the posterior of the model is
\begin{align*}
    p(\bm{\vartheta}\mid \bm{y}, \mathcal{X}, \Uppsi) = & \mathcal{L}(\bm{\beta}, \bm{\rho}_0, \bm{\rho_1},..., \bm{\rho_U}\mid \bm{y}, \mathcal{X}, \Uppsi) \cdot \prod_{k = 1}^{K}\prod_{r=1}^{R^{(k)}}\prod_{l = 1}^{L_r^{(k)}} \left[p\left(\bm{\beta}_{rl}^{(k)}\mid \bm{\tau}_{rl}^{(k)2}\right) p\left(\bm{\tau}_{rl}^{(k)2}\right)\right]\\
    &\quad \cdot p(\bm{\rho}_0\mid \bm{\nu}_0) p(\bm{\nu}_0)\cdot \prod_{u = 1}^{U} \left[p(\bm{\rho}_u\mid \bm{\nu}_u)p(\bm{\nu}_u)\right],
\end{align*}
where $p(\bm{\tau}_{rl}^{(k)2})$, $p(\bm{\nu}_0)$, and $p(\bm{\nu}_u)$ are the priors of the respective variance parameters. We evaluate the posterior using derivative-based \gls{mcmc} sampling, Gibbs sampling, and slice sampling as described in detail in \cite{umlauf2018bamlss}. The derivative-based \gls{mcmc} algorithm is used for the covariate effects and scores and constructs approximate full conditionals based on second-order Taylor series expansions of the log-posterior centered at the last state of the parameters. In the $h$-th iteration, the algorithm draws blocks of parameters $\bm{\beta}_{rl}^{(k)[h]}$ from a multivariate normal proposal density with mean $\bm{\beta}_{rl}^{(k)[h-1]} - \bm{H}(\bm{\beta}_{rl}^{(k)[h-1]})^{-1}\bm{s}(\bm{\beta}_{rl}^{(k)[h-1]})$ and covariance matrix $- \bm{H}(\bm{\beta}_{rl}^{(k)[h-1]})^{-1}$ evaluated at the previous states in iteration $h-1$, where $\bm{s}(\cdot)$ denotes the score vector and $\bm{H}(\cdot)$ the Hessian of the log-posterior. \cite{umlauf2018bamlss} show that applying the chain rule allows to calculate these derivatives in modular blocks using e.g.\ the derivatives of the inverse link functions and derivatives of the additive predictors. Note that the computation of $\bm{s}(\bm{\rho})$, $\bm{H}(\bm{\rho})$, and the corresponding quantities for the other random score vectors $\bm{\rho}_u$ is similarly straightforward, as the factorization of the likelihood \eqref{eq:likelihood} reduces the calculation to summation over all individual contributions. Variance parameters are sampled using Gibbs sampling when the full conditionals are inverse Gamma distributions, which is not the case for anisotropic model terms. Then, no closed-form full conditional can be obtained and slice sampling is employed. In order to find appropriate starting values for the sampling algorithm, a backfitting algorithm cycling through all model terms using a Newton-Raphson procedure to optimize the log-posterior, approximating the posterior mode, is used, where the variance parameters are chosen by minimizing the corrected AIC in each updating step.

\subsection{Implementation Details}

The estimation of the regression model is implemented in the \textbf{R} package \textbf{bamlss} \citep{pkg:bamlss}, which provides a very flexible infrastructure for estimating regression models with flexible covariate effects in a Bayesian framework. The wide range of modeling possibilities presented in Section \ref{subsubsec:TensorProds} are made available by leveraging the widely used \textbf{mgcv} package \citep{pkg:mgcv} for matrix and penalty set-up. The derivative-based \gls{mcmc} algorithm described in Section \ref{subsubsec:Bayes} is the default sampler used in \textbf{bamlss}. We provide the \textbf{R} package \textbf{gmfamm} \citep[][available on \href{https://cran.r-project.org/package=gmfamm}{CRAN}]{pkg:gmfamm} building on \textbf{bamlss}, which contains a convenience function constructing a \textbf{bamlss}-family for any number and combination of one- or two-parameter distributional assumptions $\mathcal{D}^{(k)}$ for the (generalized) multivariate functional data. This convenience function encompasses all most commonly used one- and two-parameter distributions with available implementations in the \textbf{bamlss} package but can be easily extended by the user. We provide further examples of families with distributional assumptions with more than two parameters as a blueprint to construct a user-specific model. 

In order to fit the regression model, the \glspl{mfpc} have to be first estimated in a preliminary step by the user. The \gls{mfpc} basis is then to be evaluated at the scalar observation times in $\bm{t}$ and attached as covariates to the data set, for which we also provide a convenience function. Given the modular nature of our approach, this allows the user to compare different methods of estimating the \glspl{mfpc} or to combine different univariate \gls{fpca} approaches. Notably, the \textbf{R} package \textbf{MFPCA} \citep{pkg:MFPCA} provides an implementation of the \gls{mfpca} approach proposed by \cite{happ2018multivariate}. To the best of our knowledge, the univariate \gls{fpca} methods used in this manuscript do not have an \textbf{R} package implementation available but we provide the code to reproduce all simulations and application results in the \textbf{gmfamm} package.

\section{Simulation}
\label{sec:Simulation}

We conduct a simulation study to evaluate the impact of the sampling regime of the functional data, the quality of the \gls{mfpc} estimation, and the truncation order on the model fit. Additionally, the multivariate modeling approach is compared to univariate (generalized) functional models using a similar principal component based modeling strategy. Note that while the proposed generalized multivariate functional regression models can include several different flexible covariate effect specifications, the focus of our simulation study is on the implications of the multivariate modeling approach rather than on the estimation of nonlinear coefficient effects, which have already been evaluated extensively for the \gls{famm} framework in previous simulation studies \citep[see e.g.][]{scheipl2015functional, scheipl2016generalized}

\subsection{Simulation Design}
\label{subsec:SimulationDesign}

We simulate $200$ data sets containing $n = 150$ multivariate functions $\bm{Y}_i(t)$ with $t\in [0,1]$ and $K = 3$, which follow the binomial, Poisson, and Gaussian distributional assumptions
\begin{gather*}
Y_{it}^{(1)} \mid \mathcal{X}_i, \bm{\Lambda}_{0i} \sim \textit{Bin}\left(1, \theta_{i1}^{(1)}(t) = \left(1 + \exp\left(-\eta_{i1}^{(1)}(t)\right)\right)^{-1}\right),\\   
Y_{it}^{(2)} \mid \mathcal{X}_i, \bm{\Lambda}_{0i} \sim \textit{Poi}\left(\theta_{i1}^{(2)}(t) = \exp\left(\eta_{i1}^{(2)}(t)\right)\right),\\
Y_{it}^{(3)} \mid \mathcal{X}_i, \bm{\Lambda}_{0i} \sim N\left(\theta_{i1}^{(3)}(t) = \eta_{i1}^{(3)}(t), \left(\theta_{i2}^{(3)}(t)\right)^2 = \exp\left(\eta_{i2}^{(3)}(t)\right)^2\right)
\end{gather*}
with \gls{iid} $\mathcal{X}_i = \{x_i, z_i\}$, $x_i \sim Unif(-1, 1)$ uniformly and $z_i \sim Bin(1, 0.5)$ binomially distributed. To facilitate direct comparison of estimation quality across the different dimensions, we include the same covariate effects in the additive predictors, which are specified as 
\begin{gather*}
    \eta_{i1}^{(k)}(t) = \beta_0^{(k)}(t) + \beta_1^{(k)}(t)\cdot x_i + \Lambda_{0i}^{(k)}(t) = \cos(2\pi t) - \cos(2\pi t)\cdot x_i + \sum_{m=1}^{6} \rho_{0im} \psi_{0m}^{(k)}(t), \quad k = 1,2,3\\
    \eta_{i2}^{(3)}(t) = \gamma_0 +  \gamma_1\cdot z_i = -2 + 0.5 \cdot z_i
\end{gather*}
with \gls{iid} $\rho_{0im} \sim N(0, \nu_{0m})$, linearly decreasing $\nu_{0m} = (M_0 +1 - m) / M_0$, and multivariate eigenfunctions created by splitting each of the first $M_0 = 6$ Fourier basis functions into three parts and translating and randomly reflecting them as described in detail in \cite{happ2018multivariate}. Figure \ref{APPfig:SimMFPCs} in Appendix \ref{APPsec:Simulation} shows the six multivariate eigenfunctions used in the simulation study.

\underline{\textit{Impact of sampling regime:}} We compare three different sampling regimes for the multivariate functional data to evaluate differences in accuracy of the model fits. For each simulated data set, we draw scalar observations of the functional data on a fine equidistant grid $\bm{d}$ of length 101 and create three versions of the functional data set by subsampling to mimic sparsely, regularly, and irregularly sampled functional data. For the sparse sampling regime, the numbers $N_i^{(k)}$ of scalar observations on each dimension $k$ is drawn independently from $\{1,2,...,10\}$ with equal probability. Then, the vector of observation times $\bm{t}_i^{(k)}$ is subsampled similarly from the full grid $\bm{d}$. The regularly sampled functional data set contains only scalar observations on the grid $\{0, 0.1,...,1\}$. The sampling for the irregular regime is analoguous to the sparse case but the number of scalar observations $N_i^{(k)}$ lies between eleven and twenty. Due to this construction, the regularly sampled data sets have more scalar observations than the sparse but less than the irregular scenarios. The regularly sampled data, however, only provide information on eleven distinct time points, while both the sparsely and irregulary sampled data might better cover the entire range of the functional domain. In order to separate the model fitting process from the eigenfunction estimation in the evaluation of the impact of the sampling regime, we use the simulated eigenfunctions as \glspl{mfpc} in the regression models.

\underline{\textit{Impact of \gls{mfpc} basis:}} In order to evaluate the impact of the quality of the \gls{mfpc} basis, we estimate the \glspl{mfpc} separately for the three previously described sampling regimes and use these estimates for regression model fits on the sparse functional data setting. As \gls{mfpc} estimation is typically more accurate for larger data sets, this allows to compare the regression fit for \glspl{mfpc} bases of different quality, while leaving the \gls{mfpc} estimation approach unchanged. We apply the univariate fast generalized \gls{fpca} proposed by \cite{leroux2023fast} by first binning the data into eleven overlapping bins with equidistant centers and then fitting local scalar generalized mixed models. The local regression models use the data from the interval covering the bin center plus up to 0.3 on each side to get a local estimate of the latent process. Note that this specification of overlapping bins might induce spurious correlation in the estimated latent process but generally outperforms non-overlapping bins in simulations \citep{leroux2023fast}. The authors do not provide a guideline for choosing e.g.\ the bin width, especially not for data with less than 1000 scalar observations per function. The parameters are thus specified so that a sufficient number of observations can contribute at each bin center to the univariate \glspl{fpca}. We apply the \gls{bamlss} framework to fit the univariate binomial, Poisson, and Gaussian models, which include a random intercept as well as a fixed intercept and a linear covariate effect for $x_i$. The univariate Gaussian models also include a linear effect for $z_i$ in the additive predictor for the scale parameter. For each of these $3\cdot11$ scalar models, we generate 1000 derivate-based \gls{mcmc} samples as in Section \ref{subsubsec:Bayes} and use the posterior means of the estimated random intercepts as input for the fast \gls{fpca} of \cite{xiao2016fast}. The proportion of variance explained, which is used to determine the number of univariate \glspl{fpc} to estimate, is set to 0.99 and the maximum possible number of third order B-splines (seven) is used for the efficient covariance smoothing. The resulting univariate \glspl{fpc} can then be directly passed on as input to the \gls{mfpca}, but the univariate scores have to be re-estimated to avoid the bias introduced by the binning procedure. We thus refit the univariate models on the entire functional domain within \gls{bamlss} using the estimated univariate \glspl{fpc} as basis functions, generating 1000 \gls{mcmc} samples after a burn-in of 1000 samples and a thinning of five. The posterior score means are then used to estimate the \gls{mfpca}, and the multivariate functional regression models in these scenarios include all available estimated \glspl{mfpc}. Note that parallelizing the estimation of the \gls{mfpc} basis offers an avenue to reducing the computation times, in particular as the univariate \glspl{fpca} are calculated separately and, within one dimension, the discretized models are estimated independently. 

\underline{\textit{Impact of \gls{mfpc} truncation:}} Truncating the \gls{mfpc} basis results in a more parsimonious model but might affect the estimation accuracy of each dimension differently. We therefore compare the model containing all \glspl{mfpc} estimated from the sparse data scenario with a model truncated at $98\%$ of total multivariate variance explained for the sparse data scenario. As a further alternative, we calculate the multivariate functional regression model using the estimates of a \gls{mfpca} based on a weighted scalar product with an analogous truncation criterion. Here, the inverse of the sum of univariate eigenvalues $\sum_{m =1}^{M^{(k)}}\hat{\upsilon}_{0m}^{(k)}$ is used as weight $w_k$. While the sum of true univariate eigenvalues in our simulation is similar across dimensions ($\approx 1.16$ for dimensions 1 and 3 and $\approx 1.18$ for dimension 2), the univariate \glspl{fpca} can yield larger discrepancies in the estimated eigenvalues. Especially for the binomial functional data, we expect a less accurate estimation of the \gls{mfpc} basis, as binary data inherently carries less information about the latent process than continuous outcomes. Consequently, we anticipate different accuracies when using different weighting schemes for the \gls{mfpca}.

\underline{\textit{Univariate vs.\ multivariate:}} Finally, we compare univariate functional regression models in the \gls{bamlss} framework to our proposed multivariate regression model on the sparsely sampled data. To isolate the effect of the multivariate modeling, both the univariate and the multivariate models estimate random scores for the true simulated eigenfunctions. For similar reasons to above, we expect the accuracy of the binomial model to benefit the most from the multivariate modeling approach.

\underline{\textit{Evaluation:}} For all functional regression models, we generate 1000 \gls{mcmc} samples after a burn-in of 1000 and a thinning of five. To evaluate the model estimation of the generic model term $\bm{f}(t)\in\mathcal{L}_K^2(\mathcal{I})$, we calculate the univariate \gls{rrmse} over all true multivariate functions $\bm{f}_i(t)\in \mathcal{L}_K^2(\mathcal{I})$ with levels $i = 1,..., n$ and their estimates $\hat{\bm{f}}_i(t)$, which we define as 
\begin{align*}
    rrMSE\left(\hat{f}_i^{(k)}\right) = \sqrt{\frac{\frac{1}{n}\sum_{i=1}^{n}||f_i^{(k)} - \hat{f}_i^{(k)}||^2}{\frac{1}{n}\sum_{i=1}^{n}||f_i^{(k)}||^2}}
\end{align*}
with squared univariate norm $||\cdot||^2$. In particular, we calculate the \gls{rrmse} in each simulated data set for the $n = 150$ fitted additive predictors $\eta_{i1}^{(k)}(t)$, the latent processes $\Lambda_{0i}^{(k)}(t)$, as well as the fitted covariate effect functions $\beta_{0}^{(k)}(t)$ and $\beta_1^{(k)}(t)$, for which $n = 1$ and the index $i$ is dropped. Additionally, we evaluate the $95\%$ frequentist coverage $FC(\hat{f}_{i}^{(k)}, t) = \frac{1}{n}\sum_{i= 1}^{n} I(\hat{f}_{it}^{(k), 2.5} \leq f_{it}^{(k)}\leq \hat{f}_{it}^{(k), 97.5})$ with indicator function $I$, $f_{it}^{(k)}$ the evaluation of $f_{i}^{(k)}(t)$ at point $t$, and $\hat{f}_{it}^{(k), \alpha}$ the $\alpha\%$ quantile of \gls{mcmc} samples at evaluation point $t$. This pointwise coverage criterion is then averaged over the 200 simulated data sets per $t$. The estimation of the coefficients $\gamma_0$ and $\gamma_1$ is evaluated using scalar definitions of bias, \gls{mse}, and frequentist coverage given in Appendix \ref{APPsec:Simulation}. To isolate and quantify the differences in goodness of eigenfunction estimation, we evaluate the estimated \gls{mfpc} bases by calculating the \gls{rrmse} for the reconstructed latent processes $\hat{\Lambda}_{0i}^{(k)}(t) = \sum_{m = 1}^{M_0^{+}}\hat{\rho}_{0im}\hat{\psi}_{0m}^{(k)} (t)$ outside of the functional regression framework, where we use the method of least-squares to estimate the $\hat{\rho}_{0im}$ from the true simulated $\Lambda_{0i}^{(k)}(t)$  using the previously estimated basis functions $\hat{\psi}_{0m}, m = 1,..., M_0^{+}$.

\subsection{Results}

\begin{table}
    \centering\footnotesize
    \begin{subtable}{\textwidth}
        \subcaption{\textit{Impact of sampling regime:} Estimation based on true eigenfunctions}
        \label{subtab:Sampling}
    \begin{tabular}{p{6.3em}|rrr|rrr|rrr|rrr}
  & \multicolumn{3}{c|}{$\eta_{i1}^{(k)}(t)$} & \multicolumn{3}{c|}{$\Lambda_{0i}^{(k)}(t)$} & \multicolumn{3}{c|}{$\beta_0^{(k)}(t)$} & \multicolumn{3}{c}{$\beta_1^{(k)}(t)$} \\
Scenario & $Bin$ & $Poi$ & $N$ &  $Bin$ & $Poi$ & $N$ & $Bin$ & $Poi$ & $N$ & $Bin$ & $Poi$ & $N$ \\ \hline
Sparse & 0.557 & 0.437 & 0.246 & 0.665 & 0.545 & 0.321 & 0.294 & 0.173 & 0.125 & 0.482 & 0.297 & 0.206 \\ 
 Regular & 0.366 & 0.271 & 0.073 & 0.451 & 0.356 & 0.155 & 0.219 & 0.137 & 0.115 & 0.363 & 0.245 & 0.195 \\ 
  Irregular &  0.361 & 0.257 & 0.070 & 0.439 & 0.332 & 0.128 &  0.203 & 0.134 & 0.115 & 0.347 & 0.244 & 0.199 \\ \hline 
\end{tabular}
    \end{subtable}
\\[1em]
    \begin{subtable}{\textwidth}
        \subcaption{\textit{Impact of MFPC basis:} Estimation on sparse setting with MFPCs estimated from different settings}
        \label{subtab:MFPCest}
 \begin{tabular}{p{6.3em}|rrr|rrr|rrr|rrr}
  & \multicolumn{3}{c|}{$\eta_{i1}^{(k)}(t)$} & \multicolumn{3}{c|}{$\Lambda_{0i}^{(k)}(t)$} & \multicolumn{3}{c|}{$\beta_0^{(k)}(t)$} & \multicolumn{3}{c}{$\beta_1^{(k)}(t)$} \\
  Scenario & $Bin$ & $Poi$ & $N$ &  $Bin$ & $Poi$ & $N$ & $Bin$ & $Poi$ & $N$ & $Bin$ & $Poi$ & $N$ \\ \hline
  True & 0.557 & 0.437 & 0.246 & 0.665 & 0.545 & 0.321 & 0.294 & 0.173 & 0.125 & 0.482 & 0.297 & 0.206 \\ 
  Sparse & 0.712 & 0.518 & 0.276 & 0.864 & 0.638 & 0.357 & 0.304 & 0.225 & 0.128 & 0.478 & 0.312 & 0.214 \\ 
  Regular & 0.659 & 0.470 & 0.265 & 0.797 & 0.584 & 0.345 & 0.297 & 0.187 & 0.127 & 0.474 & 0.303 & 0.219 \\ 
  Irregular & 0.636 & 0.462 & 0.262 & 0.768 & 0.576 & 0.340 & 0.295 & 0.182 & 0.126 & 0.471 & 0.298 & 0.218 \\ \hline 
\end{tabular}
    \end{subtable}
\\[1em]
    \begin{subtable}{\textwidth}
        \subcaption{\textit{Impact of MFPC truncation:} Estimation with MFPCs truncated at $98\%$ based on different scalar products}
        \label{subtab:mfpcTrunc}
 \begin{tabular}{p{6.3em}|rrr|rrr|rrr|rrr}
  & \multicolumn{3}{c|}{$\eta_{i1}^{(k)}(t)$} & \multicolumn{3}{c|}{$\Lambda_{0i}^{(k)}(t)$} & \multicolumn{3}{c|}{$\beta_0^{(k)}(t)$} & \multicolumn{3}{c}{$\beta_1^{(k)}(t)$} \\
  Scenario & $Bin$ & $Poi$ & $N$ &  $Bin$ & $Poi$ & $N$ & $Bin$ & $Poi$ & $N$ & $Bin$ & $Poi$ & $N$ \\ \hline
  No Truncation & 0.712 & 0.518 & 0.276 & 0.864 & 0.638 & 0.357 & 0.304 & 0.225 & 0.128 & 0.478 & 0.312 & 0.214 \\
  Equal Weights & 0.726 & 0.529 & 0.281 & 0.882 & 0.650 & 0.363 & 0.307 & 0.233 & 0.128 & 0.476 & 0.319 & 0.216 \\ 
  Weighted &  0.836 & 0.872 & 0.905 & 1.024 & 1.083 & 1.138 & 0.312 & 0.242 & 0.130 & 0.481 & 0.323 & 0.216 \\  \hline 
\end{tabular}
    \end{subtable}
\\[1em]
    \begin{subtable}{\textwidth}
        \subcaption{\textit{Univariate vs.\ multivariate:} Estimation on sparse setting with true eigenfunctions}
        \label{subtab:UniMul}
 \begin{tabular}{p{6.3em}|ccc|ccc|ccc|ccc}
  & \multicolumn{3}{c|}{$\eta_{i1}^{(k)}(t)$} & \multicolumn{3}{c|}{$\Lambda_{0i}^{(k)}(t)$} & \multicolumn{3}{c|}{$\beta_0^{(k)}(t)$} & \multicolumn{3}{c}{$\beta_1^{(k)}(t)$} \\
    Scenario & $Bin$ & $Poi$ & $N$ &  $Bin$ & $Poi$ & $N$ & $Bin$ & $Poi$ & $N$ & $Bin$ & $Poi$ & $N$ \\ \hline
   Multivariate & 0.557 & 0.437 & 0.246 & 0.665 & 0.545 & 0.321 & 0.294 & 0.173 & 0.125 & 0.482 & 0.297 & 0.206 \\ 
   Univariate & 0.711 & 0.567 & 0.338 & 0.864 & 0.707 & 0.432 & 0.304 & 0.190 & 0.111 & 0.499 & 0.306 & 0.181 \\  \hline
\end{tabular}
    \end{subtable}
    
    \caption{Mean \gls{rrmse} values for different model fits and estimated model components. Note that the first line for Tables (b) - (d) repeat the values for the corresponding reference scenario. $Bin$, $Poi$, and $N$ stand for the 
    dimensions $k = 1,2,3$, and denote the different pointwise distributional assumptions (binomial, Poisson, Gaussian).}
    \label{tab:SimResults}
\end{table}

\underline{\textit{Impact of sampling regime:}} When the true eigenfunctions are used as basisfunctions in the regression model, it is apparent that the multivariate models do not fit the different dimensions equally well. Table \ref{tab:SimResults} contains the mean \gls{rrmse} values of the different evaluated model components, different dimensions, and different simulation scenarios with the first block \ref{subtab:Sampling} comparing the sampling regimes. For the overall fitted additive predictors $\eta_{i1}^{(k)}(t)$, the latent process $\Lambda_{0i}^{(k)}(t)$, and both estimated coefficient effect functions $\beta_0^{(k)}(t)$ and $\beta_1^{(k)}(t)$, we find that the \gls{rrmse} values decrease from the binomial (dimension one) to the Poisson (dimension two) to the Gaussian data (dimension three) in all sampling regimes. Given that, per construction, the dimensions exhibit similar amounts of variation and we find a coinciding trend for the estimated univariate models (last block \ref{subtab:UniMul} of Table \ref{tab:SimResults}), we can conclude that this is due to continuous data apparently carrying more information about the latent trajectory than binary realizations. 

With more scalar observation points available when moving from the sparse to the regular and irregular settings, we see that generally, the estimation accuracy of all model components increases. Notably, the largest reduction in \gls{rrmse} values can be observed between the sparse and regular settings, indicating that spreading the scalar observations over the entire functional domain for each observation might be the driving factor for this improvement, rather than the mere number of scalar observations. Only for $\beta_1^{(k)}(t)$ there is a slight increase in \gls{rrmse} values on the normal dimension between the regular and irregular settings; Figure \ref{APPfig:SimRRMSEboxplots} in Appendix \ref{APPsec:Simulation} shows, however, that the overall distributions of \gls{rrmse} values of this model component show little difference between these two settings and the improvement is likely due to sampling variation. 

Overall, the models fit the data well, see for example the exemplary fitted sparse trajectories in Figure \ref{APPfig:SimSampleFits} in Appendix \ref{APPsec:Simulation}, and the functional shape of the coefficient effect functions is well estimated (compare Figure \ref{APPfig:SimSampleBetas} in Appendix \ref{APPsec:Simulation}). We find higher variance in the coefficient effect function estimates for the Poisson and even higher for the binomial dimensions, which is mirrored in the corresponding credible intervals, which are generally wide enough to achieve good pointwise frequentist coverage (see Figure \ref{APPfig:SimCoverage} in Appendix \ref{APPsec:Simulation}). Only for the functional intercept $\beta_0^{(k)}(t)$, the $95\%$ credible intervals can be too narrow (see Table \ref{APPtab:cred_widths} in Appendix \ref{APPsec:Simulation}), especially for the normal dimension, and thus show a low pointwise frequentist coverage. Note that in general, Bayesian credible intervals do not necessarily have good frequentist coverage \citep[see e.g.][]{rousseau2016frequentist}. The computation of one regression model takes on average $80$ minutes on a Linux system with $12$ cores at $3.5$ GHz and $32$ GB memory for the sparse, $100$ minutes for the regular, and $160$ minutes for the irregular sampling regime.

\underline{\textit{Impact of \gls{mfpc} basis:}} Estimating the \gls{mfpc} basis for the different sampling regimes takes additional 60 minutes (on average, without parallelization) with most of the time spent on the univariate \glspl{fpca}, especially refitting the scores, and the \gls{mfpca} calculated in under 10 seconds. The \gls{rrmse} values for the reconstructed latent process confirm our assumption that more scalar observation points improve the estimation of the \gls{mfpc} basis, as the \gls{mfpc} basis estimated from the irregular sampling regime yields the best approximation to the true latent processes as measured by the \gls{rrmse} values of the reconstructed latent process (compare Table \ref{APPtab:MFPCreconstruction} and Figure \ref{APPfig:SimMFPCestRRMSE} in Appendix \ref{APPsec:Simulation}). In particular, in the irregular scenario, there is little difference between the quality of the \gls{mfpc} basis between the different dimensions. Overall, we find clusters for the \gls{rrmse} values of the reconstructed latent process corresponding to the number of univariate \glspl{fpc} chosen by the fast \gls{fpca} approach by \cite{xiao2016fast} based on the proportion of variance explained (compare Figure \ref{APPfig:SimMFPCestNumb} in Appendix \ref{APPsec:Simulation}). We see a tendency that more univariate \glspl{fpc} are associated with lower \gls{rrmse} values but more research is needed to establish reliable guidelines on how to apply the binning approach of \cite{leroux2023fast} to univariate functional data in different scenarios. 

The multivariate functional regression models then contain all estimated \glspl{mfpc}, resulting in, on average, nine included basis functions for the models using the estimates from the sparsely and regularly sampled data and ten for the irregularly sampled. This increases the computation times to e.g.\ on average 110 minutes for nine \glspl{mfpc} on the sparsely sampled data. Using an estimated eigenfunction basis in turn decreases the model accuracy as given by the \gls{rrmse} values (compare the first row of the second block \ref{subtab:MFPCest} of Table \ref{tab:SimResults} with the bottom three rows). The models, however, still fit the data reasonably well (see Figure \ref{APPfig:SimMFPCmodelsFits} in Appendix \ref{APPsec:Simulation}) in this challenging setting with sometimes very few data points per (generalized) function and the increase in \gls{rrmse} values for the coefficient effects is hardly noticeable in the coefficient effect estimates (compare Figure \ref{APPfig:SimMFPCBetas} in Appendix \ref{APPsec:Simulation}). We find that a more accurate \gls{mfpc} basis leads to considerably lower \gls{rrmse} values of the additive predictors $\eta_{i1}^{(k)}(t)$ and the latent process $\Lambda_{0i}^{(k)}(t)$ as reported in Table \ref{subtab:MFPCest}, while the coefficient effect function estimates are much more robust: The distributions of \gls{rrmse} values for $\beta_0^{(k)}(t)$ and $\beta_0^{(k)}(t)$ are very similar and differences between the true and differently estimated eigenfunction bases are small (compare Figure \ref{APPfig:SimRRMSEboxplots} in Appendix \ref{APPsec:Simulation}). Given that the models do not explicitly account for the additional uncertainty in estimating the empirical basis and the estimation accuracy for the overall additive predictors and the latent process drops, it is not surprising that the frequentist coverage of $\eta_{i1}^{(k)}(t)$ and $\Lambda_{0i}^{(k)}(t)$ decreases, particularly for the binary data with \glspl{mfpc} estimated from the sparse sampling scenario, where the pointwise frequentist coverage can be as low as $60\%$ (see Figure \ref{APPfig:SimCoverage} in Appendix \ref{APPsec:Simulation}). We again emphasize that this is a very challenging scenario and the frequentist coverage is ten to twenty percent higher for the \gls{mfpc} bases from the other sampling regimes. The frequentist coverage of the coefficient effect functions, however, is similar to the coverage in the models using the true eigenfunction basis.

\underline{\textit{Impact of \gls{mfpc} truncation:}} By excluding trailing \glspl{mfpc} from the regression model, we reduce the flexibility of the model and introduce a further level of regularization on the noisiest components of the covariance structure. This stabilizes the model estimation as fewer parameters have to be estimated and it has a negligible effect on the estimation accuracy of the examined model components as given in the third block \ref{subtab:mfpcTrunc} of Table \ref{tab:SimResults}. Using a truncation order of $98\%$ explained multivariate variance based on a scalar product with equal weights results in models containing between five and eight (median six) estimated \glspl{mfpc}. For all model components, this increases the mean \gls{rrmse} values slightly but the overall distributions as well as the pointwise frequentist coverage are very similar to the models without truncation (see Figures \ref{APPfig:SimRRMSEboxplots} and \ref{APPfig:SimCoverage} in Appendix \ref{APPsec:Simulation}).

On the other hand, using a truncated \gls{mfpc} basis estimated from an \gls{mfpca} with a weighted scalar product puts a high weight on the first dimension (mean 1.36) and low weights on the second (mean 0.15) and third (mean 0.14) dimensions in our simulation scenario. This focus on the binary dimension leads to lower \gls{rrmse} values of the additive predictor $\eta_{i1}^{(k)}(t)$ and the latent process $\Lambda_{0i}^{(k)}(t)$ for the binary functional data compared to the count and Gaussian functional data as indicated by Table \ref{subtab:mfpcTrunc}. Note that the weights are chosen to be the inverse of estimates of the total amount of variation on each dimension. As the true sums of eigenvalues are much more similar than the obtained estimates, such an extreme upweighting of dimension one is not beneficial for the analysis and constitutes a considerable overall deterioration of the estimation accuracy and frequentist coverage for the additive predictor and the latent process. The covariate effect functions, however, are again very robust and see little change in their \gls{rrmse} values and frequentist coverage (see Figures \ref{APPfig:SimRRMSEboxplots} and \ref{APPfig:SimCoverage} in Appendix \ref{APPsec:Simulation}). We thus find that using weights in the \gls{mfpca} can have a considerable effect on the model estimates, improving the model fits for dimensions with high weights compared to those with lower weights, but we conclude that the binary \gls{fpca} approach applied here seems to seriously underestimate the true variance of the latent process and weights should be carefully considered for plausibility before the analysis. In particular, the local scalar regression models feature a substantially lower posterior mean of the random effects variance for the binomial data (0.52) than for the Poisson (0.94) and Gaussian (0.98) data, which leads to less variation in the discretized realizations of the latent process on the binomial dimension and consequently a lower total sum of univariate eigenvalues in the approach proposed by \cite{leroux2023fast}. This suggests that the generalized \gls{fpca} approach by \cite{leroux2023fast} is not ideal for estimating \glspl{mfpc} for data of mixed type where at least one dimension is binomially distributed, and substituting their method by a different approach or at least not upweighting the binomial data as strongly as the estimated variances seem to imply should be considered, when basis truncation is desired.

\underline{\textit{Multivariate vs.\ univariate:}} We note that the univariate modeling approach has a clear computational advantage: even without parallel computing of the univariate models, the total computation time is on average only 50 minutes compared to the 80 minutes for the multivariate approach. The last block \ref{subtab:UniMul} in Table \ref{tab:SimResults} contains the mean \gls{rrmse} values for a univariate functional regression approach on the sparse data using the true eigenfunctions. We find that the multivariate models clearly fit the additive predictor $\eta_{i1}^{(k)}(t)$ and the latent process $\Lambda_{0i}^{(k)}(t)$ better on all dimensions, as the covariance between the functional outcomes is incorporated into the analysis. The worse estimation of the latent process for the univariate models also negatively impacts the frequentist coverage of resulting credible intervals, leading to a considerably lower coverage than the nominal value for the univariate models, especially on the Gaussian dimension (compare Figure \ref{APPfig:SimCoverage} in Appendix \ref{APPsec:Simulation}). The estimated coefficient effect functions in the univariate models are overall similar to the estimates of the multivariate functions and we find only a slight tendency for better estimates of the coefficient effect functions for the multivariate approach on the binomial and Poisson dimensions: \gls{rrmse} values are slightly smaller (see Fig.\ \ref{tab:SimResults}) and there is a slightly lager number of iterations, where the multivariate approach yields lower \gls{rrmse} values for $\beta_0^{(k)}(t)$ and $\beta_1^{(k)}(t)$ (see Table \ref{APPtab:MulUniBetaRRMSE} in Appendix \ref{APPsec:Simulation}) for these two dimensions, with a reverse trend for the third Gaussian dimension. Only a small difference in these numbers suggests a trade-off for the multivariate approach between the first two dimensions, which are slightly better estimated compared to the univariate models, and the Gaussian dimension, which shows slightly higher \gls{rrmse} values.

We thus conclude that the multivariate regression approach leads to overall better model fits and uncertainty quantification by combining the information across dimensions. This becomes especially apparent when comparing the univariate approach with true eigenfunctions to the more realistic scenario of using a multivariate regression with an estimated \gls{mfpc} basis (without truncation) as even then, the multivariate analysis results in lower \gls{rrmse} values for most components. The multivariate approach also leads to similar estimates of the additive predictor of the scale parameter $\eta_{i2}^{(3)}(t)$, which is generally estimated well but is also slightly negatively affected by estimating the \gls{mfpc} basis and potential basis truncation (see Table \ref{APPtab:sigma_evals} and Figure \ref{APPfig:sim_sigmas} in Appendix \ref{APPsec:Simulation}). The coefficient effect function estimates for $\beta_0^{(k)}(t)$ and $\beta_1^{(k)}(t)$, however, are very robust over all examined scenarios, regardless of whether the true eigenfunctions are known or estimated (and truncated), or whether a multivariate or univariate analysis approach is followed. The main benefit of the multivariate models thus is the more accurate estimation of the latent process and we find that, for a reasonable number of scalar observations, our proposed \gls{mfpca} approach  yields a good empirical eigenfunction basis, even for the (arguably more difficult) binomial data.

\section{Application}
\label{sec:Application}

In the following, we apply our proposed approach to modeling multivariate functional data of mixed type to new and publicly available data on traffic flows in the German capital. Over the last couple of years, the federal state of Berlin has taken several measures to reform its concept of urban mobility, such as the \href{https://www.berlin.de/sen/uvk/en/mobility-and-transport/transport-policy/berlin-mobility-act/}{Berlin Mobility Act} in 2018, in order to reduce emissions and traffic fatalities. As part of data-based policy making, Berlin uses infrared detectors to collect data on the number and mean speed of vehicles passing more than 240 sites at main roads spread out over the municipal area. The provided data collection (\url{https://daten.berlin.de/datensaetze/verkehrsdetektion-berlin}) contains hourly information for each location starting from January 2015. With four potential outcomes of interest given by the hourly number and mean speed of cars and trucks, this constitutes a rich and interesting data base, which, to the best of our knowledge, has seen little scientific attention: \cite{kock2023truly} restrict their analysis to one available site and follow a scalar time-series approach using lagged values and scalar copula regression. Instead, we view the outcomes as number and mean speed of vehicles over the last hour $(t-1, t]$, allowing the interpretation as functional data over the course of one day with $t \in [0, 24]$. Adopting this view of the data, we have four dimensional multivariate functional data with two functional count outcomes (absolute number of cars, denoted as \textit{qCar}, and number of trucks, \textit{qTruck}) and two continuous and positive-valued functions (mean speed of cars in km/h, \textit{vCar}, and trucks, \textit{vTruck}).

\subsection{Data Set}

\begin{figure}
    \centering
    \begin{minipage}{0.49\textwidth}
        \includegraphics[width = \textwidth]{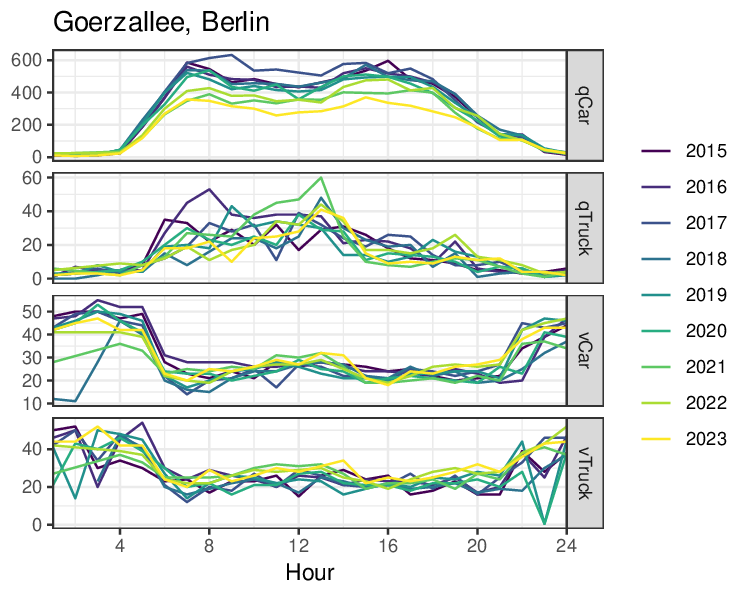}
    \end{minipage}\hfill
    \begin{minipage}{0.49\textwidth}
    \includegraphics[width=\textwidth]{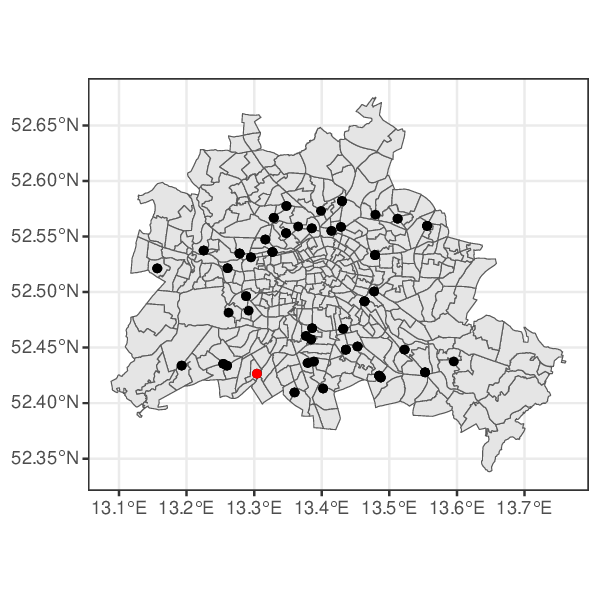}
    \end{minipage}
    \caption{\textit{Left:} Multivariate functional observations for the detection site ``Goerzallee''. \textit{Right:} Locations of all 45 selected detection sites with ``Goerzallee'' highlighted in red.}
    \label{fig:ApplicationData}
\end{figure}

In our analysis, we are interested in modeling the traffic flow of vehicles into the city center of Berlin given as the area within the Berlin S-Bahn (rapid transit system) circle line. We therefore manually select all available traffic detection sites outside of the Berlin S-Bahn circle line which are placed on streets leading towards the city center. A map of the 45 selected locations can be found in Figure \ref{fig:ApplicationData}. We include data collected from 2015 up to 2023 in our analysis. For this first analysis, in order to simplify the data structure, reduce the autocorrelation of the multivariate functional observations, and remove seasonality, we focus on the traffic on the first Monday of each September. This removes weekday and holiday effects as well as seasonal trends from the analysis and results in reasonable computing times. Figure \ref{fig:ApplicationData} shows the observed multivariate functional data for one randomly selected site over the different years.

We visually inspect the entire time-series of the sites to spot potential detector failures and consequently remove time intervals for selected sites where failure is apparent (see Figure \ref{APPfig:AppMuellerstr} in Appendix \ref{APPsec:Application} for an example). Such intervals, ranging from one multivariate observation to consecutive observations for multiple years, can be identified for 8 sites. The final data set then contains 354 multivariate functional observations with 8,024 scalar measurements. Note that missing values for \textit{vCar} and \textit{vTruck} are generated when no vehicles are measured so that only 8,022 and 7,793 scalar observations are available for these outcomes. The mean speed variables are provided as rounded integer values so we set 0 mean speed values to 0.5 (for 66 scalar observations of outcome \textit{vTruck}).

\subsection{Model Specification}

We assume negative binomial distributions for the functional counts, Gamma distributions for the positive-valued functional outcomes, and log-links to connect all additive predictors to the distributional parameters for all outcomes. The negative binomial distribution $Y^{(k')}\sim NB(\theta_1, \theta_2), k'\in \{qCar, qTruck\}$ with $\mathbb{E}(Y^{(k')})= \theta_1$ and $\mathrm{Var}(Y^{(k')}) = \theta_1 + \frac{\theta_1^2}{\theta_2}$ is often used to model count data as it allows to incorporate overdispersion in the analysis and has been shown to adequately fit the Berlin traffic data \citep[see][]{kock2023truly}. In contrast to \cite{kock2023truly}, who assume the symmetric Student t distribution for the marginal mean speed distributions and find ``significant deviations from the distributional assumption'', we choose a Gamma distribution $Y^{(k'')} \sim Ga(\theta_1, \theta_2),k'' \in \{vCar, vTruck\}$ with $\mathbb{E}(Y^{(k'')}) = \theta_1$ and $\mathrm{Var}(Y^{(k'')}) = \frac{\theta_1^2}{\theta_2}$ as this is an adequate assumption for skewed and positive data. We further assume constant scale parameters for the distributions and include covariate effects only in the location parameters. The dependency between the functional outcomes is modeled using a multilevel structure with a multivariate latent Gaussian process for the site and a nested year-specific multivariate functional random effect per site.

This gives the following additive predictors for the location parameters $\eta_{ij1}^{(k')}(t), k' \in \{qCar, qTruck\}$ and $\eta_{ij1}^{(k'')}(t), k'' \in \{vCar, vTruck\}$, and the scale parameters $\eta_{ij2}^{(k)}(t)$, $k \in \{qCar, qTruck, vCar, vTruck\}$
\begin{gather*}
    \eta_{ij1}^{(k')}(t) =     \beta_0^{(k')}(t) + f^{(k')}(year_{j}, t)+ lanes1_{i}\beta_1^{(k')}(t) + lanes3_{i}\beta_2^{(k')}(t) + \Lambda_{1i}^{(k')}(t) + \Lambda_{0ij}^{(k')}(t), \\
    \eta_{ij1}^{(k'')}(t) =  
    \beta_0^{(k'')}(t) + f^{(k'')}(year_{j}, t)+ limit30_{i}\beta_1^{(k'')}(t) + limit60_{i}\beta_2^{(k'')}(t) +\Lambda_{1i}^{(k'')}(t) + \Lambda_{0ij}^{(k'')}(t), \\
    \eta_{ij2}^{(k)}(t) = \gamma_0^{(k)},
\end{gather*}
with site $i = 1,...,45$, year $j = 2015,...,2023$, and continuous covariate $year$ for the year of the observation. We also include the dummy covariates $lanes1$, $lanes3$, $limit30$, and $limit60$ indicating the number of lanes at site $i$, which are either one (seven sites), two (reference), or three (seven sites), and whether the speed limit of the given site, determined in February 2024, is at $30$ km/h (seven sites), at $50$ km/h (reference), or at $60$ km/h or above (four sites). Slightly adapting the notation in \eqref{eq:additive_eta}, $\Lambda_{1i}^{(k)}(t)$ is the $k$-th dimension of the latent site-specific and $\Lambda_{0ij}^{(k)}(t)$ the site-year-specific random process with \gls{iid} $\bm{\Lambda}_{1i}(t) \sim \mathcal{GP}(\bm{0}(t), \bm{\mathcal{K}}_{1}(t, \cdot))$ and \gls{iid} $\bm{\Lambda}_{0ij}(t) \sim \mathcal{GP}(\bm{0}(t), \bm{\mathcal{K}}_0(t, \cdot))$. Note that the specified model showcases different possible effect specifications: we include constant effects $\gamma_0^{(k)}$, intercept functions $\beta_0^{(k)}(t)$, categorical functional effects $\beta_1^{(k)}(t)$ and $\beta_2^{(k)}(t)$, as well as 
nonlinear functional effects $f^{(k)}(year_j, t)$ in addition to nested multivariate functional random intercepts.

\subsection{Model Estimation}

To account for the multilevel structure of the data set, we follow \cite{zhou2023analysis} to estimate the univariate generalized \glspl{fpc}. Their proposed method follows the binning and fast \gls{fpca} approach presented in Section \ref{subsec:SimulationDesign} but allows to estimate \glspl{fpc} for a nested study design. A detailed description of how the approach by \cite{zhou2023analysis} is used to estimate the univariate eigenfunctions can be found in Appendix \ref{APPsubsec:APPUniFPCA}. We emphasize again, that while this approach represents a relatively fast generalized \gls{fpca} for large data sets, other methods for the univariate \gls{fpc} estimation might be suitable and could be easily substituted. The estimation of the univariate generalized \glspl{fpca} takes about 75 hours without parallelization on a Linux system with $12$ cores at $3.5$ GHz and $32$ GB memory and results in three or six (for dimension \textit{qTruck}) univariate \glspl{fpc} for the site-specific, and five (for dimension \textit{qTruck}) or six univariate \glspl{fpc} for the site-year-specific random processes as given by Figure \ref{APPApplicationFig:estimatedUFPCs} in Appendix \ref{APPsec:Application}.

We then use the posterior means of the scores in the estimation of the \gls{mfpca}. In order to get holistic representations of the site-specific and site-year-specific processes, we choose unequal weights for the multivariate scalar product, which forms the basis of the \gls{mfpca}. In fact, the estimated univariate eigenvalues given in Table \ref{APPApplicationTab:EstimatedEV} in Appendix \ref{APPsec:Application} indicate that the amount of variation of both processes differs considerably across dimensions, with the dimension $qTruck$ exhibiting most variation (sums of univariate eigenvalues $0.504$ for the site-specific and $0.174$ for the site-year-specific process) and $vCar$ the least amount of variation ($0.045$ and $0.001$, respectively). Using equal weights in the scalar product would ignore this and the corresponding \gls{mfpca} would show leading eigenfunctions that mainly capture variation on the dimension $qTruck$. Given that we also find different relative contributions of the four dimensions to the total amount of variation for both processes (see Table \ref{APPApplicationTab:EstimatedEV} in Appendix \ref{APPsec:Application}), we additionally decide to use different scalar products for the two multivariate Gaussian processes $\bm{\Lambda}_{1i}(t)$ and $\bm{\Lambda}_{0ij}(t)$. Consequently, we conduct the two separate \glspl{mfpca} with differing weights using the inverse sums of eigenvalues for each process respectively. As we use all available univariate \glspl{fpc} to avoid losing important information about the variation, we obtain $15$ \glspl{mfpc} for the site-specific random process and $23$ for the site-year-specific process.

When considering the eigenvalues of the estimated \gls{mfpca}, it becomes apparent that for both processes, the trailing \glspl{mfpc} are associated with very small contributions to the variation of the random processes (see Table \ref{APPApplicationtab:EstimatedMEVweights} in Appendix \ref{APPsec:Application}). Given that for each \gls{mfpc} included in the multivariate functional regression model a total of $45$ site-specific and, respectively, $354$ site-year-specific random scores plus variance parameter need to be estimated, we chose to introduce a truncation of the \gls{mfpc} basis at $98\%$ explained total variation of the random processes. This reduces the number of included \glspl{mfpc} to $9$ and $14$ basis functions, which implies a considerable reduction in model complexity and computation time compared to including all available \glspl{mfpc}.

The functional intercepts $\beta_0^{(k)}(t)$ and the categorical functional effects $\beta_1^{(k)}(t)$ and $\beta_2^{(k)}(t)$  are represented using $14$ cubic P-splines while the bivariate interaction terms $f^{(k)}(year_j, t)$ are modeled using anisotropic tensor product splines of $7$ (for $year$) and $14$ (for $t$) marginal cubic P-splines allowing different variance parameters for the different margins. These specifications lead to a model with around 5900 parameters, requiring about 16 GB of RAM for model fitting. After the $38$-hour backfitting algorithm to find appropriate starting values as described in Section \ref{subsubsec:Bayes}, we generate $2000$ \gls{mcmc} burn-in samples, which takes an additional $122$ hours. We use the final burn-in sample as a starting point for ten separate \gls{mcmc} chains, each generating 100 samples after a thinning of 10, resulting in a total of 1000 \gls{mcmc} samples after generating 12000 draws from the posterior. This set-up allows to parallelize the model fitting process over several adequate platforms and reduces the computation time to 61 hours per parallel chain.

\subsection{Results}

\begin{figure}
    \centering
        \includegraphics[width = \textwidth]{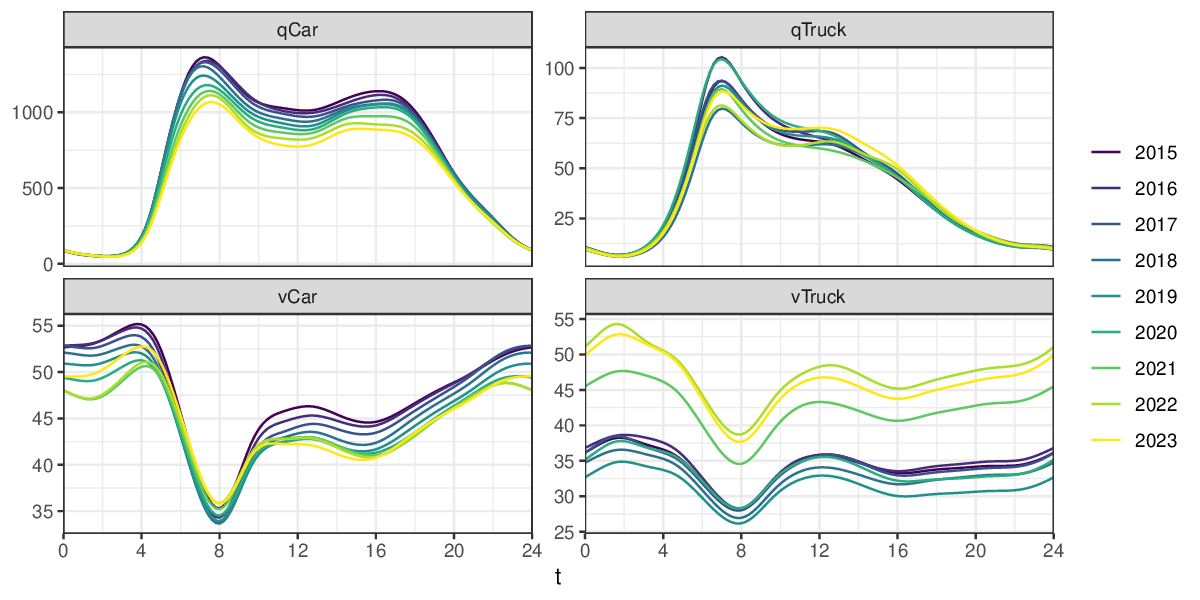}
    \caption{Estimated mean functions (in absolute numbers of vehicles for $qCar$ and $qTruck$, and in mean speed in km/h for $vCar$ and $vTruck$) for a site with two lanes and speed limit of 50 km/h over the examined years.}
    \label{fig:ApplEstimYears}
\end{figure}

In contrast to \cite{kock2023truly}, our analysis deliberately focuses on daily patterns in the data and takes the functional nature of the traffic outcomes into account. We are thus the first to describe a clear ``rush hour'' effect for traffic into Berlin as illustrated in Figure \ref{fig:ApplEstimYears} using predicted curves for sites with two lanes and speed limit $50$, which manifests itself as a prominent peak around 7 a.m.\ for the number of vehicles $qCar$ and $qTruck$ and as a slightly shifted low point of the mean speed of vehicles $vCar$ and $vTruck$ around 8 a.m. Overall, we find that the estimated mean number can be more than ten times higher for cars than for trucks during this peak. Figure \ref{fig:ApplEstimYears} also shows that the amount of early morning traffic has decreased over the observed time period, i.e.\ the peak number of cars has declined from 2015 to 2023. For the mean number of trucks, the change over the years does not follow a similar linear trend but shows more variation, which reflects the raw data: the lowest estimated rush hour peak for outcome $qTrucks$ is in 2018 and corresponds to the minimum when taking the simple (marginal) mean number of trucks at 7 a.m.\ in the data set. On the other hand, the estimated mean speed of trucks during the morning rush hour is relatively stable up until 2020 and has considerably increased since. The estimated mean speed of cars, on the other hand, exhibits smaller variation around 35 km/h over the years during the morning rush hour with the mean speed first decreasing until 2019 and increasing in later years. 

After this morning traffic, the number of vehicles slightly decreases and stabilizes around noon with a second, less pronounced peak for cars around 4 to 5 p.m.\, after which the numbers for trucks and cars diminish. During the night, there is an overall low number of vehicles, which only starts to increase roughly around 3 a.m.\ for cars and 2 a.m.\ for trucks. Over the years, the functional form of the estimated mean $qCar$ functions shows an overall downwards shift together with a flattening of the second rush hour peak. For $qTruck$, we see that the functional forms throughout the day vary more over the analyzed years. The nonlinear functional effect $f^{(k)}(year_j, t)$ is estimated using the typical \gls{famm} identifiability constraint \citep{scheipl2015functional}, which enforces a sum-to-zero constraint for all $t$. This allows to interpret the estimated effect as a local deviation from the global functional intercept and the credible interval of the estimated effect can be used for pointwise testing. We find that, for example at 4 p.m., the $95\%$ credible intervals of the estimated effect for $qTruck$ include zero over all years (without adjusting for multiple testing) and only differences in the morning hours of truck traffic would be considered statistically significant at the $5\%$ level (see Figure \ref{APPApplicationfig:Inference} in Appendix \ref{APPsec:Application}).

For the mean speed outcomes, there is an increase after the morning rush hour to a slight peak around noon for both cars and trucks. During the second peak in the number of cars at 4 p.m., we again see a slight drop in the mean speed outcomes for cars and trucks, followed by a steady increase throughout the night, reaching a high point around 4 a.m.\ for cars and 2 a.m.\ for trucks. We find that the overall change in mean speed over the course of the day is much more pronounced for cars than for trucks, as for example in 2015, the mean speed for cars ranges from about 35 km/h to 55 km/h, while the range for trucks is  roughly 28 km/h to 38 km/h. Over the years, this range has somewhat decreased for cars and increased for trucks. Additionally, for trucks, there is an almost constant shift (across $t$) of the estimated mean functions over the years: first slightly downwards from 2015 to 2019 followed by a stark upwards shift until 2022. For cars, the estimated mean functions for a location with two lanes and speed limit of 50 km/h show a relatively uniform downwards trend over the observed years except during the morning rush hour as described above. We thus conclude that over the last couple of years, especially car traffic has reduced and slowed down in Berlin while truck traffic might have only slightly decreased during the morning rush hour and even increased its mean speed over the entire day. For cars, at least, this development seems to be in line with the proclaimed efforts and goals of the Berlin government to reduce emissions. Note, however, that our analysis does not imply a causal effect of any particular policy measure.

\begin{figure}
    \centering
        \includegraphics[width = 0.8\textwidth]{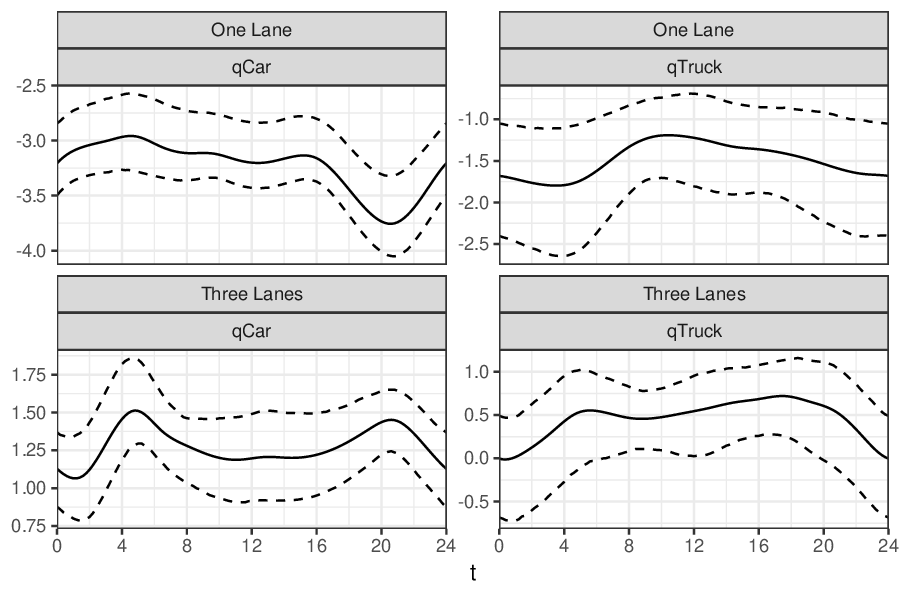}
    \caption{Estimated coefficient effect functions $\hat{\beta}_1^{(k')}(t)$ and $\hat{\beta}_2^{(k')}(t), k' \in \{qCar, qTruck\}$ for sites with one or three lanes, respectively, on the linear predictor scale. The dashed lines show the pointwise $95\%$ credible intervals.}
    \label{fig:ApplEstimLanes}
\end{figure}

The estimated coefficient effect functions for the considered covariates might also be interesting to policy makers as they can help to identify possible pathways for influencing traffic flows. Figure \ref{fig:ApplEstimLanes} shows the estimated coefficient effect functions $\hat{\beta}_1^{(k')}(t)$ and $\hat{\beta}_2^{(k')}(t), k' \in \{qCar, qTruck\}$ depicting the difference for sites with one or three lanes, respectively, compared to two lanes on the additive predictor scale. Keep in mind that e.g.\ a positive effect of $\hat{\beta}_1^{(k')}(t)$ on the additive predictor scale, using a log link, corresponds to a multiplicative effect on the scale of the distributional parameter by the factor $\exp{\hat{\beta}_1^{(k')}(t)}$, which is then larger than one. We see that, all other things being equal, a site with one lane has considerably less traffic than a site with two lanes, corresponding to the estimated effects for both $qCar$ and $qTruck$ being significantly negative (compare the indicated pointwise $95\%$ credible intervals) over the entire day. The estimated mean function for $qCar$ is particularly decreased between 8 p.m.\ and 9 p.m., leading to even less cars in that hour compared to a site with two lanes. Note, however, that a credible region constructed from the pointwise credible intervals does not correctly account for the functional correlation and should not formally be used to draw conclusions about the functional form such as the linearity of the estimate. The effect estimate $\hat{\beta}_2^{(k')}(t)$ (bottom row in Figure \ref{fig:ApplEstimLanes}) allows to assess the difference between a site with three and two lanes. We find a strong positive effect on the number of cars over the entire day, while the number of trucks sees a significantly positive effect only during daytime. For both estimated coefficient effect functions, the effect sizes are much more pronounced for cars than for trucks, suggesting that changing the number of lanes would have a much larger impact on the number of cars passing the respective site. 

Figure \ref{APPApplicationfig:BetaLimit} in Appendix \ref{APPsec:Application} shows the estimated coefficient effect functions for the covariate speed limit, which show few areas with significant effects on the estimated mean functions of $vCar$ and $vTruck$, except for a significant increase of mean speed of trucks during the day for sites with a 60 km/h speed limit (or above) compared to 50 km/h. The multivariate nature of our regression approach allows to directly formulate and test joint hypotheses across dimensions, such as $H_0: \hat{\beta}_1^{(vCar)}(t) = \hat{\beta}_1^{(vTruck)}(t) = 0$ for a given $t$, which are helpful to assess whether a covariate might be dropped from the model altogether. Figure \ref{APPApplicationfig:JointLimit} in Appendix \ref{APPsec:Application} demonstrates one such test for $t = 12$ using empirical two-dimensional credible regions based on a kernel density estimate of the \gls{mcmc} samples, indicating that the null hypothesis could only be rejected at a significance level of $10\%$ but not at the $5\%$ level. This illustrates that the multivariate functional regression model provides a useful framework for efficiently combining information across dimensions and generating additional insights not available from univariate models.

\begin{figure}
    \centering
    \begin{minipage}{0.49\textwidth}
        \includegraphics[width = \textwidth]{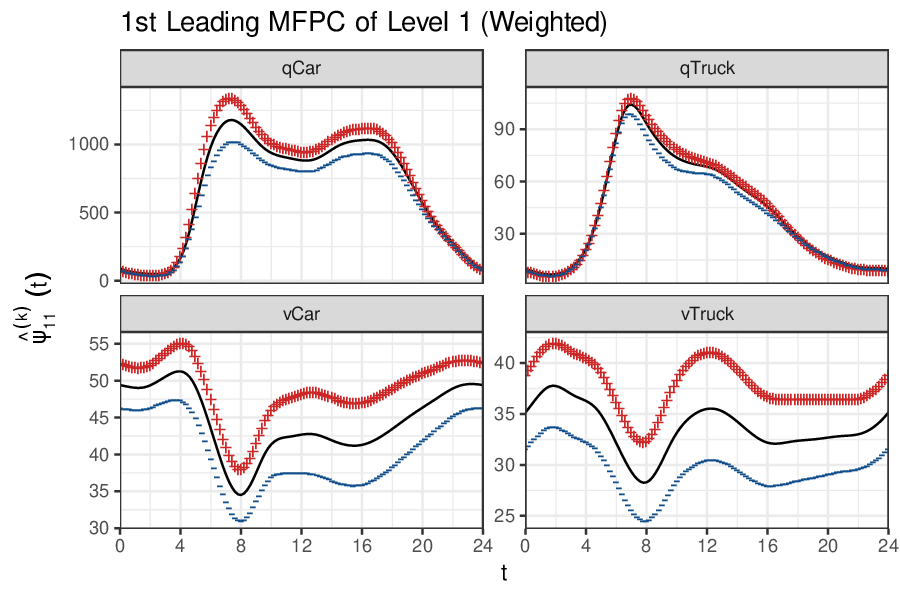}
    \end{minipage}\hfill
    \begin{minipage}{0.49\textwidth}
        \includegraphics[width=\textwidth]{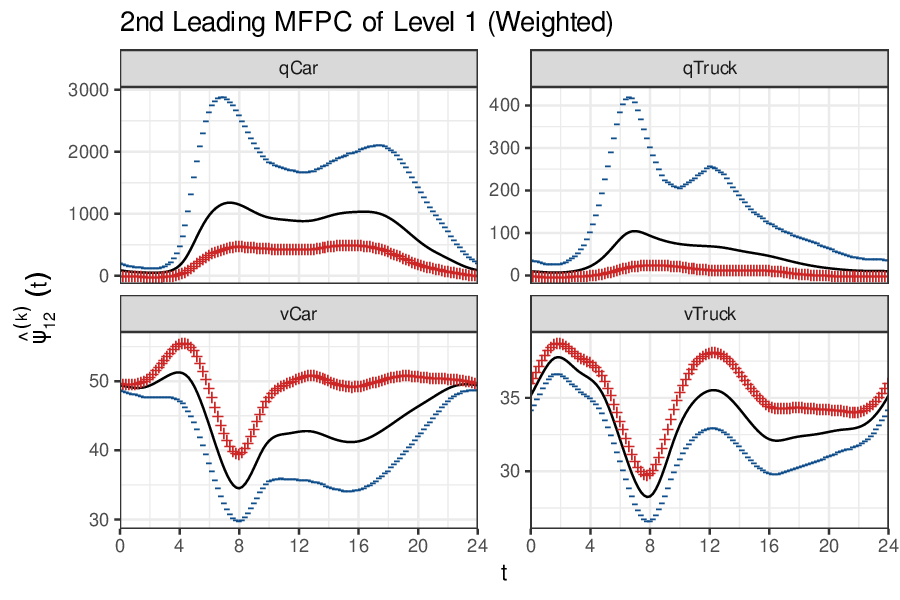}
    \end{minipage}
    \begin{minipage}{0.49\textwidth}
        \includegraphics[width=\textwidth]{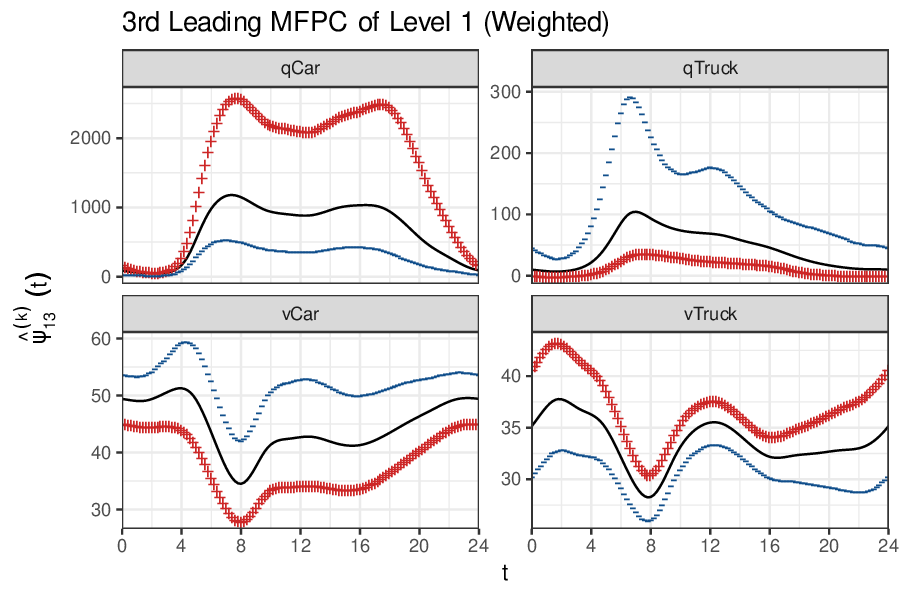}
    \end{minipage}
    \caption{Leading three estimated \glspl{mfpc} based on the weighted scalar product for the site-specific (level 1) random process. The black line corresponds to the estimated mean of a site with two lanes and 50 km/h speed limit in 2020 with plus (red +) or minus (blue -) $2\sqrt{\hat{\nu}_{1m}}, m = 1, 2, 3$ times the estimated \gls{mfpc}.}
    \label{fig:EstimatedWMFPCs}
\end{figure}

Similarly, our multivariate approach provides empirical basis functions, which capture the multivariate functional dependency structure present in the data. Figure \ref{fig:EstimatedWMFPCs} shows the first three estimated eigenfunctions for the site-specific process as the mean function for a site with two lanes and a speed limit of 50 km/h in 2020 (black line) plus (red +) or minus (blue -) $2\sqrt{\hat{\nu}_{1m}}$ times the estimated \gls{mfpc} $\hat{\psi}_{1m}^{(k)}(t)$, where $\hat{\nu}_{1m}, m = 1, 2,3$ is the posterior variance of the random scores. The first leading \gls{mfpc} $\hat{\psi}_{11}^{(k)}(t)$ accounts for about $3\%$ of the total (unweighted) site-specific (posterior) variation and captures differences between light and heavy traffic sites, i.e.\ more or less vehicles together with higher or lower mean speeds, without really changing the functional form of the mean function. The second estimated \gls{mfpc} $\hat{\psi}_{12}^{(k)}(t)$ shows a negative association between the number of vehicles and the mean speed so that e.g.\ a higher amount of traffic reduces the mean speed of the vehicles. This mode of site-specific variation is the second leading mode in the preliminary (weighted) \gls{mfpc} estimation but shows the highest posterior variance of the site-specific \gls{mfpc} basis functions, accounting for about $40\%$ of the (posterior unweighted) variation. This illustrates that the \gls{mfpca} here estimates multivariate eigenfunctions using weighted variances, upweighing dimensions with smaller variances, and ordering of \gls{mfpc}s may flip when evaluating importance in terms of unweighted variances: The first \gls{mfpc} mostly loads on the two speed dimensions with small absolute variation, which are upweighted in the \gls{mfpca}, and thus corresponds to a relatively small amount of explained unweighted variance, while the second \gls{mfpc} with large contributions to the number of vehicles dimensions is more important in these terms. The third leading \gls{mfpc} again accounts for a large proportion, about $38\%$, of the total site-specific (posterior unweighted) variation and depicts a contrast between the types of vehicles: For example, a site with a large score on this component would feature an overall higher number of cars with lower mean speed but simultaneously fewer trucks with higher speed.

Generally speaking, the interpretation of the leading \glspl{mfpc} is easiest since they capture the main modes of variation in the data and all following \glspl{mfpc} must reflect modes orthogonal to them. For the site-year-specific random process, the \gls{mfpc} accounting for the the most (posterior unweighted) variation ($43\%$) is only the fifth, which somewhat complicates its interpretation (compare Figure \ref{APPApplicationfig:EstimatedWMFPCsLev2} in Appendix \ref{APPsec:Application}). One apparent feature of this \gls{mfpc}, however, is its distinctive impact on the number of trucks late in the day, leading to a stark increase (or decrease) of traffic in the afternoon. The leading \gls{mfpc} $\hat{\psi}_{01}^{(k)}(t)$ only accounts for $8\%$ of the total site-year-specific (posterior unweighted) variation, which constitutes the second largest contribution to the total (posterior unweighted) variation (compare Table \ref{APPApplicationtab:EstimatedMEVPost} in Appendix \ref{APPsec:Application}). It captures a similar mode to $\hat{\psi}_{11}^{(k)}(t)$ in Figure \ref{fig:EstimatedWMFPCs} with only slight differences (see Figure \ref{APPApplicationfig:EstimatedWMFPCsLev2} in Appendix \ref{APPsec:Application}). Note that a direct comparison between the processes is not straightforward as the scalar products used in the respective \gls{mfpca} are different. Overall, we find that the qualitative examination of the empirical basis functions can provide first insights into the dependency structure in the data even though the proposed two-step approach produces ``only'' an empirical basis.

The presented model fits the data reasonably well (see Figure \ref{APPApplicationfig:ModelFits} in Appendix \ref{APPsec:Application} for exemplary fitted curves of the location parameter for the detection site ``Goerzallee'') but analyzing the normalized (randomized) quantile residuals \citep{dunn1996randomized} suggests that the model might be further improved (compare Figure \ref{APPApplicationfig:WormPlots} in Appendix \ref{APPsec:Application} for outcome-specific worm plots \citep{buuren2001worm}). While the presented plots seem to indicate a strong deviation from the distributional assumption, in particular for the mean speeds, we would like to point out that by estimating the \glspl{mfpc}, we might find the worm plots affected by e.g.\ residual auto-correlation. Increasing the flexibility of the model for the scale parameters is also straightforward in our modeling framework and might improve the residual diagnostics at the expense of additional computation time. Our analysis, however, has already shown that the proposed (generalized) multivariate functional regression models allows to identify daily characteristics of traffic flows and model their change over several years, as well as their dependence on different covariate effects, while incorporating information from various dimensions simultaneously. This provides policy makers with the means to better understand traffic flows in the Berlin capital, evaluate the effect of specific measures, or identify new avenues of potential traffic management.

\section{Discussion and Outlook}
\label{sec:Discussion}

We propose a general flexible regression framework to handle generalized multivariate functional data of mixed type. In order to incorporate the covariance between and within the functional dimensions, we assume independent latent multivariate Gaussian processes and estimate their covariance functions from the data via \gls{mfpca}. This allows a parsimonious representation of the latent processes and provides a direct tool for introducing further regularization in the model by truncating the multivariate \gls{kl} representation of the latent processes. We further extend this approach to generalized multivariate functional data which exhibit a multilevel structure such as repeated observations for independent subjects. As the model builds on the \gls{famm} framework, a wide range of covariate effects, e.g.\ linear, non-linear functional as well as interaction effects, can be included in the regression models. Our simulation study shows that a multivariate functional analysis approach is beneficial for fitting the functional data compared to separate univariate functional models, and provides robust estimation of covariate effects over several scenarios, including for challenging sparse functional data. Furthermore, we showcase our regression framework in an application to model traffic data in Berlin. To the best of our knowledge, this is the first time that an analysis of this complex data source has incorporated the (multivariate) functional nature of traffic flows. We are thus able to generate new insight into characteristic traffic patterns as well as changes over time or influences of additional characteristics. The analysis simultaneously models all available functional dimensions consisting of the number and mean speed of different types of vehicles and promotes the understanding of dependency structures within and between these dimensions.

While we present one avenue of estimating the \gls{mfpc} basis, our two-step procedure can be easily adapted to account for new and improved \gls{mfpca} strategies, as only the estimated \glspl{mfpc} are required for the model fit. Generally, we find that the estimation of the \glspl{mfpc} can have a substantial direct effect on the quality of the model fit and should thus be handled with care. In recent years, there have been some considerable developments for generalized univariate \gls{fpca}, but well-thought-out large-scale comparison studies might be helpful to establish practicable guidelines and identify open issues. Simultaneous estimation of the \glspl{mfpc}, scores, and covariate effects in one multivariate functional regression model would arguably be more user-friendly and eliminate the implication that the resulting inference is conditional on the estimated \gls{mfpc} basis. Such an approach, however, is challenging in our complex setting: it would require, for example, the extension of existing methods for univariate multilevel \gls{fpca} for generalized functional data \citep{goldsmith2015generalized} to multivariate functional data while also supplying the powerful modeling possibilities of our flexible regression framework.

One important aspect to consider in such endeavors is providing efficient implementations of newly developed methods. Our analysis of the Berlin traffic data is restricted to a small subset of the original data base, in part due to the computational demands of the generalized multivariate functional regression model. It will therefore be of primary interest in the future to further improve the method's efficiency to handle large scale data sets by e.g.\ fully exploiting sparse structures in the design matrices and using more efficient programming. Improving the scalability of our regression framework is already a work in progress within a new \textbf{R} package. Alternatively, using approximate Bayesian methods such as Integrated Nested Laplace Approximation \citep{rue2009approximate} or penalized frequentist likelihood approaches \citep{rigby2005generalized} present promising alternatives to our computationally quite intensive \gls{mcmc} sampling algorithm. 

Improving our implementation will make the full range of our proposed framework accessible to the statistician and provide several opportunities to improve the model fit in our analysis. Including more \glspl{mfpc} in the application model can improve the residual diagnostics by supplying more flexibility to the location parameter. Introducing covariate effects in the additive predictors of the scale parameters further increases the flexibility of the model. Similarly, other pointwise distributional assumptions for the generalized multivariate functional data are generally available in our framework, yet our experience shows that distributions with more distributional parameters can lead to noticeably increased computation times. Further combining the available data with external data sources might also lead to a better model fit and provide deeper insight into the evolution of daily traffic patterns in Berlin. Including weather data or data about policy measures with respect to infrastructure or epidemiologically motivated sanctions might provide further relevant information to accurately model the data and interpret the estimated effects.

Throughout this paper, we assume that the latent processes are Gaussian -- a common assumption, but the effects of a misspecification in this aspect are not yet fully understood. Further simulations might also focus on the impact of misspecifying the pointwise distributional assumptions of the (generalized) functional data. Despite these remaining questions and avenues for further improvements and research, we are able to considerably expand the previously available toolkit for and improve the modeling of generalized multivariate functional data of mixed types.

\bibliography{gmfamm}

\newpage
\appendix

\section{Simulation}
\label{APPsec:Simulation}

\subsection{Simulation Design}
\begin{figure}
    \centering
    \includegraphics[width=\textwidth]{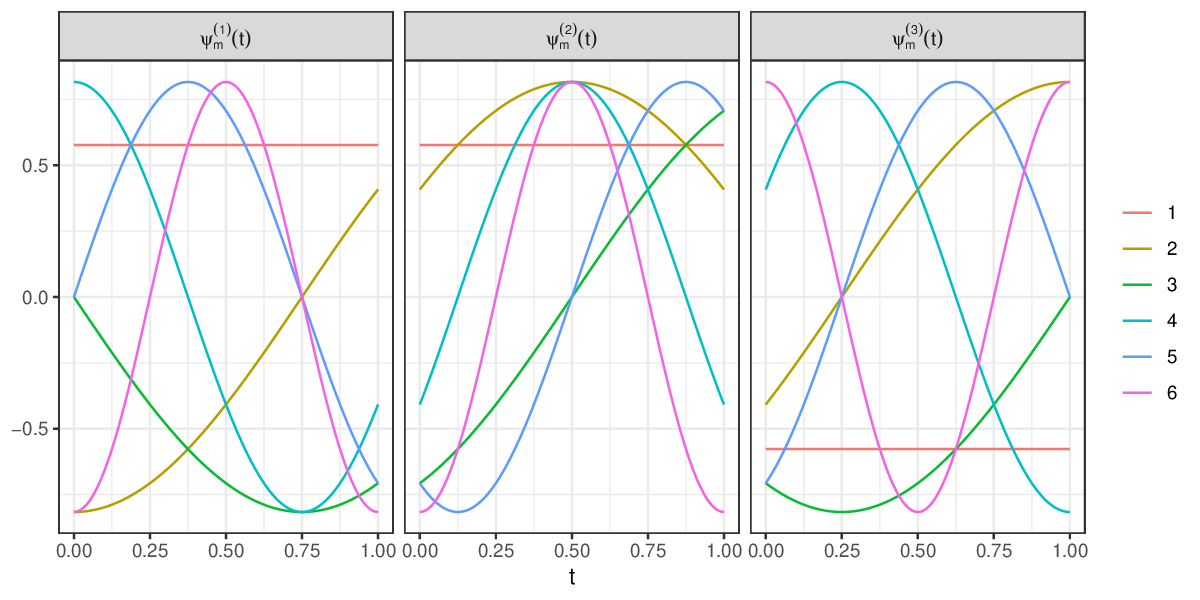}
    \caption{Multivariate eigenfunction basis used to simulate the latent process $\bm{\Lambda}_i(t)$.}
    \label{APPfig:SimMFPCs}
\end{figure}

\begin{figure}
    \centering
    \includegraphics[width=0.8\textwidth]{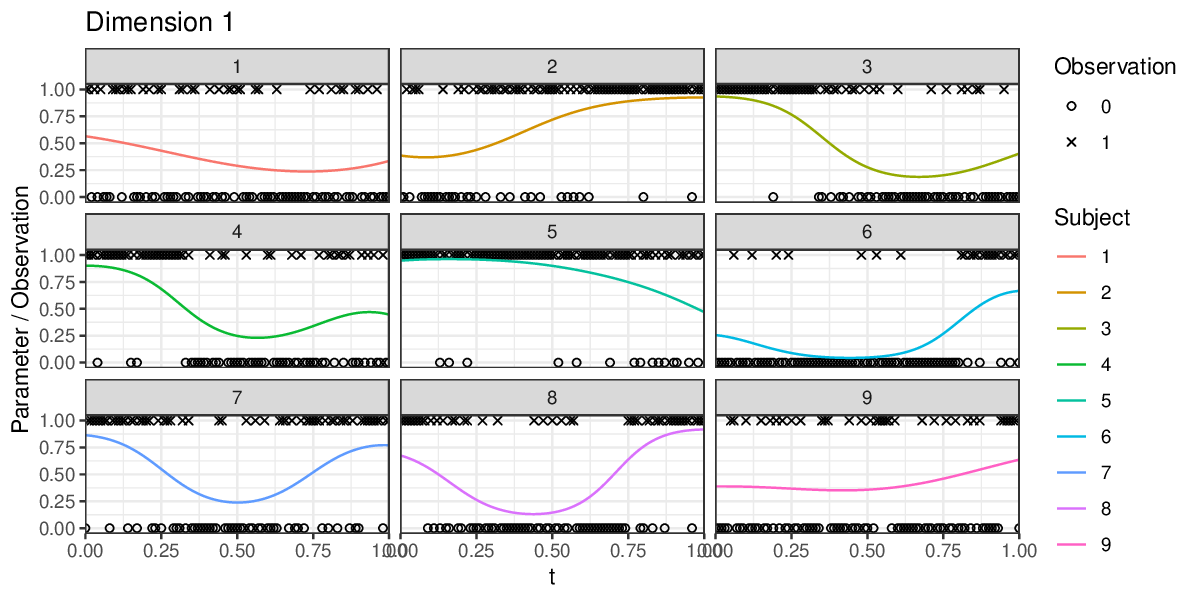}
    \includegraphics[width=0.8\textwidth]{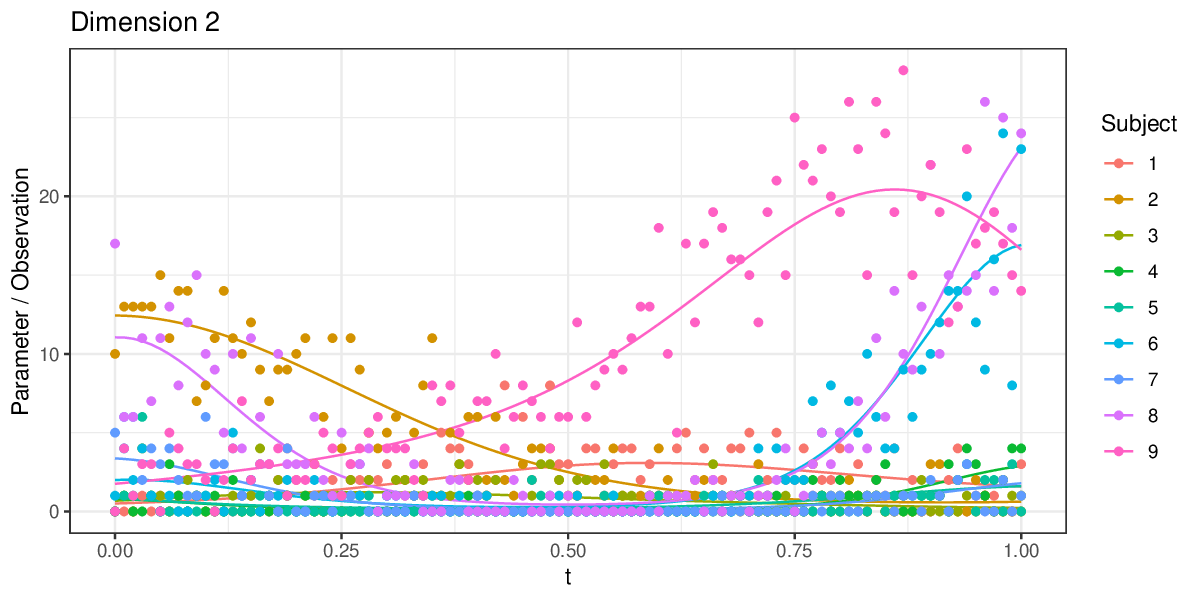}
    \includegraphics[width=0.8\textwidth]{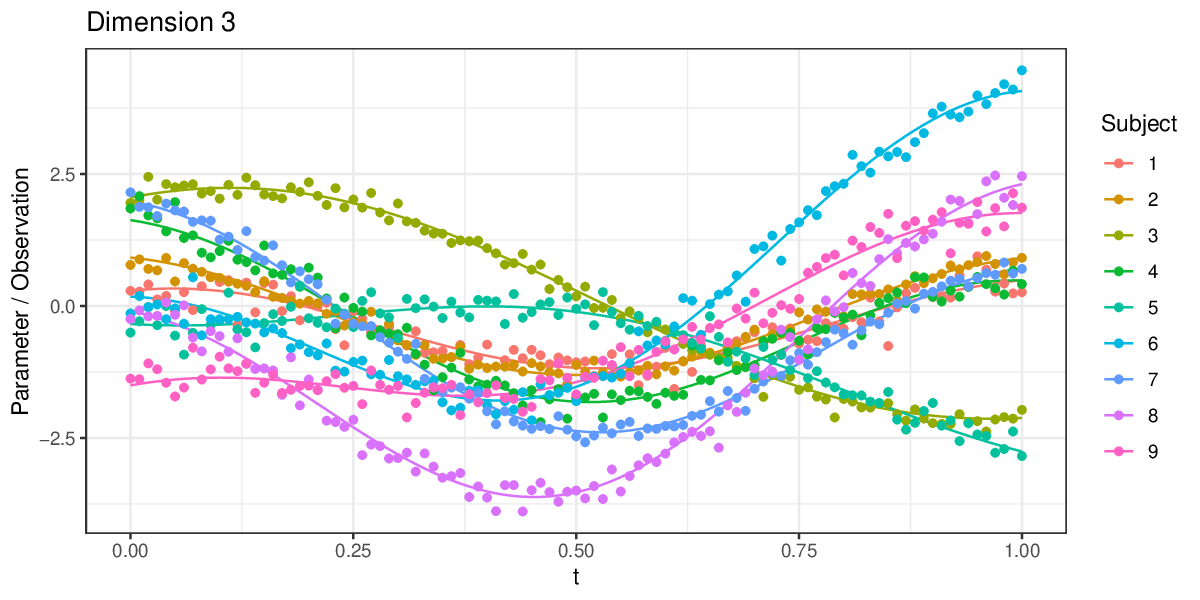}
    \caption{Nine exemplary generalized multivariate functions with observations on a dense, equidistant grid of distance 0.01. The plots show the distributional parameters $\theta_{i1}^{(k)}(t), k = 1,2,3$ as lines together with the observed values $y_{it}^{(k)}$. Data used in the simulations consist of sparsified versions of these observations.}
    \label{APPfig:SimSampleObs}
\end{figure}

Figure \ref{APPfig:SimMFPCs} displays the six eigenfunctions used to simulate the generalized multivariate functional data with Figure \ref{APPfig:SimSampleObs} showing nine exemplary simulated multivariate functions including scalar observations.

\subsection{Additional Results}

\begin{figure}
    \centering
    \begin{minipage}{0.49\textwidth}
        \includegraphics[width=\textwidth]{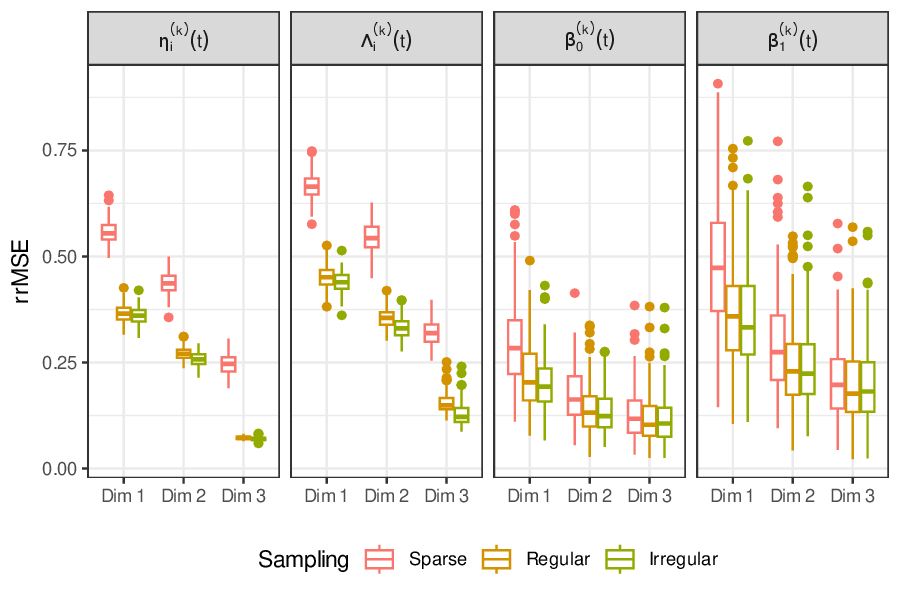}
        \includegraphics[width=\textwidth]{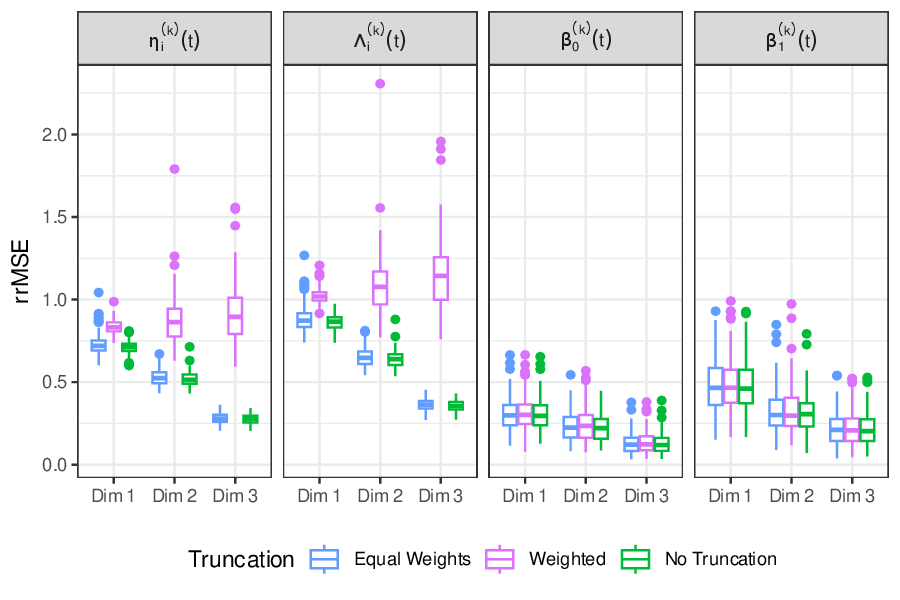}
    \end{minipage}\hfill
    \begin{minipage}{0.49\textwidth}
        \includegraphics[width=\textwidth]{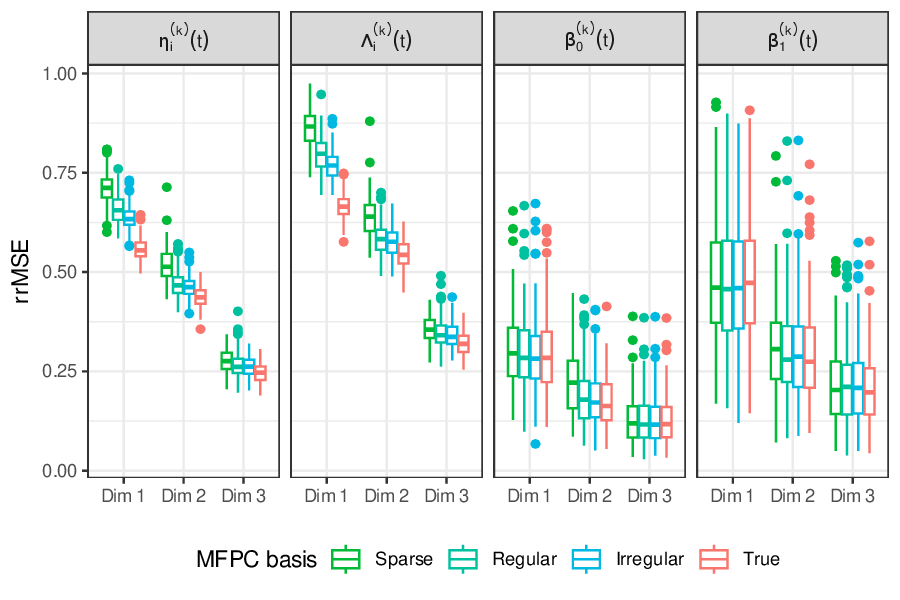}
        \includegraphics[width=\textwidth]{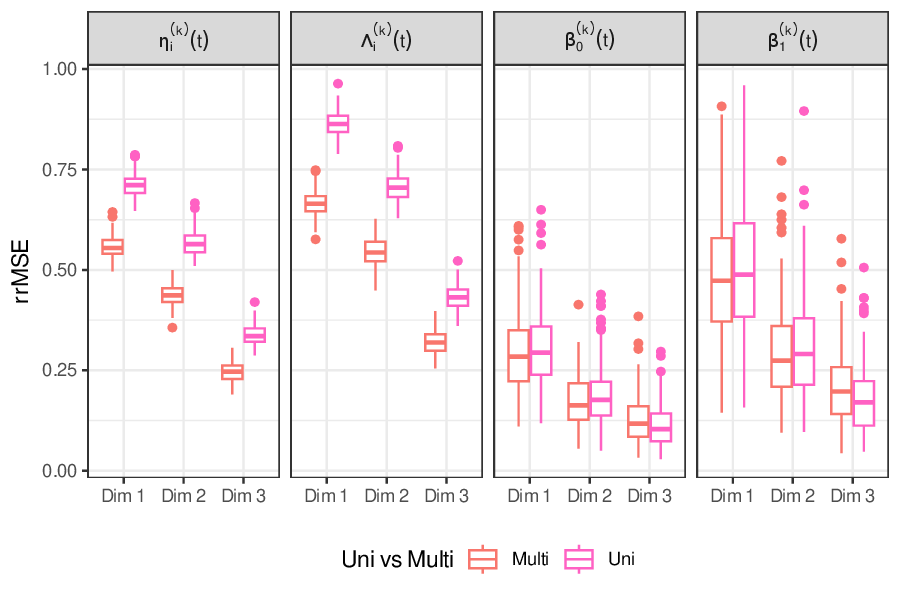}
    \end{minipage}  
        \caption{Boxplots of \gls{rrmse} values for different scenarios. \textit{Top left:} Impact of sampling regime, corresponding to Table \ref{subtab:Sampling}. \textit{Top right:} Impact of \gls{mfpc} basis, corresponding to Table \ref{subtab:MFPCest}. \textit{Bottom left:} Impact of \gls{mfpc} truncation, corresponding to Table \ref{subtab:mfpcTrunc}. \textit{Bottom right:} Univariate vs.\ multivariate, corresponding to Table \ref{subtab:UniMul}.}
    \label{APPfig:SimRRMSEboxplots}
\end{figure}

\begin{figure}
    \centering
    \begin{minipage}{0.49\textwidth}
        \includegraphics[width = \textwidth]{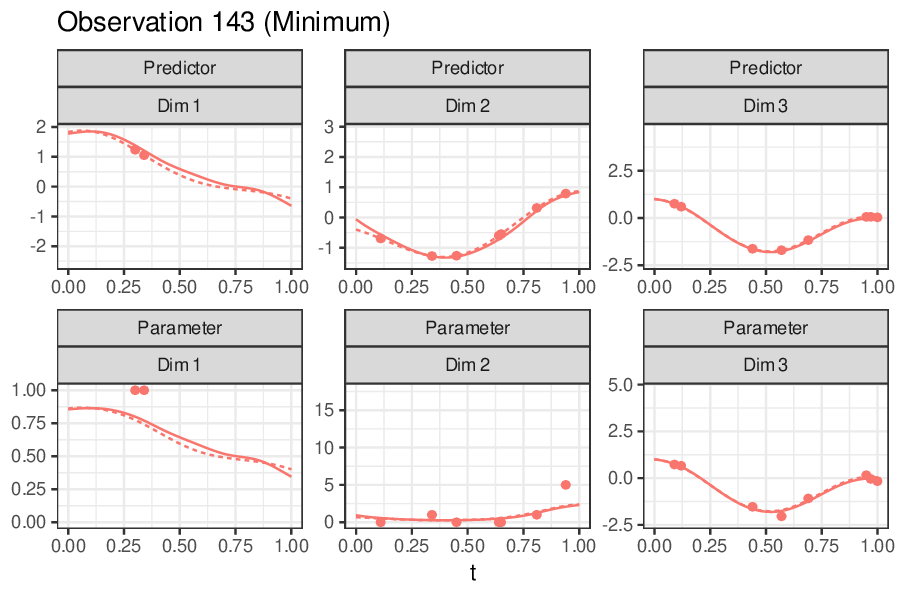}
        \includegraphics[width = \textwidth]{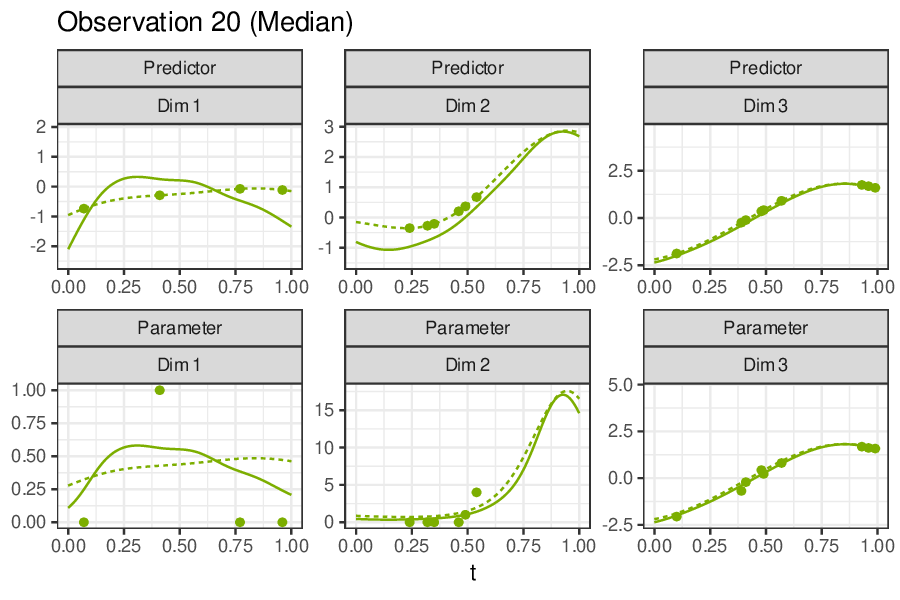}
    \end{minipage}\hfill
    \begin{minipage}{0.49\textwidth}
        \includegraphics[width = \textwidth]{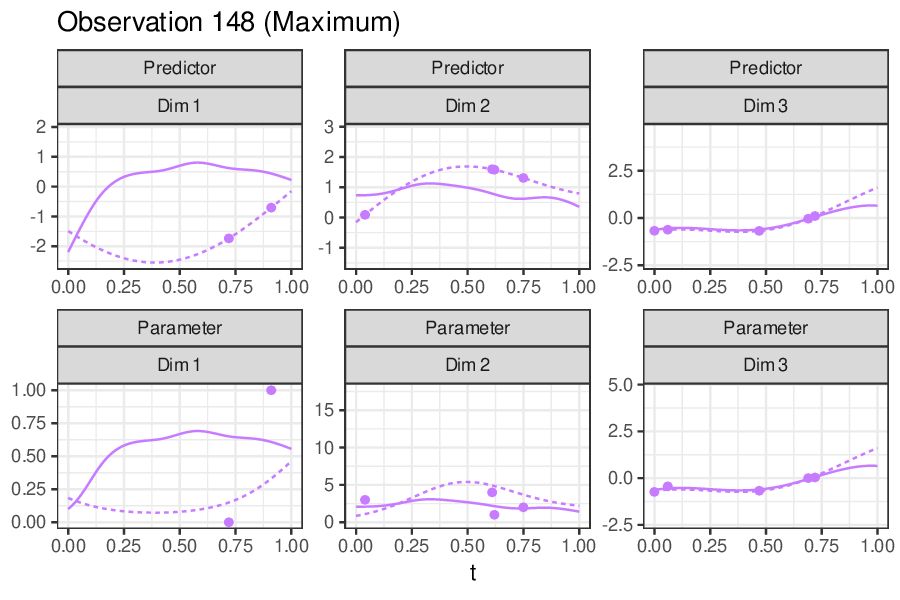}
        \includegraphics[width = \textwidth]{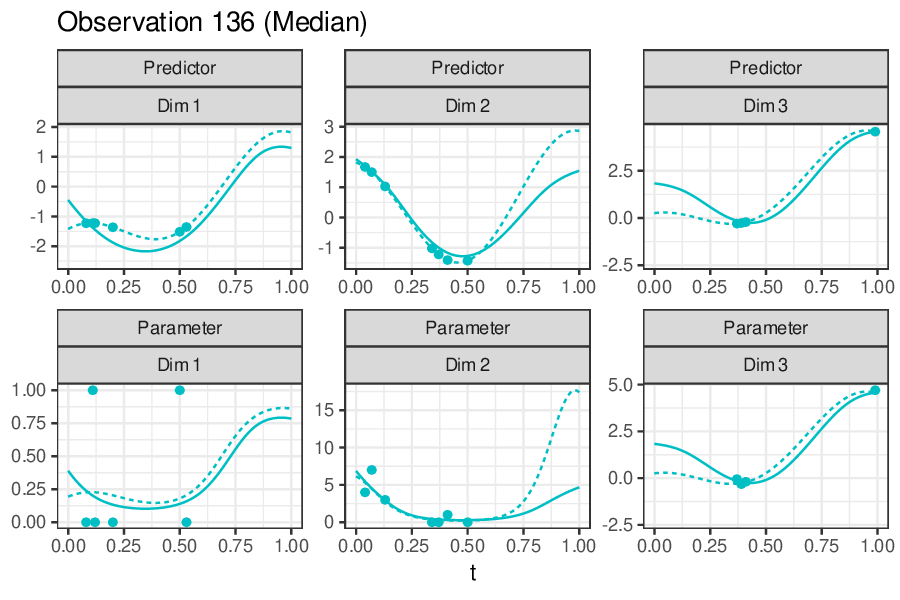}
    \end{minipage}
    \caption{Exemplary model fits for selected observations in an iteration of the sparse sampling scenario with true eigenfunctions. Observations are chosen corresponding to the minimum, maximum, and median values of $|||\eta_{i1} - \hat{\eta}_{i1}|||$ over all $n = 150$ observations. The dashed lines are the true data generating $\eta_{i1}^{(k)}(t)$ (\textit{Predictor}) and $\theta_{i1}^{(k)}$ (\textit{Parameter}) with dots indicating observations.}
    \label{APPfig:SimSampleFits}
\end{figure}

\begin{figure}
    \centering
    \begin{minipage}{0.49\textwidth}
        \includegraphics[width = \textwidth]{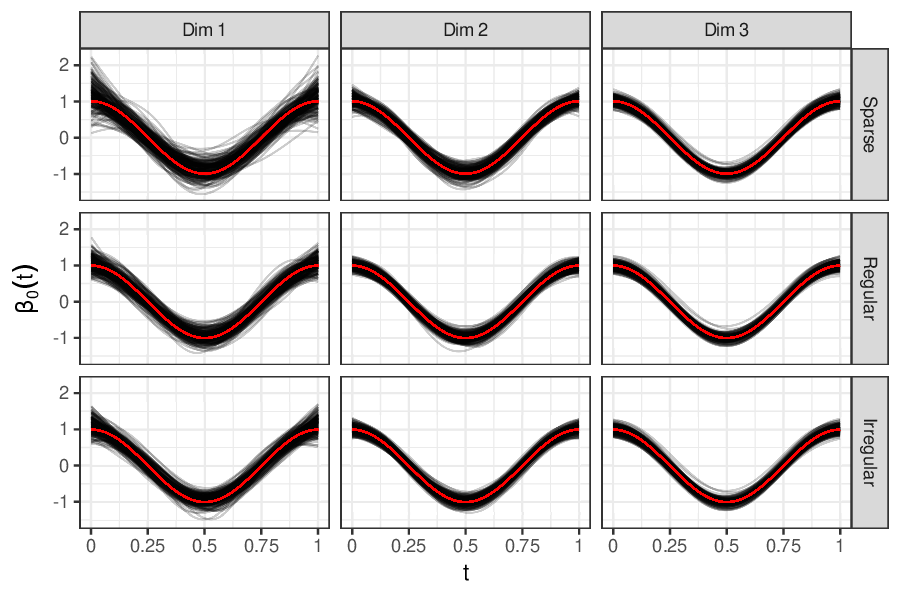}
    \end{minipage}\hfill
    \begin{minipage}{0.49\textwidth}
        \includegraphics[width = \textwidth]{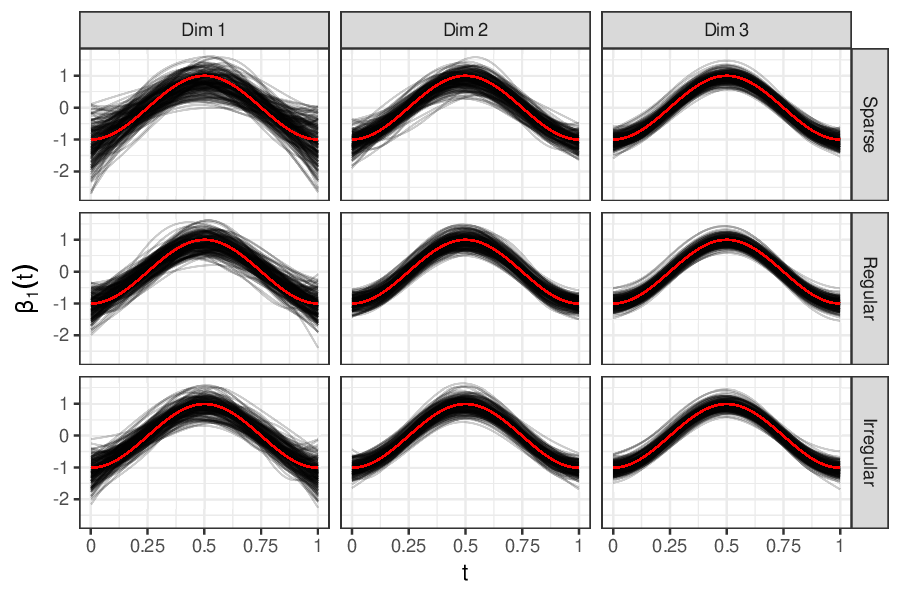}
    \end{minipage}
    \caption{Estimated coefficient effect functions $\beta_0^{(k)}(t)$ (left) and $\beta_1^{(k)}(t)$ (right) for different $k$ and different sampling regimes.}
    \label{APPfig:SimSampleBetas}
\end{figure}

\begin{figure}
    \centering
    \begin{minipage}{0.49\textwidth}
        \includegraphics[width=\textwidth]{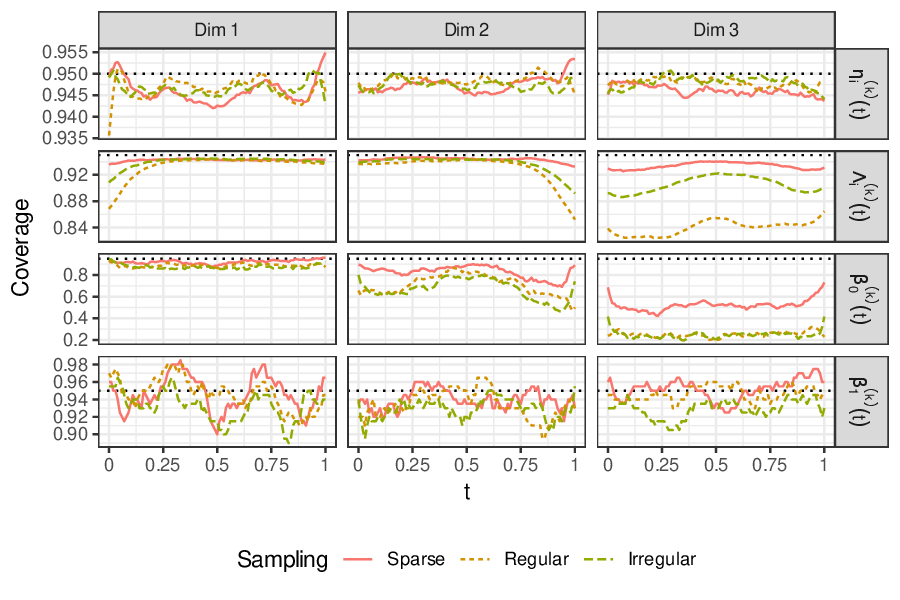}
        \includegraphics[width=\textwidth]{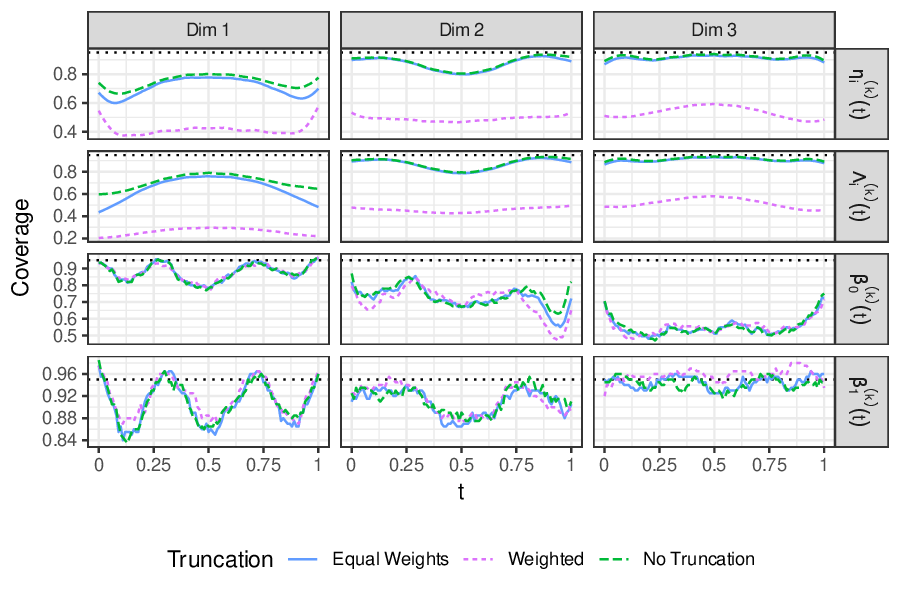}
    \end{minipage}\hfill
    \begin{minipage}{0.49\textwidth}
        \includegraphics[width=\textwidth]{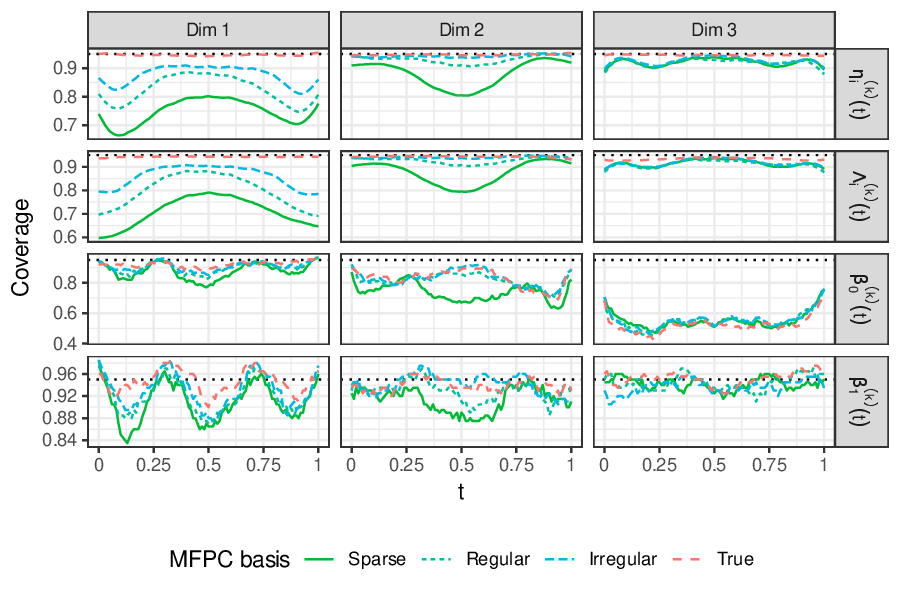}
        \includegraphics[width=\textwidth]{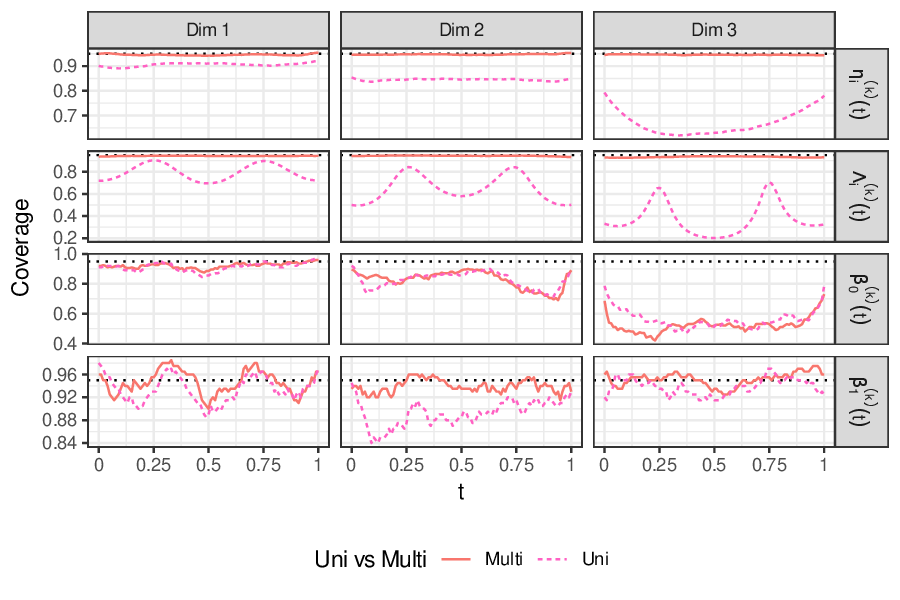}
    \end{minipage}  
        \caption{Pointwise evaluation of the frequentist coverage of Bayesian credible intervals of the different model components for different scenarios. The black dotted line corresponds to the nominal coverage of $0.95$. \textit{Top left:} Impact of sampling regime, corresponding to Table \ref{subtab:Sampling}. \textit{Top right:} Impact of \gls{mfpc} basis, corresponding to Table \ref{subtab:MFPCest}. \textit{Bottom left:} Impact of \gls{mfpc} truncation, corresponding to Table \ref{subtab:mfpcTrunc}. \textit{Bottom right:} Univariate vs.\ multivariate, corresponding to Table \ref{subtab:UniMul}.}
    \label{APPfig:SimCoverage}
\end{figure}

Figure \ref{APPfig:SimRRMSEboxplots} shows the \gls{rrmse} values of the additive predictor $\eta_{i1}^{(k)}(t)$ of the location parameter, the latent process $\Lambda_{0i}^{(k)}(t)$, and the coefficient effect functions $\beta_{0}^{(k)}(t)$ and $\beta_1^{(k)}(t)$ for all examined scenarios and all 200 simulation iterations. Figure \ref{APPfig:SimSampleFits} shows  exemplary model fits for a simulation iteration with \gls{rrmse} values $\eta_{i1}^{(1)}(t) = 0.556$, $\eta_{i1}^{(2)}(t) = 0.439$, $\eta_{i1}^{(3)}(t) = 0.211$, which is close to the overall mean \gls{rrmse} values, in the sparse sampling scenario using the true eigenfunctions. Figure \ref{APPfig:SimSampleBetas} shows the estimated coefficient effect functions of all multivariate regression models using the true eigenfunctions as an \gls{mfpc} basis with data from the different sampling regimes. Both the overall model fits and the coefficient effect functions are well estimated. Figure \ref{APPfig:SimCoverage} shows the pointwise frequentist coverage of the Bayesian credible intervals for all model components over all examined simulation scenarios. Table \ref{APPtab:cred_widths} shows the mean width of credible intervals for the coefficient effect functions $\beta_0^{(k)}(t)$ and $\beta_1^{(k)}(t)$. It is apparent that especially for $\beta_0^{(3)}(t)$ on the Gaussian dimension, the posterior credible intervals are very narrow, leading to a low frequentist coverage of the estimates.

\begin{figure}
    \centering
    \includegraphics[width = 0.6\textwidth]{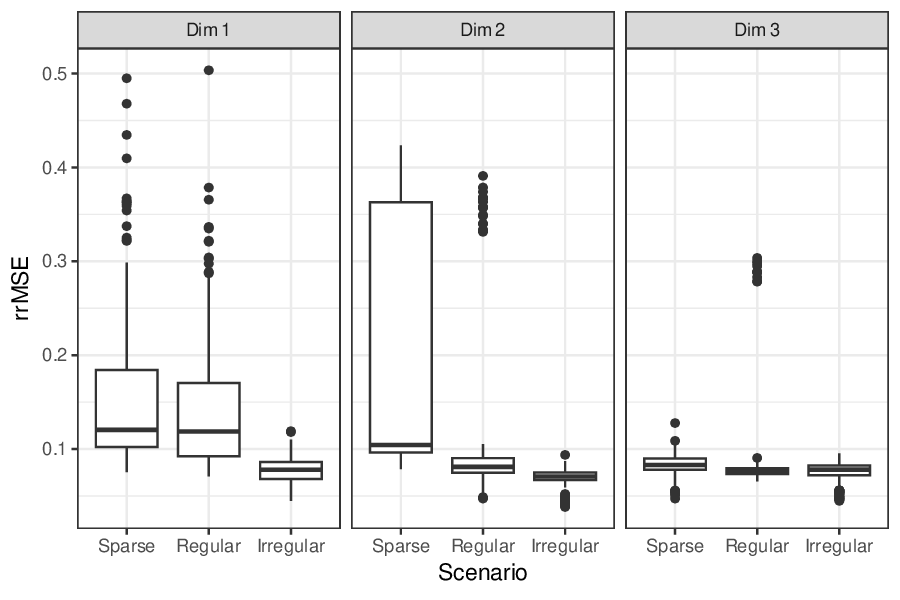}
    \caption{Boxplots of \gls{rrmse} values for the reconstructed latent process for different sampling regimes.}
    \label{APPfig:SimMFPCestRRMSE}
\end{figure}

\begin{figure}
    \centering
    \includegraphics[width = 0.6\textwidth]{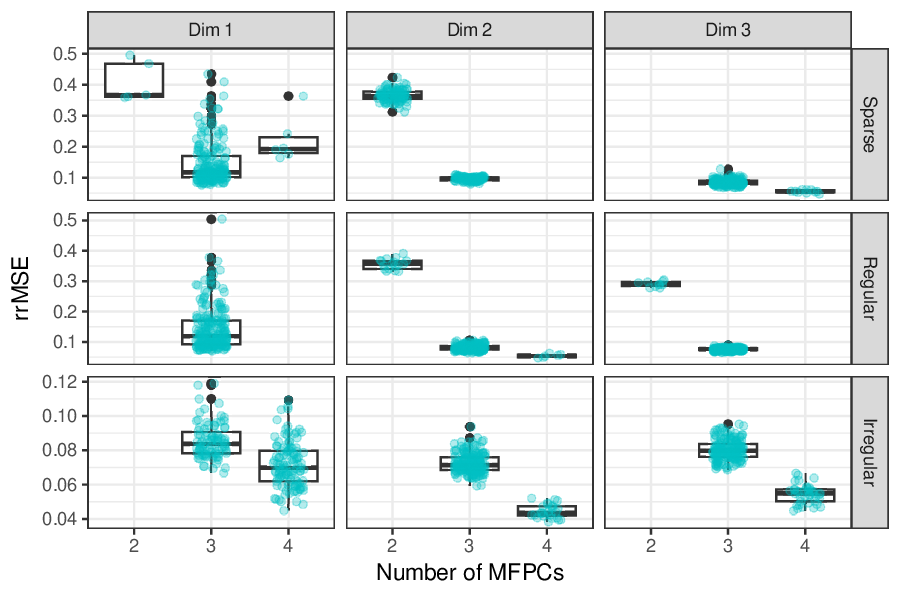}
    \caption{Boxplots of the \gls{rrmse} values for the reconstructed latent process for different sampling regimes stratified by the number of estimated univariate \glspl{fpc}. The \gls{rrmse} values of the 200 simulation iterations are overlaid in blue to show the numbers of iterations contributing to the boxplots.}
    \label{APPfig:SimMFPCestNumb}
\end{figure}

\begin{figure}
    \centering
    \begin{minipage}{0.49\textwidth}
        \includegraphics[width = \textwidth]{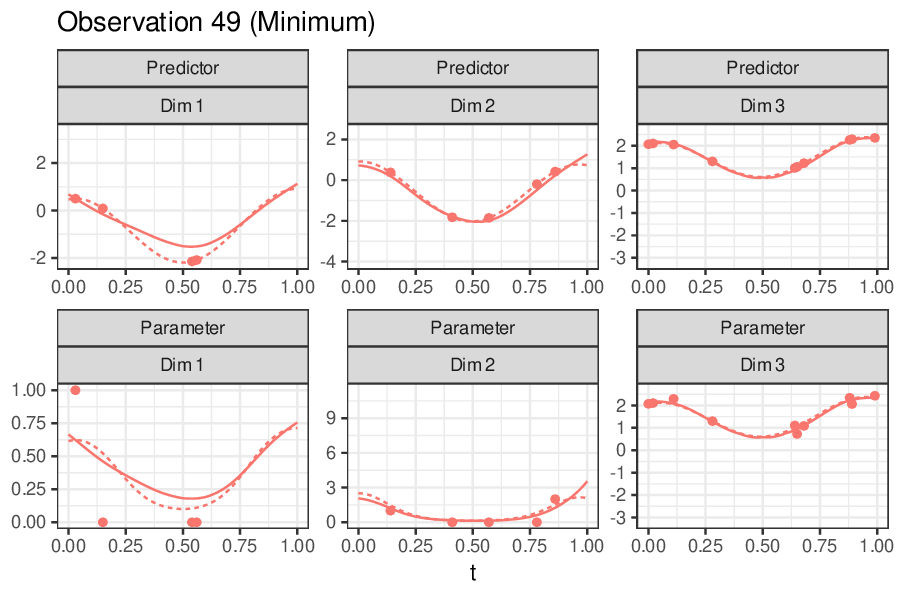}
        \includegraphics[width = \textwidth]{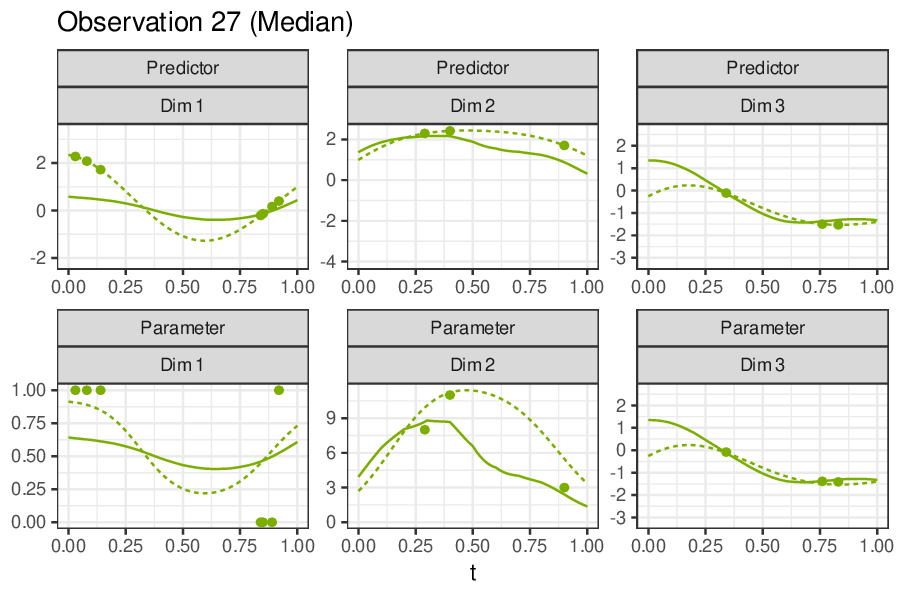}
    \end{minipage}\hfill
    \begin{minipage}{0.49\textwidth}
        \includegraphics[width = \textwidth]{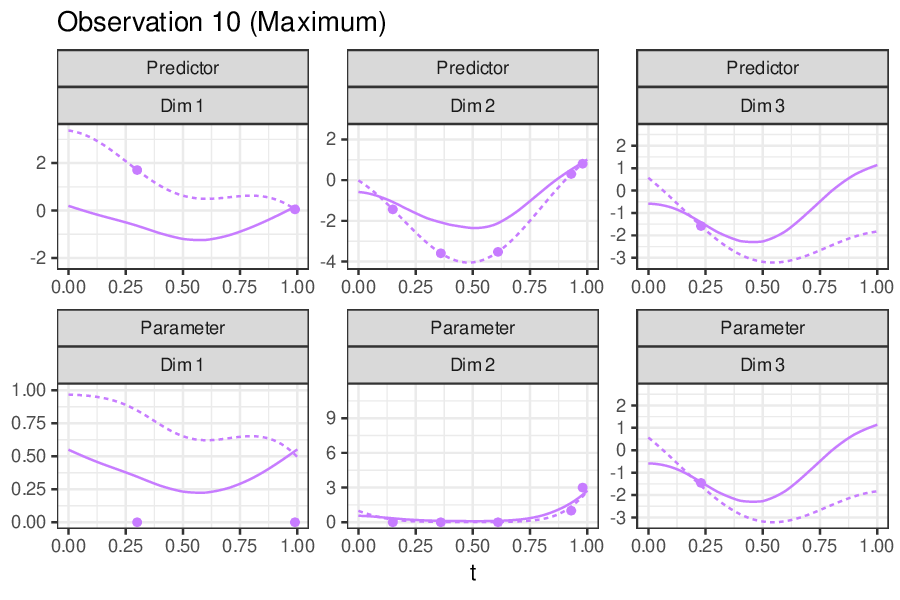}
        \includegraphics[width = \textwidth]{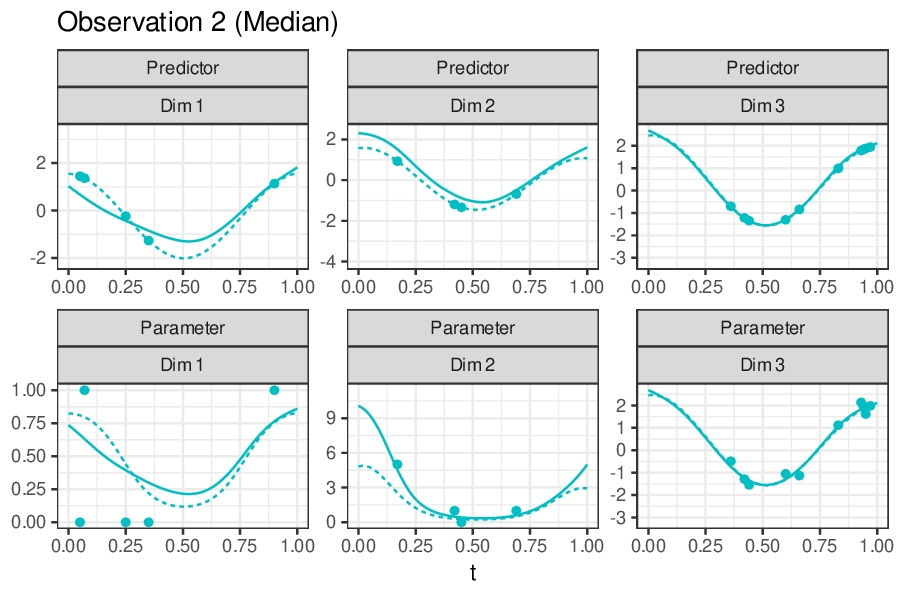}
    \end{minipage}
    \caption{Exemplary model fits for selected observations in an iteration with estimated eigenfunctions from the sparse sampling scenario. Observations are chosen corresponding to the minimum, maximum, and median values of $|||\eta_{i1} - \hat{\eta}_{i1}|||$ over all $n = 150$ observations. The dashed lines are the true data generating $\eta_{i1}^{(k)}(t)$ (\textit{Predictor}) and $\theta_{i1}^{(k)}$ (\textit{Parameter}) with dots indicating observations.}
    \label{APPfig:SimMFPCmodelsFits}
\end{figure}

\begin{figure}
    \centering
    \begin{minipage}{0.49\textwidth}
        \includegraphics[width = \textwidth]{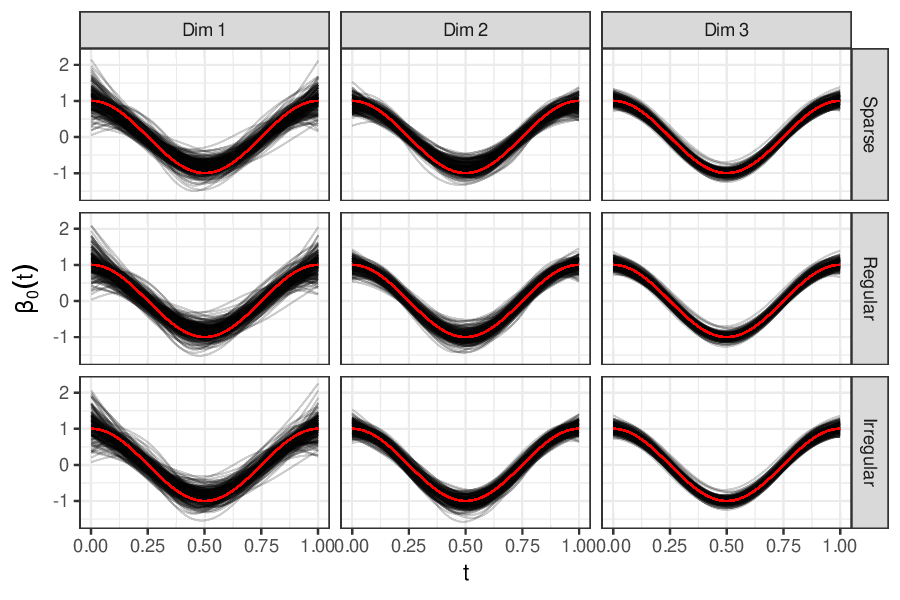}
    \end{minipage}\hfill
    \begin{minipage}{0.49\textwidth}
        \includegraphics[width = \textwidth]{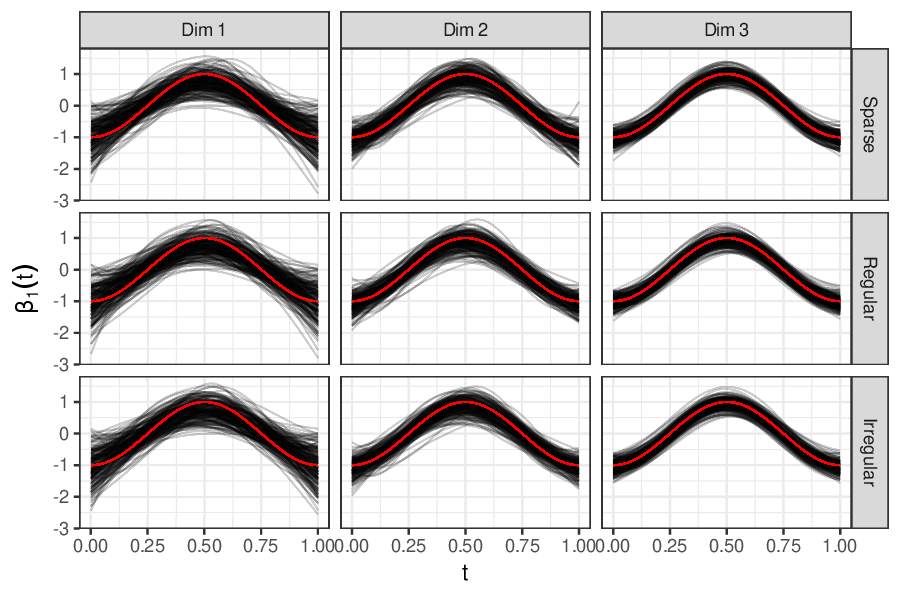}
    \end{minipage}
    \caption{Estimated coefficient effect functions $\beta_0^{(k)}(t)$ (left) and $\beta_1^{(k)}(t)$ (right) for sparsely sampled data based on models using an estimated \gls{mfpc} basis from different sampling regimes.}
    \label{APPfig:SimMFPCBetas}
\end{figure}

Table \ref{APPtab:MFPCreconstruction} shows the mean \gls{rrmse} values and their Monte Carlo error of the reconstructed latent processes for the \gls{mfpc} bases estimated from different sampling scenarios. Note that this is different from the fitted latent process as for the reconstruction, the estimated basis functions are directly fit to the true latent process $\Lambda_{0i}(t)$ using a least-squares approach. For some scenarios and dimensions, we find rather large Monte Carlo errors corresponding to wide boxplots of the \gls{rrmse} values as shown in Figure \ref{APPfig:SimMFPCestRRMSE}. This can be attributed to the number of estimated univariate \glspl{fpc}, see Figure \ref{APPfig:SimMFPCestNumb}, which shows an overall tendency that a higher number of univariate \glspl{fpc} is associated with lower \gls{rrmse} values for the reconstructed latent process $\Lambda_{0i}^{(k)}(t)$. Figure \ref{APPfig:SimMFPCmodelsFits} shows exemplary model fits for a simulation iteration with estimated eigenfunction basis resulting in \gls{rrmse} values $\eta_{i1}^{(1)}(t) = 0.712$, $\eta_{i1}^{(2)}(t) = 0.511$, $\eta_{i1}^{(3)}(t) = 0.310$, which is close to the overall mean \gls{rrmse} values, in the sparsely sampled data. Figure \ref{APPfig:SimMFPCBetas} shows the estimated coefficient effect functions of the models using the differently estimated \glspl{mfpc} on the sparsely sampled data. Comparing the estimates to the first row of Figure \ref{APPfig:SimSampleBetas} suggests that the coefficient function estimation is robust against the estimation of the \gls{mfpc} basis.

Table \ref{APPtab:MulUniBetaRRMSE} illustrates that the univariate and multivariate modelling approach result in comparable \gls{rrmse} values. The rows (columns) of the crosstables report the number of simulation iterations where the univariate models yield smaller \gls{rrmse} values than the multivariate models, denoted as \textit{Uni}, or vice versa, denoted as \textit{Multi}, for $\beta_0^{(k)}(t)$ (or $\beta_1^{(k)}(t)$ for the columns). 

\begin{figure}
    \centering
    \includegraphics[width=\textwidth]{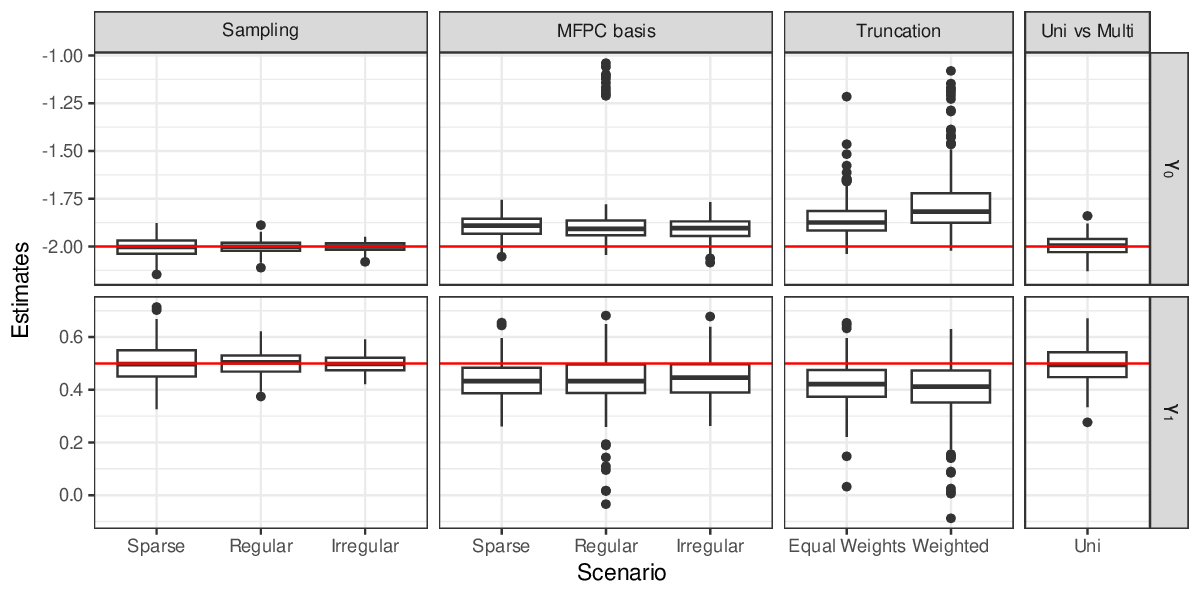}
    \caption{Estimated values of $\gamma_0$ and $\gamma_1$ of the additive predictor of the scale parameter of dimension three for the different examined scenarios. The red line corresponds to the true values.}
    \label{APPfig:sim_sigmas}
\end{figure}

In order to evaluate the estimated scalar coefficient effects $\gamma_l, l\in \{0, 1\}$ of the additive predictor $\eta_{i2}^{(3)}(t)$ of the scale parameter on the Gaussian dimension, we use the bias $Bias(\hat\gamma_l) = \frac{1}{200}\sum_{i= 1}^{200} \hat{\gamma}_{l}^{[i]} - \gamma_{l}$ with $\hat\gamma_{l}^{[i]}$ denoting the posterior mean estimate of the true value $\gamma_{l}$ in iteration $i = 1, ..., 200$, the root \gls{mse} $rMSE(\hat\gamma_l) = \left(\frac{1}{200}\sum_{i= 1}^{200} \left(\hat{\gamma}_{l}^{[i]} - \gamma_{l}\right)^2\right)^{\frac{1}{2}}$, and the mean frequentist coverage $FC(\hat\gamma_l) = \frac{1}{200}\sum_{i=1}^{200}I(\hat{\gamma}_{l}^{[i], 2.5} \leq \gamma_{l} \leq \hat{\gamma}_{l}^{[i], 97.5})$ with indicator function $I$ and $\hat{\gamma}_{l}^{[i], \alpha}$ the $\alpha\%$ quantile of \gls{mcmc} samples in iteration $i$. Figure \ref{APPfig:sim_sigmas} shows the estimates of $\gamma_0$ and $\gamma_1$ together with the true values and Table \ref{APPtab:sigma_evals} contains the previously defined evaluation criteria for the examined simulation scenarios. We find that while the estimation of the scale parameter is good when the true eigenfunctions are used, estimating the \gls{mfpc} basis introduces some bias and variance in the estimates, resulting in lower frequentist coverage.

\begin{table}
    \centering
\begin{tabular}{l|rrr|rrr}

 & \multicolumn{3}{c|}{$\beta_0^{(k)}(t)$} & \multicolumn{3}{c}{$\beta_1^{(k)}(t)$} \\ 
  \hline
  Scenario & $Bin$ & $Poi$ & $N$ & $Bin$ & $Poi$ & $N$ \\  \hline
Sparse & 0.764 & 0.371 & 0.139 & 1.397 & 0.861 & 0.636 \\ 
  Regular & 0.525 & 0.231 & 0.058 & 1.040 & 0.705 & 0.578 \\ 
  Irregular & 0.471 & 0.206 & 0.053 & 0.973 & 0.681 & 0.564 \\  
  \hline 
\end{tabular}
    \caption{Mean width of credible intervals of the coefficient effect functions $\beta_0(t)$ and $\beta_1(t)$ for different sampling regimes, averaged over the functional domain $t$.}
    \label{APPtab:cred_widths}
\end{table}

\begin{table}
    \centering
\begin{tabular}{lccc}
  \hline
  Scenario & $Bin$ & $Poi$ & $N$ \\ 
  \hline
Sparse & 0.155 (0.081) & 0.220 (0.135) & 0.083 (0.011)\\ 
  Regular & 0.144 (0.071)& 0.104 (0.078)& 0.087 (0.047)\\ 
  Irregular &  0.078 (0.014) & 0.069 (0.011) & 0.075 (0.012) \\ 
   \hline
\end{tabular}
    \caption{Mean \gls{rrmse} values with Monte Carlo error in brackets for the reconstructed latent process using the estimated \glspl{mfpc} as basis functions for the different sampling regimes. The scores of the multivariate \gls{kl} representation are estimated using least-squares.}
    \label{APPtab:MFPCreconstruction}
\end{table}

\begin{table}
    \centering
    \begin{minipage}{0.33\textwidth}
    \small
        \begin{tabular}{l|cc|c}
        \textbf{Dim 1} & Multi & Uni & $\beta_0^{(1)}(t)$ \\\hline
        Multi & 73 & 49 & 122 \\
        Uni & 47 & 31 & 78 \\\hline
        $\beta_1^{(1)}(t)$ & 120 &  80 & 200\\
    \end{tabular}
    \end{minipage}\hfill
        \begin{minipage}{0.33\textwidth}
        \small
        \begin{tabular}{l|cc|c}
        \textbf{Dim 2} & Multi & Uni & $\beta_0^{(2)}(t)$ \\\hline
        Multi & 53 & 52 & 105\\
        Uni & 51& 44 & 95\\\hline
        $\beta_1^{(2)}(t)$ & 104 & 96 & 200 \\
    \end{tabular}
    \end{minipage}\hfill
        \begin{minipage}{0.33\textwidth}
        \small
        \begin{tabular}{l|cc|c}
        \textbf{Dim 3} & Multi & Uni & $\beta_0^{(3)}(t)$ \\\hline
        Multi & 36 & 39 & 75\\
        Uni & 40 & 85 & 125\\\hline
        $\beta_1^{(3)}(t)$ & 76 & 124 & 200 \\
    \end{tabular}
    \end{minipage}
    \caption{Crosstables for the number of simulation runs, where the univariate (\textit{Uni}) or the multivariate (\textit{Multi}) estimation approach result in lower \gls{rrmse} values for the coefficient effect functions $\beta_0(t)$ and $\beta_1(t)$ over the three dimensions in the data. The rows correspond to $\beta_0(t)$ with the last column showing the marginal frequencies and analogous for the columns and last row for $\beta_1(t)$.}
    \label{APPtab:MulUniBetaRRMSE}
\end{table}

\begin{table}
    \centering
    \begin{subtable}{\textwidth}
    \centering
        \subcaption{\textit{Impact of sampling regime:} Estimation based on true eigenfunctions}
        \label{APPsubtab:Sampling}
   \begin{tabular}{p{6.3em}|rrr|rrr}

 & \multicolumn{3}{c|}{$\gamma_0$} & \multicolumn{3}{c}{$\gamma_1$} \\ 
  \hline
  Sampling & Bias & rMSE & FC & Bias & rMSE & FC \\ \hline
Sparse & -0.003 & 0.051 & 0.960 & -0.000 & 0.073 & 0.930 \\
  Regular & -0.002 & 0.030 & 0.965 & -0.000 & 0.043 & 0.960 \\ 
  Irregular & 0.000 & 0.024 & 0.960 & -0.001 & 0.033 & 0.940 \\ \hline
\end{tabular}
    \end{subtable}
\\[1em]
    \begin{subtable}{\textwidth}
    \centering
        \subcaption{\textit{Impact of MFPC basis:} Estimation on sparse setting with MFPCs estimated from different settings}
        \label{APPsubtab:MFPCest}
\begin{tabular}{p{6.3em}|rrr|rrr}

 & \multicolumn{3}{c|}{$\gamma_0$} & \multicolumn{3}{c}{$\gamma_1$} \\ 
  \hline
  Sampling & Bias & rMSE & FC & Bias & rMSE & FC \\ \hline
  True & -0.003 & 0.051 & 0.960 & -0.000 & 0.073 & 0.930 \\
 Sparse & 0.108 & 0.122 & 0.405 & -0.064 & 0.097 & 0.825 \\
  Regular & 0.131 & 0.220 & 0.495 & -0.070 & 0.124 & 0.795 \\ 
  Irregular & 0.093 & 0.110& 0.520 & -0.057 & 0.093 & 0.870 \\  \hline 
\end{tabular}
    \end{subtable}
\\[1em]
    \begin{subtable}{\textwidth}
    \centering
        \subcaption{\textit{Impact of MFPC truncation:} Estimation with MFPCs truncated at $98\%$ based on different scalar products}
        \label{APPsubtab:mfpcTrunc}
 \begin{tabular}{p{6.3em}|rrr|rrr}

 & \multicolumn{3}{c|}{$\gamma_0$} & \multicolumn{3}{c}{$\gamma_1$} \\ 
  \hline
  Sampling & Bias & rMSE & FC & Bias & rMSE & FC \\ \hline
  No Truncation & 0.108 & 0.122 & 0.405 & -0.064 & 0.097 & 0.825 \\
  Equal Weights & 0.145 & 0.177 & 0.300 & -0.078 & 0.177 & 0.775 \\  
  Weighted & 0.232 & 0.292 & 0.160 & -0.105 & 0.156 & 0.730 \\\hline 
\end{tabular}
    \end{subtable}
\\[1em]
    \begin{subtable}{\textwidth}
    \centering
        \subcaption{\textit{Univariate vs.\ multivariate:} Estimation on sparse setting with true eigenfunctions}
        \label{APPsubtab:UniMul}
 \begin{tabular}{p{6.3em}|rrr|rrr}

 & \multicolumn{3}{c|}{$\gamma_0$} & \multicolumn{3}{c}{$\gamma_1$} \\ 
  \hline
  Sampling & Bias & rMSE & FC & Bias & rMSE & FC \\ \hline
   Multivariate & -0.003 & 0.051 & 0.960 & -0.000 & 0.073 & 0.930 \\ 
   Univariate & 0.005 & 0.052 & 0.955 & -0.003 & 0.073 & 0.935 \\   \hline
\end{tabular}
    \end{subtable}
    
    \caption{Evaluation of the estimation of the coefficients $\gamma_0$ and $\gamma_1$ in the additive predictor $\eta_{i2}^{(3)}$ of the scale parameter of dimension three for the different simulation scenarios.}
    \label{APPtab:sigma_evals}
\end{table}

\section{Application}
\label{APPsec:Application}

\subsection{Data Preprocessing}

\begin{figure}
    \centering
    \includegraphics[width=0.8\textwidth]{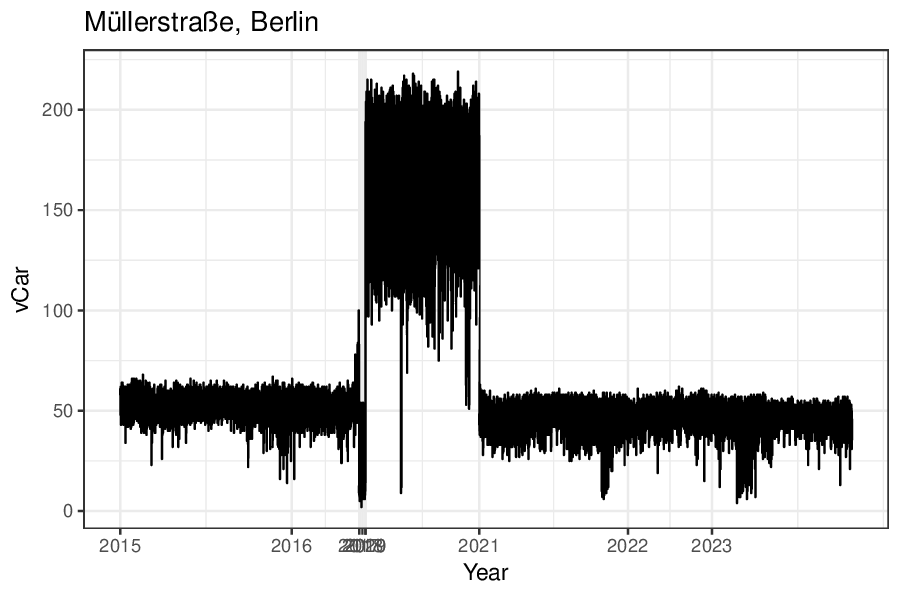}
    \caption{Example of detector failure for site ``Müllerstraße'' in the Berlin traffic data. It is apparent that the time-series for outcome $vCar$ is incomplete and shows an unrealistic jump for the observations from 2017 to 2021.}
    \label{APPfig:AppMuellerstr}
\end{figure}

Figure \ref{APPfig:AppMuellerstr} provides an example of a detector failure in the Berlin traffic data. The site ``Müllerstraße'' is not fully observed on the time interval and features an unrealistic jump in the outcome $vCar$ between 2017 and 2021. As other outcomes are not affected, only this section is removed from the final data set. 

\subsection{Estimation of Univariate Generalized FPCs}
\label{APPsubsec:APPUniFPCA}

We follow the outlined approach by \cite{zhou2023analysis} to estimate the univariate generalized \glspl{fpc} and adapt it to the application at hand. In the first step, the data is grouped into $s = 1,..., 24$ overlapping bins with one bin containing the measurement data $t_{ij} \in [s-2, s+2]$. Note that we use data over the ends of the domains, e.g.\ the first bin contains all observations between 11 p.m.\ and 3 a.m., as the traffic data are of a cyclic nature. For each of these bins, four univariate \gls{bamlss} regression models are estimated with the distributional assumption analoguous to the multivariate functional data. These models ignore the functional nature of the data and estimate site-specific and site-year-specific (scalar) random intercepts while accounting for the covariate effects. This results in additive predictors for the respective subset of data as
\begin{gather*}
    \eta_{ij1}^{s,k'}(t) =  %\bar{N}_{i,k'} + 
    \beta_0^{s,k'} + f^{s, k'}(year_j) + lanes1_i\beta_1^{s,k'} + lanes3_i\beta_2^{s,k'} + b_{1i}^{s, k'} + b_{0ij}^{s, k'}, \\
     \eta_{ij1}^{s,k''}(t) =  \beta_0^{s,k''} + f^{s, k'}(year_j) + limit30_i\beta_1^{s, k''} + limit60_i\beta_1^{s, k''} + b_{1i}^{s, k'} + b_{0ij}^{s, k'}\\
     t = s-2,..., s+2, \quad k'\in \{qCar, qTruck\},\quad k''\in \{vCar, vTruck\},
\end{gather*}
where the random intercepts are denoted as $b_{1i}^{s,k}$ and $b_{0ij}^{s,k}$. The models are estimated with 1000 \gls{mcmc} samples after a burn-in of 200 using seven cubic P-spline basis functions with second order difference penalty for the nonlinear effect of $year$. Estimating one such univariate regression model takes five to ten minutes on a Linux system with $12$ cores at $3.5$ GHz and $32$ GB memory, which amounts to 11 hours total without parallelization.

\begin{figure}
    \centering
    \begin{minipage}{0.49\textwidth}
        \includegraphics[width=\textwidth]{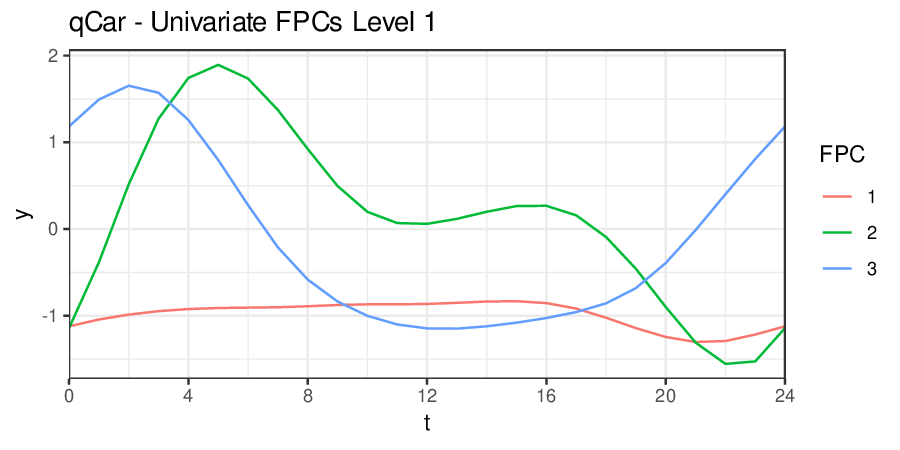}\\
         \includegraphics[width=\textwidth]{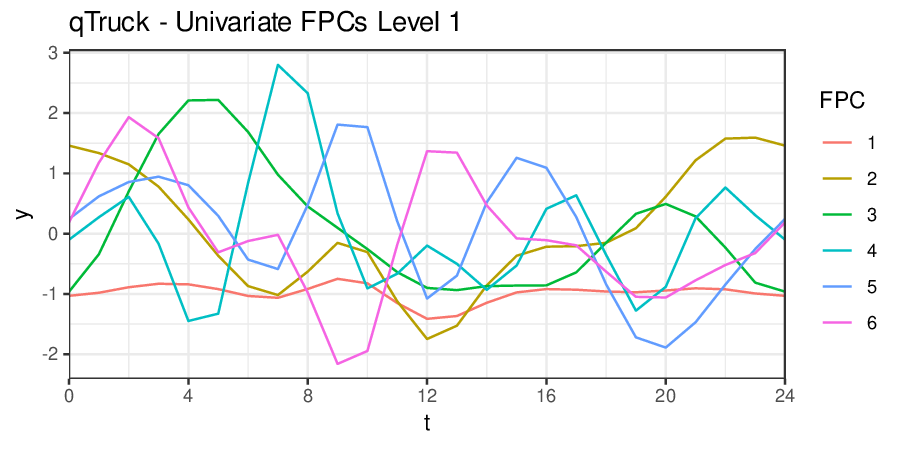}\\
         \includegraphics[width=\textwidth]{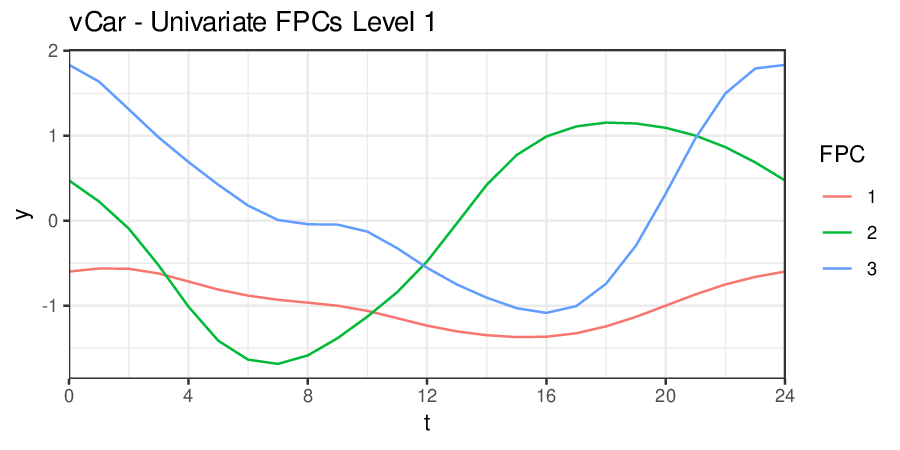}\\
         \includegraphics[width=\textwidth]{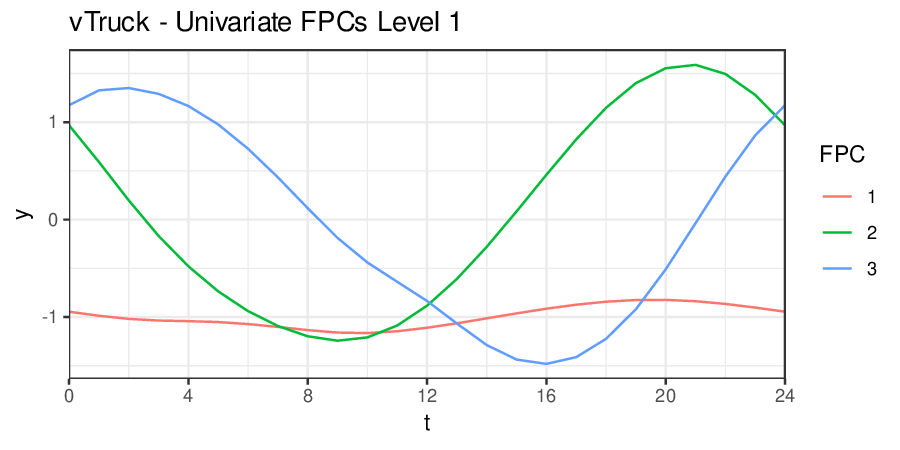}
    \end{minipage}\hfill
    \begin{minipage}{0.49\textwidth}
    \includegraphics[width=\textwidth]{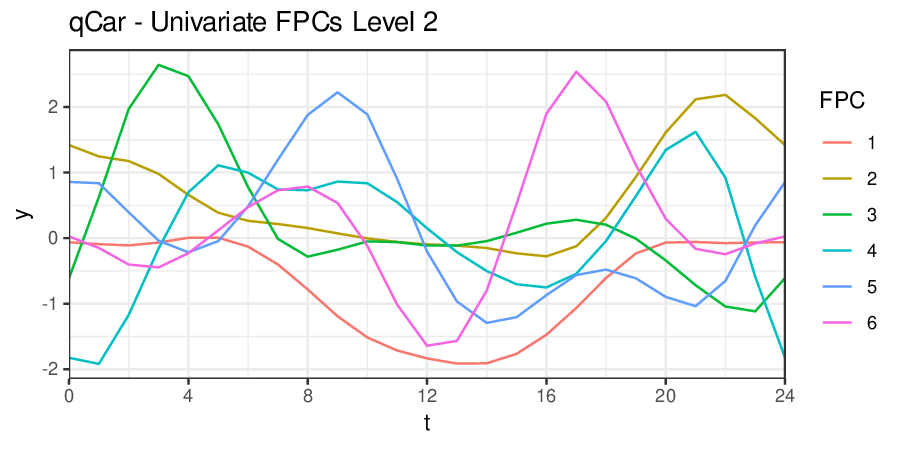}\\
         \includegraphics[width=\textwidth]{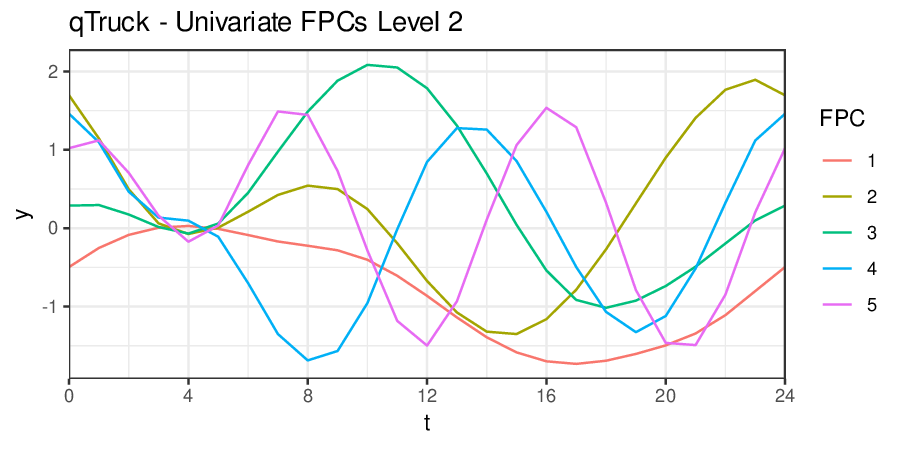}\\
         \includegraphics[width=\textwidth]{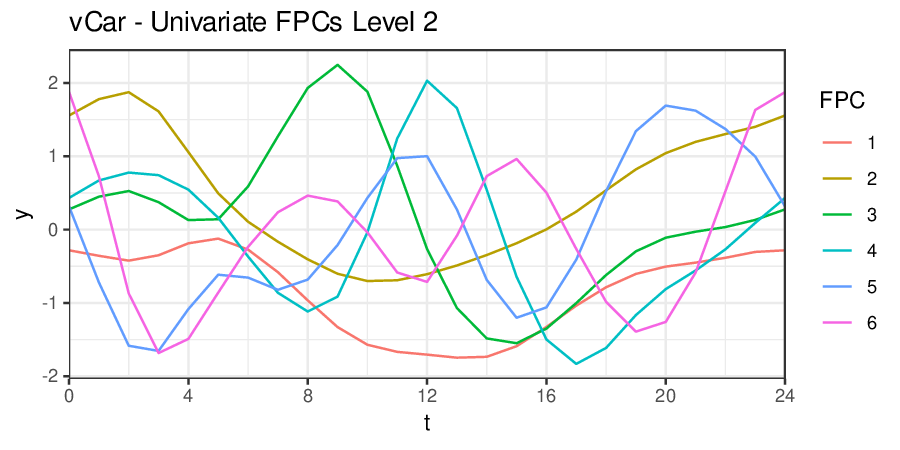}\\
         \includegraphics[width=\textwidth]{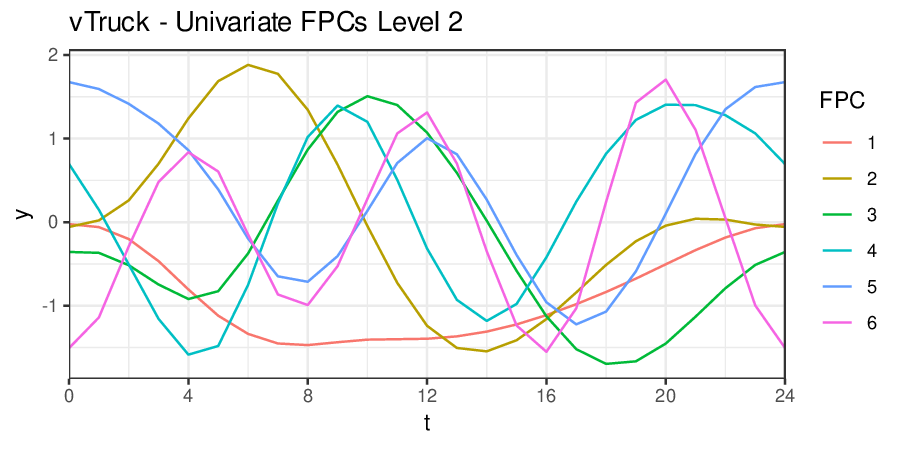}
    \end{minipage}
    \caption{Estimated univariate FPCs for the different outcomes with level 1 denoting the estimates $\hat{\phi}_{1m}^{(k)}(t), m = 1,..., M_1^k$ corresponding to the site-specific random process. Level 2 denotes the estimates $\hat{\phi}_{m}^{(k)}(t), m = 1,..., M^k$ for the site-year-specific random process. The estimates are available at the bin centers and are linearly interpolated only for this representation.}
    \label{APPApplicationFig:estimatedUFPCs}
\end{figure}

The estimated random intercepts $\hat{b}_{1i}^{s,k}$ and $\hat{b}_{0ij}^{s,k}, k \in \{qCar, qTruck, vCar, vTruck\}$ are then used in the fast multilevel \gls{fpca} method proposed by \cite{cui2023fast} to obtain smooth estimates of the univariate eigenfunctions $\phi_{1m}^{(k)}(t)$ and $\phi_{0m}^{(k)}$ of the site-specific and site-year-specific processes, taking only a few seconds. For the smooth estimation of eigenfunctions, we choose a cyclic cubic P-spline basis with second order difference penalty and ten knots, as we find a larger number of basis functions to yield undersmoothed results \citep[as described similarly by][]{leroux2023fast}. The number of eigenfunctions $M_1^k$ and $M_0^k$ estimated by the fast \gls{fpca} approach is determined based on the cumulative percentage of variance explained with a threshold of 0.99. For the site-specific random process, $\left(M_1^{qCar}, M_1^{qTruck}, M_1^{vCar}, M_1^{vTruck}\right) = (3, 6, 3, 3)$, and for the site-year-specific random process  $\left(M_0^{qCar}, M_0^{qTruck}, M_0^{vCar}, M_0^{vTruck}\right) = (6, 5, 6, 6)$ \glspl{fpc} are estimated with the estimates given in Figure \ref{APPApplicationFig:estimatedUFPCs}.

In order to avoid bias introduced by the binning step, we re-estimate the univariate scores in separate univariate generalized functional additive mixed models with additive predictors
\begin{align*}
    \eta_{ij1}^{(k')}\left(t \mid \left\{\hat{\phi}_{1m}^{(k')}(t)\right\}_{m=1}^{M_1^{k'}}, \left\{\hat{\phi}_{0m}^{(k')}(t)\right\}_{m=1}^{M_0^{k'}}\right) = & \beta_0^{(k')}(t) + f^{(k')}(year_j, t) \\
    & + lanes1_i\beta_1^{(k')}(t) + lanes3_i\beta_2^{(k')}(t) \\
    &  + \sum_{m=1}^{M_1^{k'}}\xi_{1im}^{(k')}\hat{\phi}_{1m}^{(k')}(t) + \sum_{m=1}^{M_0^{k'}}\xi_{0ijm}^{(k')}\hat{\phi}_{0m}^{(k')}(t)\\
    \eta_{ij1}^{(k'')}\left(t \mid \left\{\hat{\phi}_{1m}^{(k'')}(t)\right\}_{m=1}^{M_1^{k''}}, \left\{\hat{\phi}_{0m}^{(k'')}(t)\right\}_{m=1}^{M_0^{k''}}\right) = &\beta_0^{(k'')}(t) + f^{(k'')}(year_j, t) + \\ 
    & + limit30_i\beta_1^{(k'')}(t) 
     + limit60_i\beta_2^{(k'')}(t)\\ 
     &  + \sum_{m=1}^{M_1^{k''}}\xi_{im}^{(k'')}\hat{\phi}_{1m}^{(k'')}(t) + \sum_{m=1}^{M_0^{k''}}\xi_{ijm}^{(k'')}\hat{\phi}_{0m}^{(k'')}(t),
\end{align*}
for $k'\in \{qCar, qTruck\}$ and $k''\in \{vCar, vTruck\}$, conditioning on the estimated univariate \glspl{fpc} $\hat{\phi}_{1m}^{(k)}(t)$ and $\hat{\phi}_{0m}^{(k)}(t), k \in \{qCar, qTruck, vCar, vTruck\}$. The functional intercepts $\beta_0^{(k)}(t)$ and the categorical functional effects $\beta_1^{(k)}(t)$ and $\beta_2^{(k)}(t)$  are represented using $14$ cubic P-splines while the bivariate interaction terms $f^{(k)}(year_j, t)$ are modeled using anisotropic tensor product splines of $7$ (for $year$) and $14$ (for $t$) marginal cubic P-splines allowing different variance parameters for the different margins.
These four univariate models are again estimated in the \gls{bamlss} framework with 1,000 \gls{mcmc} samples after a burn-in of 1,000 and a thinning of 5 in order to get more stable estimates of the univariate scores. The computation of the univariate generalized functional additive mixed models takes 12 to 18 hours per outcome. We then use the posterior mean of the scores as estimates of the univariate scores and Table \ref{APPApplicationTab:EstimatedEV} reports the estimated univariate eigenvalues from this univariate generalized \gls{fpca} approach.

\subsection{Additional Results}

\begin{figure}
    \centering
    \includegraphics[width=0.8\textwidth]{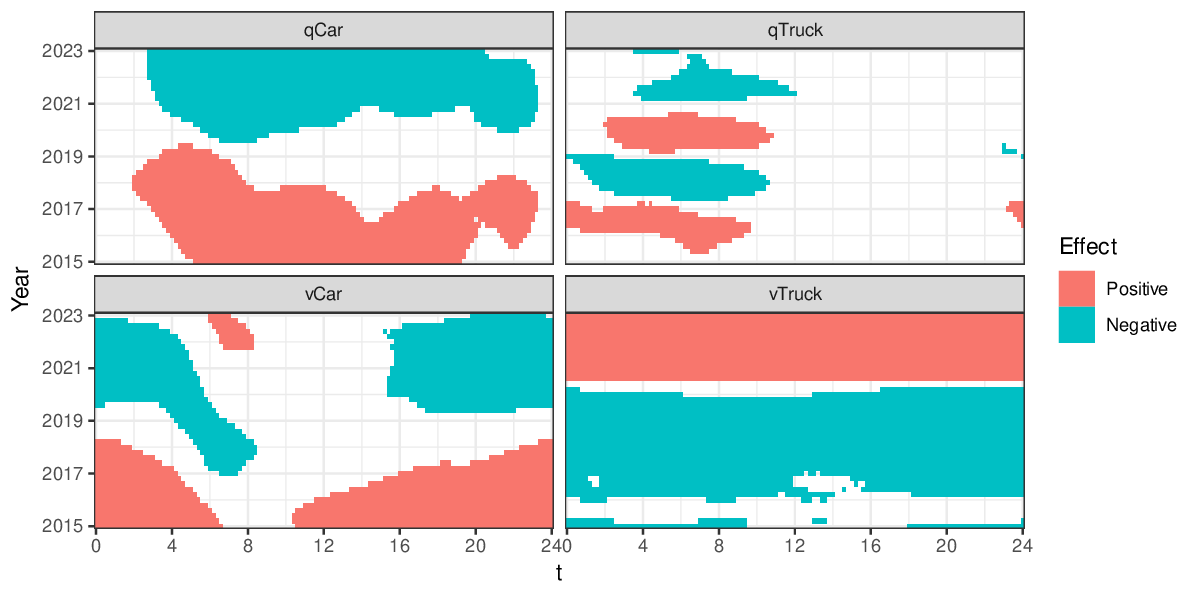}
    \caption{Areas where the pointwise $95\%$ credible intervals of the estimated bivariate interaction effect $f^{(k)}(year_{j}, t)$ do not include 0. Areas where the estimated effect is larger than 0 are colored in red, effects lower than 0 in blue.}
    \label{APPApplicationfig:Inference}
\end{figure}

\begin{figure}
    \centering
    \includegraphics[width=0.6\textwidth]{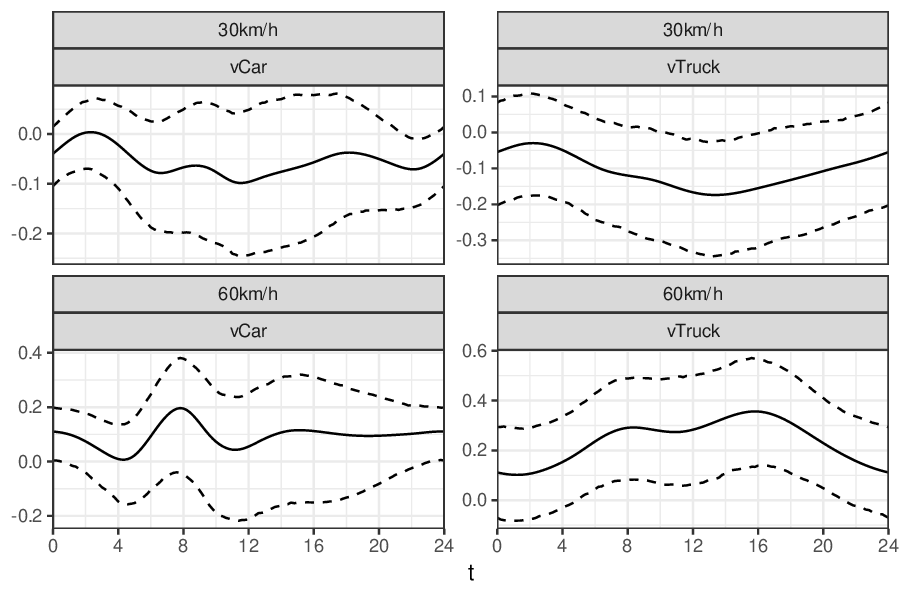}
   \caption{Estimated coefficient effect functions $\hat{\beta}_1^{(k'')}(t)$ and $\hat{\beta}_2^{(k'')}(t), k'' \in \{vCar, vTruck\}$ for sites with a 30 km/h or 60 km/h (or higher) speed limit, respectively, on the linear predictor scale. The dashed lines show the pointwise $95\%$ credible intervals.}
    \label{APPApplicationfig:BetaLimit}
\end{figure}

\begin{figure}
    \centering
    \includegraphics[width=0.6\textwidth]{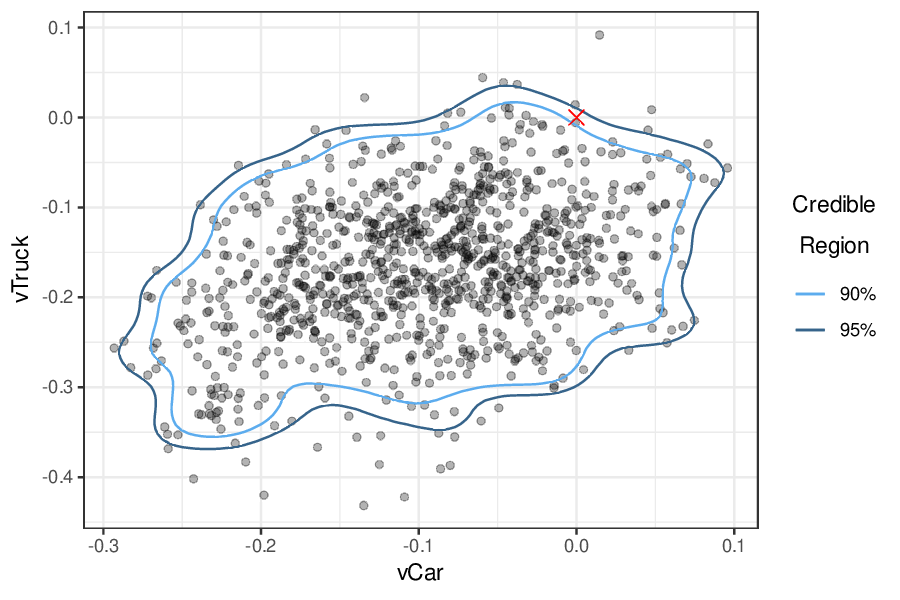}
     \caption{Estimated coefficient effects of $\hat{\beta}_1^{(vCar)}(t)$ and $\hat{\beta}_1^{(vTruck)}(t)$ at time point $t = 12$ based on the \gls{mcmc} samples. Empirical credible regions are included as contour lines with the origin marked in red.}
    \label{APPApplicationfig:JointLimit}
\end{figure}

\begin{figure}
    \centering
    \begin{minipage}{0.49\textwidth}
        \includegraphics[width = \textwidth]{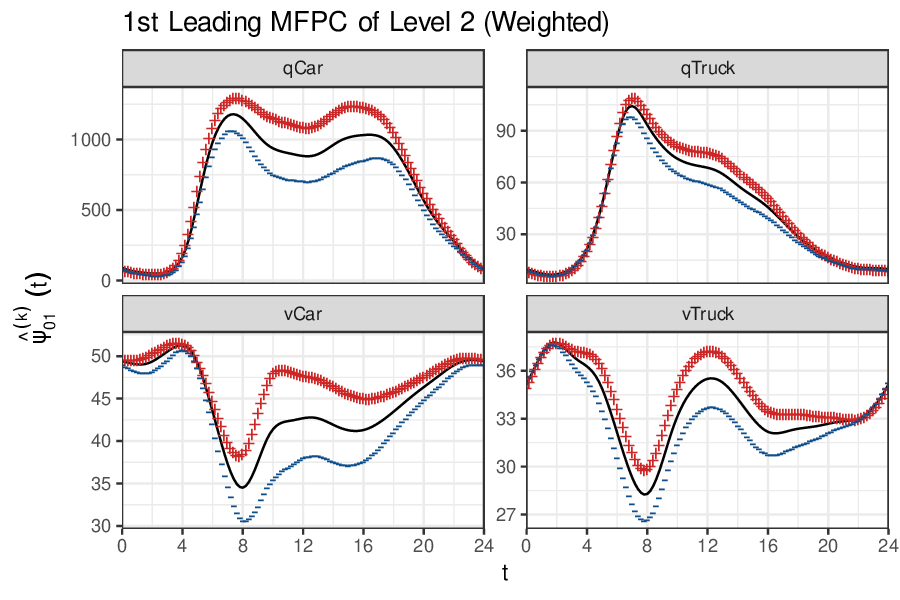}
        \includegraphics[width = \textwidth]{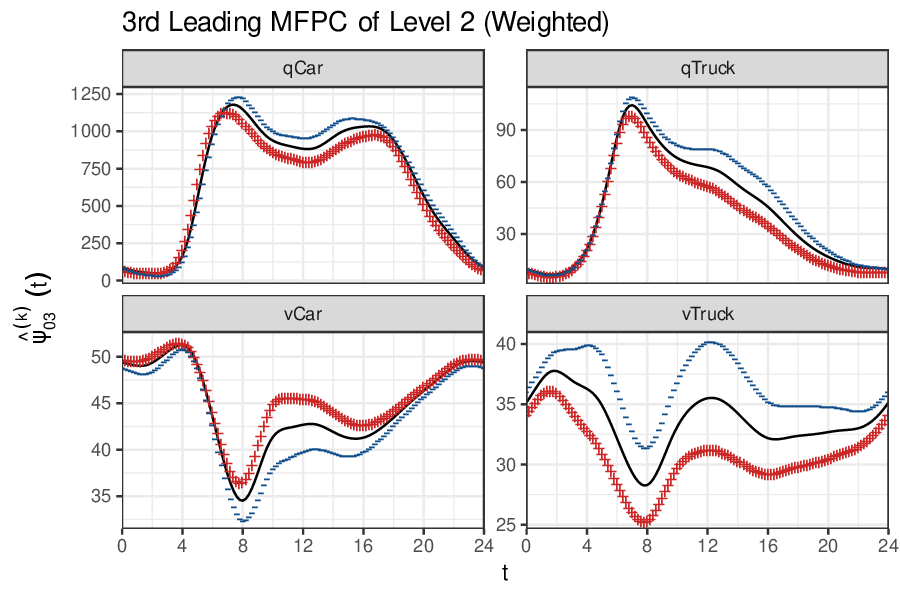}
    \end{minipage}\hfill
    \begin{minipage}{0.49\textwidth}
    \includegraphics[width = \textwidth]{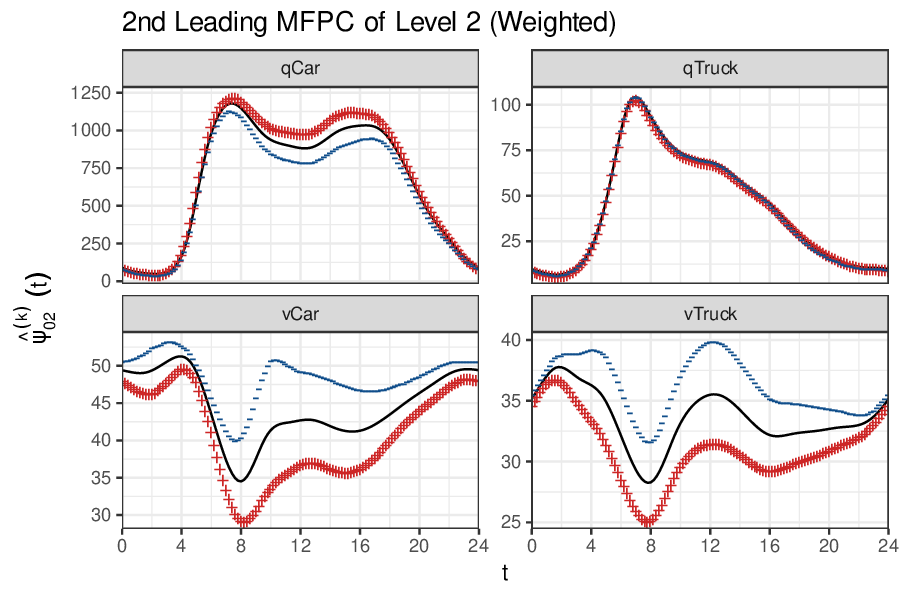}
    \includegraphics[width = \textwidth]{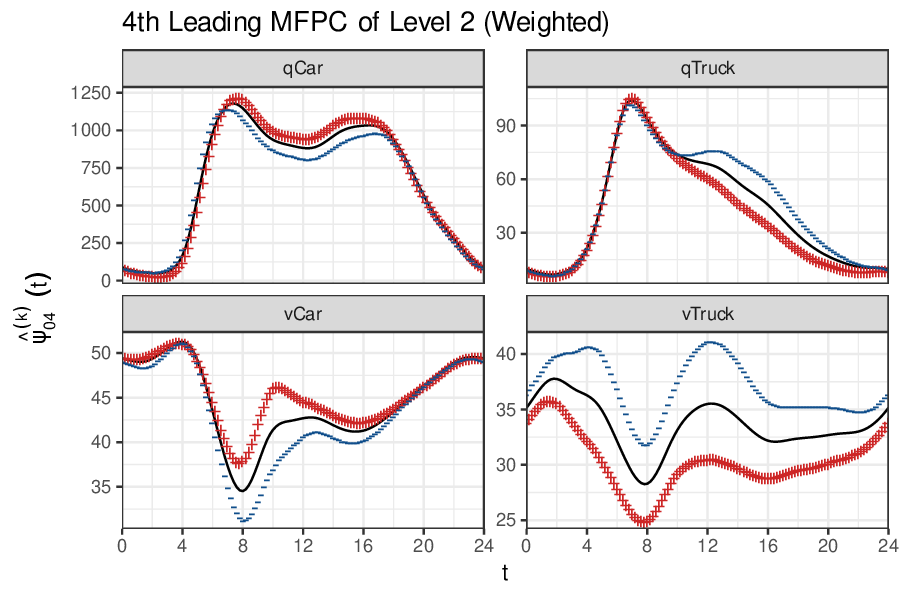}
    \end{minipage}
    \begin{minipage}{0.49\textwidth}
    \includegraphics[width = \textwidth]{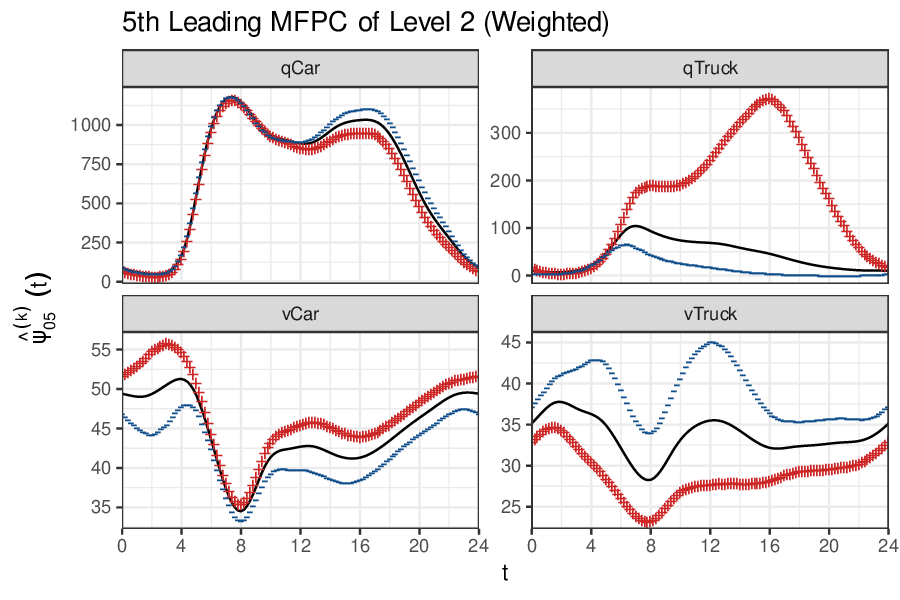}
    \end{minipage}
    \caption{Leading five estimated \glspl{mfpc} based on a weighted scalar product for the site-year-specific (level 2) random process. The black line corresponds to the estimated mean of a site with two lanes and 50 km/h speed limit in 2020 with plus (red +) or minus (blue -) $2\sqrt{\hat{\nu}_{2m}}, m = 1, 2, 3$ times the estimated \gls{mfpc}.}
    \label{APPApplicationfig:EstimatedWMFPCsLev2}
\end{figure}

\begin{figure}
    \centering
    \includegraphics[width=\textwidth]{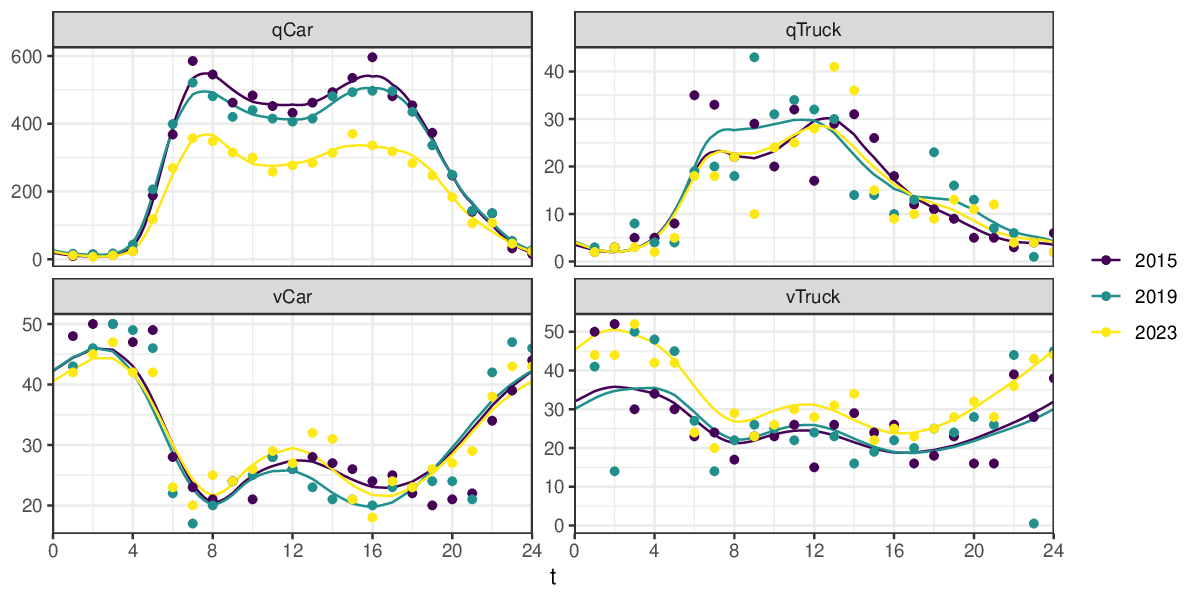}
    \caption{Estimated mean function (line) and scalar observations (points) for site ``Goerzallee'' for three different years.}
    \label{APPApplicationfig:ModelFits}
\end{figure}

\begin{figure}
    \centering
    \includegraphics[width=\textwidth]{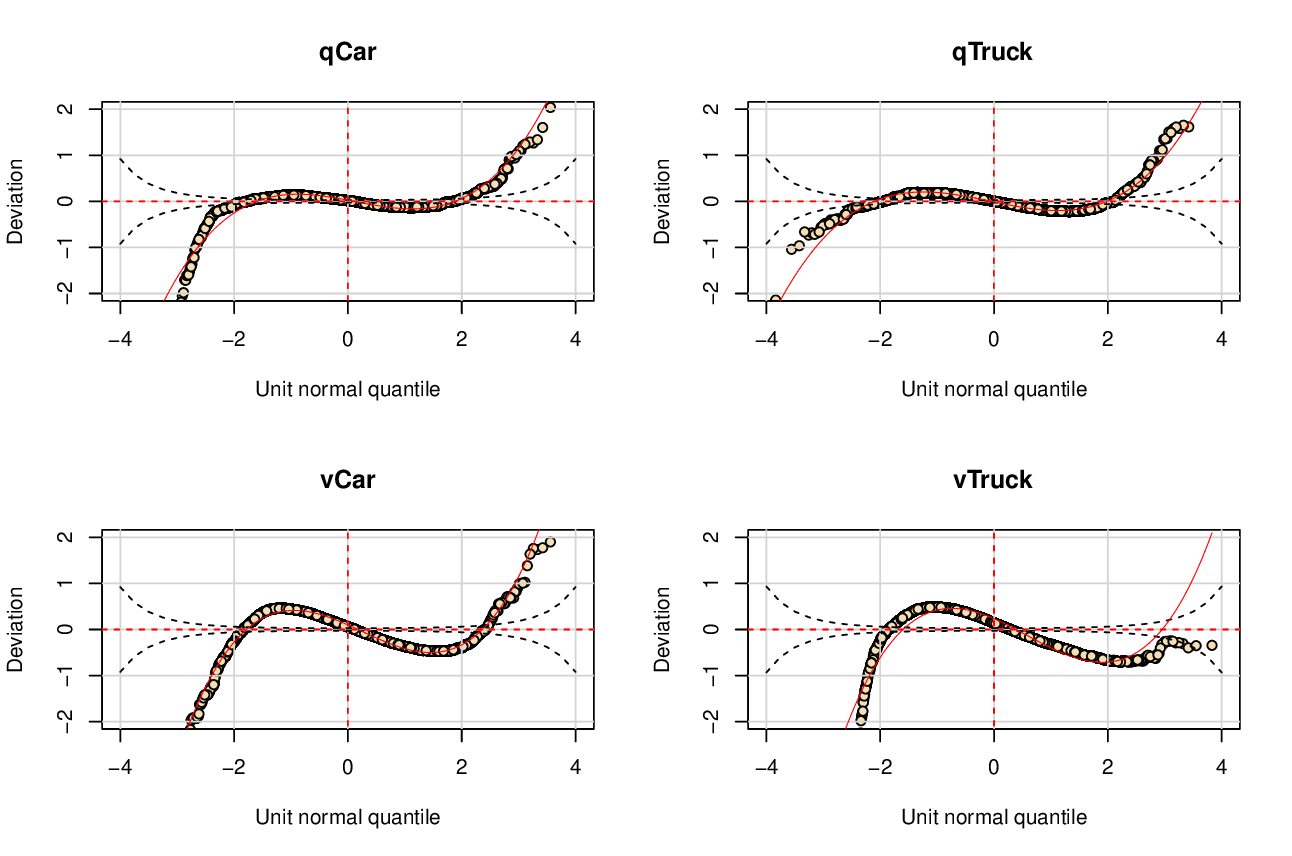}
    \caption{Dimension specific worm plots corresponding to de-trended QQ-plots of normalized (randomized) quantile residuals.}
    \label{APPApplicationfig:WormPlots}
\end{figure}

Table \ref{APPApplicationtab:EstimatedMEVweights} reports the estimated multivariate eigenvalues based on a weighted scalar product. Figure \ref{APPApplicationfig:Inference} presents the areas where pointwise $95\%$ credible intervals of the bivariate interaction effect $f^{(k)}(year_{j}, t)$ do not include zero. Areas that do not include zero are highlighted by their effect (either positive or negative) compared to the global functional intercept. Figure \ref{APPApplicationfig:BetaLimit} shows the estimated coefficient effect functions $\hat{\beta}_1^{(k'')}(t)$ and $\hat{\beta}_2^{(k'')}(t), k'' \in \{vCar, vTruck\}$ for sites with a 30 km/h or 60 km/h (or higher) speed limit, respectively, depicting the difference compared to a site with a 50 km/h speed limit on the linear predictor scale, all other things being equal. The dashed lines show the pointwise $95\%$ credible intervals. Figure \ref{APPApplicationfig:JointLimit} constructs a 90\% and a 95\% credible region for the joint estimate of $\left(\hat{\beta}_1^{(vCar)}(t), \hat{\beta}_1^{(vTruck)}(t)\right)^\top$ at time point $t = 12$ based on a two-dimensional kernel density estimate on the 1000 \gls{mcmc} samples. This can be used to test the null hypothesis $H_0: \hat{\beta}_1^{(vCar)}(12) = \hat{\beta}_1^{(vTruck)}(12) = 0$ by checking whether the origin $(0, 0)^{\top}$ (marked in red) is included in the $1-\alpha$ credible region. Figure \ref{APPApplicationfig:EstimatedWMFPCsLev2} shows the impact of the five leading estimated \glspl{mfpc} on the mean for the site-year-specific (level 2) random process. The black line corresponds to the estimated mean of a site with two lanes and 50 km/h speed limit in 2020 with plus (red +) or minus (blue -) $2\sqrt{\hat{\nu}_{2m}}, m = 1, 2, 3$ times the estimated \gls{mfpc}. Table \ref{APPApplicationtab:EstimatedMEVPost} reports the relative contributions of the estimated posterior score variances to the total (posterior unweighted) variation for the site-specific random process (level 1, $\nu_{1,m}, m = 1,...,9$) and the site-year-specific random process (level 2, $\nu_{0,m}, m = 1,...,$) based on an \gls{mfpca} with scalar product with unequal weights. Figure \ref{APPApplicationfig:ModelFits} shows the fitted location parameter together with the scalar observations for three different years at the traffic detector site ``Goerzallee''. Figure \ref{APPApplicationfig:WormPlots} presents outcome specific worm plots \citep{buuren2001worm} of normalized (randomized) quantile residuals \citep{dunn1996randomized}.

\begin{table}[tb]
    \centering
    \begin{minipage}{0.45\textwidth}
        \begin{tabular}{crrrr}
 \multicolumn{5}{c}{Level 1}\\
  \hline
 Absolute & $qCar$ & $qTruck$ & $vCar$ & $vTruck$ \\ 
  \hline
$\upsilon_{11}^k$ & 0.227 & 0.393 & 0.041 & 0.056 \\ 
  $\upsilon_{12}^k$ & 0.019 & 0.073 & 0.002 & 0.010 \\ 
  $\upsilon_{13}^k$ & 0.013 & 0.016 & 0.001 & 0.001 \\ 
  $\upsilon_{14}^k$ & -- & 0.010 & -- & -- \\ 
  $\upsilon_{15}^k$ & -- & 0.009 & -- & -- \\ 
  $\upsilon_{16}^k$ & -- & 0.004 & -- & -- \\ \hline
  $\sum_{M^k}\upsilon_{1m}^k$ & 0.259 & 0.504 & 0.045 & 0.067 \\ 
   \hline
\end{tabular}
\\ \vspace{2em} \\
\begin{tabular}{crrrr}
 \multicolumn{5}{c}{Level 2}\\
  \hline
Absolute & $qCar$ & $qTruck$ & $vCar$ & $vTruck$ \\ 
  \hline
$\upsilon_{01}^k$ & 0.0079 & 0.1344 & 0.0070 & 0.0112 \\ 
  $\upsilon_{02}^k$ & 0.0022 & 0.0190 & 0.0013 & 0.0019 \\ 
  $\upsilon_{03}^k$ & 0.0010 & 0.0145 & 0.0008 & 0.0013 \\ 
  $\upsilon_{04}^k$ & 0.0007 & 0.0041 & 0.0003 & 0.0005 \\ 
  $\upsilon_{05}^k$ & 0.0006 & 0.0020 & 0.0002 & 0.0004 \\ 
  $\upsilon_{06}^k$ & 0.0003 & -- & 0.0001 & 0.0002 \\ \hline
  $\sum_{M^k}\upsilon_{0m}^k$ & 0.0127 & 0.1739 & 0.0097 & 0.0155 \\ 
   \hline
\end{tabular}
    \end{minipage}
    \hfill
    \begin{minipage}{0.45\textwidth}
        \begin{tabular}{crrrr}
         \multicolumn{5}{c}{Level 1}\\
  \hline
Relative & $qCar$ & $qTruck$ & $vCar$ & $vTruck$ \\ 
  \hline
$\upsilon_{11}^k$ & 0.877 & 0.779 & 0.924 & 0.838 \\ 
  $\upsilon_{12}^k$ & 0.074 & 0.145 & 0.043 & 0.143 \\ 
  $\upsilon_{13}^k$ & 0.049 & 0.031 & 0.033 & 0.019 \\ 
  $\upsilon_{14}^k$ & -- & 0.019 & -- & -- \\ 
  $\upsilon_{15}^k$ & -- & 0.017 & -- & -- \\ 
  $\upsilon_{16}^k$ & -- & 0.008 & -- & -- \\ \hline
  $\sum_{M^k}\upsilon_m^k$ & 0.296 & 0.576 & 0.051 & 0.076 \\ 
   \hline
\end{tabular}
\\ \vspace{2em} \\
\begin{tabular}{crrrr}
 \multicolumn{5}{c}{Level 2}\\
  \hline
Relative & $qCar$ & $qTruck$ & $vCar$ & $vTruck$ \\ 
  \hline
$\upsilon_{01}^k$ & 0.624 & 0.773 & 0.727 & 0.723 \\ 
  $\upsilon_{02}^k$ & 0.176 & 0.109 & 0.132 & 0.125 \\ 
  $\upsilon_{03}^k$ & 0.078 & 0.083 & 0.084 & 0.083 \\ 
  $\upsilon_{04}^k$ & 0.057 & 0.024 & 0.027 & 0.035 \\ 
  $\upsilon_{05}^k$ & 0.044 & 0.011 & 0.020 & 0.023 \\ 
  $\upsilon_{06}^k$ & 0.020 & -- & 0.011 & 0.012 \\ \hline
  $\sum_{M^k}\upsilon_m^k$ & 0.060 & 0.821 & 0.046 & 0.073 \\ 
   \hline
\end{tabular}
    \end{minipage}
    \vspace{1em}
    \caption{Estimated univariate eigenvalues for the site-specific random process (level 1) and the site-year-specific random process (level 2). The tables on the left report the absolute values with the last row containing the sum of the estimated eigenvalues per dimension. The tables on the right report the corresponding percent of explained variation within the dimension and the last row contains the percent of explained variation per dimension over the multivariate functional process.}
    \label{APPApplicationTab:EstimatedEV}
\end{table}

\begin{table}
    \centering
    \begin{minipage}[t]{0.49\textwidth}
        \centering
        \begin{tabular}[t]{rrrr}
  \multicolumn{4}{c}{Level 1}\\
  \hline
 & Absolute & Relative & Cumulative \\ 
  \hline
$\nu_{1,1}$ & 15382.305 & 0.429 & 0.429 \\ 
  $\nu_{1,2}$ & 9432.398 & 0.263 & 0.692 \\ 
  $\nu_{1,3}$ & 5163.201 & 0.144 & 0.837 \\ 
  $\nu_{1,4}$ & 1820.428 & 0.051 & 0.887 \\ 
  $\nu_{1,5}$ & 1525.451 & 0.043 & 0.930 \\ 
  $\nu_{1,6}$ & 838.268 & 0.023 & 0.953 \\ 
  $\nu_{1,7}$ & 498.967 & 0.014 & 0.967 \\ 
  $\nu_{1,8}$ & 430.993 & 0.012 & 0.979 \\ 
  $\nu_{1,9}$ & 270.716 & 0.008 & 0.987 \\ 
  $\nu_{1,10}$ & 187.071 & 0.005 & 0.992 \\ 
  $\nu_{1,11}$ & 129.847 & 0.004 & 0.996 \\ 
  $\nu_{1,12}$ & 60.462 & 0.002 & 0.997 \\ 
  $\nu_{1,13}$ & 42.722 & 0.001 & 0.999 \\ 
  $\nu_{1,14}$ & 26.987 & 0.001 & 0.999 \\ 
  $\nu_{1,15}$ & 25.547 & 0.001 & 1.000 \\ 
   \hline
\end{tabular}
    \end{minipage}\hfill
    \begin{minipage}[t]{0.49\textwidth}
    \centering
        \begin{tabular}[t]{rrrr}
          \multicolumn{4}{c}{Level 2}\\
  \hline
 & Absolute & Relative & Cumulative \\ 
  \hline
$\nu_{0,1}$ & 28455.212 & 0.272 & 0.272 \\ 
  $\nu_{0,2}$ & 22257.918 & 0.212 & 0.484 \\ 
  $\nu_{0,3}$ & 11770.427 & 0.112 & 0.596 \\ 
  $\nu_{0,4}$ & 10453.869 & 0.100 & 0.696 \\ 
  $\nu_{0,5}$ & 7540.123 & 0.072 & 0.768 \\ 
  $\nu_{0,6}$ & 5301.006 & 0.051 & 0.819 \\ 
  $\nu_{0,7}$ & 4994.258 & 0.048 & 0.866 \\ 
  $\nu_{0,8}$ & 3057.358 & 0.029 & 0.896 \\ 
  $\nu_{0,9}$ & 2603.484 & 0.025 & 0.920 \\ 
  $\nu_{0,10}$ & 1846.931 & 0.018 & 0.938 \\ 
  $\nu_{0,11}$ & 1350.284 & 0.013 & 0.951 \\ 
  $\nu_{0,12}$ & 1269.752 & 0.012 & 0.963 \\ 
  $\nu_{0,13}$ & 963.401 & 0.009 & 0.972 \\ 
  $\nu_{0,14}$ & 893.100 & 0.009 & 0.981 \\ 
  $\nu_{0,15}$ & 756.275 & 0.007 & 0.988 \\ 
  $\nu_{0,16}$ & 384.172 & 0.004 & 0.992 \\ 
  $\nu_{0,17}$ & 352.843 & 0.003 & 0.995 \\ 
  $\nu_{0,18}$ & 317.335 & 0.003 & 0.998 \\ 
  $\nu_{0,19}$ & 169.490 & 0.002 & 1.000 \\ 
  $\nu_{0,20}$ & 30.391 & 0.000 & 1.000 \\ 
  $\nu_{0,21}$ & 8.652 & 0.000 & 1.000 \\ 
  $\nu_{0,22}$ & 0.038 & 0.000 & 1.000 \\ 
  $\nu_{0,23}$ & 0.031 & 0.000 & 1.000 \\
   \hline
\end{tabular}
    \end{minipage}
    \caption{Estimated multivariate eigenvalues for the site-specific random process (level 1, $\nu_{1,m}, m = 1,...,15$) and the site-year-specific random process (level 2, $\nu_{0,m}, m = 1,...,23$) for an \gls{mfpca} with weighted scalar product.}
    \label{APPApplicationtab:EstimatedMEVweights}
\end{table}

\begin{table}
    \centering
    \begin{minipage}[t]{0.49\textwidth}
        \centering
        \begin{tabular}[t]{rr}
          \multicolumn{2}{c}{Level 1}\\
  \hline
 & Relative \\ 
  \hline
$\hat{\nu}_{1,1}$ & 0.032 \\ 
  $\hat{\nu}_{1,2}$ & 0.404 \\ 
  $\hat{\nu}_{1,3}$ & 0.385 \\ 
  $\hat{\nu}_{1,4}$ & 0.040 \\ 
  $\hat{\nu}_{1,5}$ & 0.067 \\ 
  $\hat{\nu}_{1,6}$ & 0.019 \\ 
  $\hat{\nu}_{1,7}$ & 0.008 \\ 
  $\hat{\nu}_{1,8}$ & 0.025 \\ 
  $\hat{\nu}_{1,9}$ & 0.020 \\ 
   \hline
\end{tabular}
    \end{minipage}\hfill
    \begin{minipage}[t]{0.49\textwidth}
    \centering
        \begin{tabular}[t]{rr}
          \multicolumn{2}{c}{Level 2}\\
  \hline
 & Relative \\ 
  \hline
$\hat{\nu}_{0,1}$ & 0.084 \\ 
  $\hat{\nu}_{0,2}$ & 0.068 \\ 
  $\hat{\nu}_{0,3}$ & 0.059 \\ 
  $\hat{\nu}_{0,4}$ & 0.058 \\ 
  $\hat{\nu}_{0,5}$ & 0.431 \\ 
  $\hat{\nu}_{0,6}$ & 0.017 \\ 
  $\hat{\nu}_{0,7}$ & 0.018 \\ 
  $\hat{\nu}_{0,8}$ & 0.074 \\ 
  $\hat{\nu}_{0,9}$ & 0.017 \\ 
  $\hat{\nu}_{0,10}$ & 0.047 \\ 
  $\hat{\nu}_{0,11}$ & 0.010 \\ 
  $\hat{\nu}_{0,12}$ & 0.030 \\ 
  $\hat{\nu}_{0,13}$ & 0.031 \\ 
  $\hat{\nu}_{0,14}$ & 0.056 \\ 
   \hline
\end{tabular}
    \end{minipage}
    \caption{Relative contributions of the estimated posterior score variances to the total (posterior unweighted) variation for the site-specific random process (level 1, $\nu_{1,m}, m = 1,...,9$) and the site-year-specific random process (level 2, $\nu_{0,m}, m = 1,...,14$) based on an \gls{mfpca} with scalar product with unequal weights.}
\label{APPApplicationtab:EstimatedMEVPost}
\end{table}

\end{document}